\documentclass[12pt, oneside]{article} 	% use "amsart" instead of "article" for AMSLaTeX format
\usepackage[left=1in,right=1in,top=1in,bottom=1in]{geometry} 		% See geometry.pdf to learn the layout options. There are lots.
\usepackage{graphicx}				% Use pdf, png, jpg, or epsÂ§ with pdflatex; use eps in DVI mode
\usepackage{amssymb,amsmath,bm,accents}
\usepackage{csquotes}
\usepackage{setspace}
\usepackage[colorlinks=true,linkcolor=blue,citecolor=blue]{hyperref}
\usepackage{natbib}
\usepackage{subcaption}
\usepackage{multirow}
\usepackage{tabularx}
\usepackage{adjustbox}
\usepackage[justification=centering]{caption}
\usepackage{fancyhdr}
\usepackage{clipboard}

\renewcommand{\small}{\fontsize{10pt}{11pt}\selectfont}

\title{\bf{New Facts and Data about \\ Professors and their Research}}
\author{
\begin{tabular}{c c c c}
Kyle Myers\thanks{Harvard Business School} \thanks{D$^3$ Institute and the Laboratory for Innovation Science at Harvard} & Wei Yang Tham$^*$$^\dagger$ & Jerry Thursby$^\dagger$ & Marie Thursby$^\dagger$\thanks{Georgia Institute of Technology} \\
Nina Cohodes$^\dagger$ & Karim Lakhani$^*$$^\dagger$ & Rachel Mural\thanks{Harvard Kennedy School \\ Corresponding author: \href{mailto:kmyers@hbs.edu}{kmyers@hbs.edu}. This project received financial support from The Alfred P. Sloan Foundation and the Harvard Business School and excellent research assistance from Kate Powell. We are grateful for comments from Teresa Amabile, Song Ma, Peter Schiffer, participants of the Innovation Information Initiative,  and many others.} & Yilun Xu$^\dagger$ 
\end{tabular}
}
\date{\today \\ \-\\ $[$\emph{Draft --- comments welcome}$]$}

\begin{document}
\maketitle 
\begin{abstract}\normalsize
We introduce a new survey of professors at roughly 150 of the most research-intensive institutions of higher education in the US. We document seven new features of how research-active professors are compensated, how they spend their time, and how they perceive their research pursuits: (1) there is more inequality in earnings within fields than there is across fields; (2) institutions, ranks, tasks, and sources of earnings can account for roughly half of the total variation in earnings; (3) there is significant variation across fields in the correlations between earnings and different kinds of research output, but these account for a small amount of earnings variation; (4) measuring professors' productivity in terms of output-per-year versus output-per-research-hour can yield substantial differences; (5) professors' beliefs about the riskiness of their research are best predicted by their fundraising intensity, their risk-aversion in their personal lives, and the degree to which their research involves generating new hypotheses;  (6) older and younger professors have very different research outputs and time allocations, but their intended audiences are quite similar; (7) personal risk-taking is highly predictive of professors' orientation towards applied, commercially-relevant research.%An anonymous version of the individual-level responses are publicly available at: \href{TBD}{[data preparation still in progress]}.
\end{abstract}
\thispagestyle{empty}
%%%%%%%%%%%%%%%%%%%%%%%%%%%%%%%%%%%%%%%%%%%%%%%%%%%%%%%%%%%%%%%%%%%%%%
%----------------------------------------------------------------------------------------%

%%%%%%%%%%%%%%%%%%%%%%%%%%%%%%%%%%%
%----------------------------------------------------------------------------------------%
\clearpage
\setcounter{page}{1}
\onehalfspacing
%\doublespacing
%\setlength{\parskip}{4pt}
%\setlength{\parindent}{1em}

\section{Introduction}
%----------------------------------------------------------------------------------------%
%%%%%%%%%%%%%%%%%%%%%%%%%%%%%%%%%%%
Researchers are at the core of economic growth. The quantity and quality of their new ideas are what enable technological change (\citealt{jones2022past}). In the U.S., the most common employer of Ph.D.-level researchers is academic institutions,\footnote{According to the 2021 National Science Foundation Survey of Doctoral Recipients: 42\% of Ph.D. scientists and engineers work at educational institutions; 37\% work at private, for-profit businesses; 15\% work at non-profit or government agencies; and the remainder are employed by other organization types or are self-employed (\citealt{nsf2021sdr}).} and a long body of evidence has documented the importance of academic science for industrial progress.\footnote{See, for example, \cite{cohen2010fifty} or \cite{perkmann2021academic} for excellent reviews on this topic.} Amidst recent concerns of declining R\&D productivity (\citealt{jones2009burden,bloom2017ideas}) and an increasing division of labor between industrial firms and professors in academia (\citealt{arora2020changing}), it has become increasingly important to understand professors' roles in the scientific workforce.

Once dominated by philosophers (e.g., \citealt{popper1934logic,kuhn1962structure}), the ``science of science'' has become increasingly populated by quantitative analyses by economists, sociologists, physicists, and other scholars looking to turn their empirical tool-kits inwards (\citealt{azoulay2018toward,fortunato2018science}). However, most of these empirical studies make use of a small number of data sources that are byproducts of conducting science, which were often not designed as intentional sources of data to study science. Inputs are usually measured using federal grant databases, which covers only a fraction of funding flows and leaves unobserved one of the most important input in the scientific production function, researchers' time. Outputs are usually measured using publication databases, which generates serious measurement error for any researcher whose output is not codified in print and makes across-field comparisons difficult to interpret given the range in publication norms. Many other important aspects of researchers' work and lives are left only partly visible at best: their professional position (e.g., tenure status, administrative duties); their incentives (e.g., sources of income); and the objectives of their research (e.g., their intended audience).\footnote{Still, impressive efforts are still underway to improve the fidelity and interoperability of existing meta-science datasets (e.g., \citealt{marx2020reliance,lin2023sciscinet}, UMETRICS at \href{https://iris.isr.umich.edu/}{https://iris.isr.umich.edu/}, and the Innovation Information Initiative at \href{https://iii.pubpub.org/datasets}{https://iii.pubpub.org/datasets}).}

In this paper, we document a new nationally representative survey of research-active professors at roughly 150 of the largest institutions of higher education in the US. The population and sample includes professors from all fields of science, broadly construed: engineering, math, and related fields; humanities and related fields; medicine and health; natural sciences; social sciences. The survey instrument includes a number of novel elements related to professors' rank and tenure status, time use, funding, salaries and sources thereof, the nature of their research, and a battery of socio-demographic and household-related factors. For the majority of our analyses, we focus on professors who report a non-zero amount of their time being spent on research activities (95\% of respondents).

Our approach follows a long line of prior work that has used surveys of academic researchers to uncover their otherwise unobservable features, choices, preferences, or beliefs (e.g., \citealt{levin1991research,kahn1993gender,ginther1999gender,fox2001careers,thursby2002selling,stern2004scientists,fleming2004science,walsh2007collaboration,sauermann2010makes,roach2010taste,walsh2015bureaucratization,curty2017attitudes,levecque2017work,shortlidge2018trade,ganguli2019postdocs,cohen2020not,philipps2022research}). However, unlike many prior surveys of researchers, a guiding principle of our effort was ``breadth over depth'' such that many design choices reflect an objective of shedding new light on features of this market that have been largely ignored by empiricists. Our hope is that the summary statistics and correlations in this survey will spark more detailed, focused, and rigorous investigations into the causal effects underlying the patterns we see.

Before reporting the major findings of the survey, we describe the population, sampling methodology, recruitment protocol, and summary statistics. To test for representativeness, we use multiple dimensions of data that are observable for respondents and non-respondents. At the institutional level, we use data from the National Science Foundation's HERD survey (\citealt{nsf2023herd}) to show that respondents come from institutions that receive relatively equal amounts of research funding from a variety of sources. At the individual level, we use data from publication and grant records to show a high degree of similarity in these metrics between respondents and non-respondents. The only key difference between our sample and the population is an under-response from professors at medical schools, which we discuss further below. We sometimes observe other statistically significant degrees of non-response bias, but the practical magnitudes are often relatively small and on the scale of 5 percent of means. Additionally, as one test of attention and honesty, we match a sub-sample of respondents to their publicly reported salaries and find a high degree of alignment.

We document seven key new findings about the research-active professor workforce:
\begin{itemize}
\item[] \hyperlink{sec:figssalfield}{Finding \ref{figs_sal_field}}: \Copy{clipboard:finding1}{There is much more inequality in earnings within fields than across. This holds true at the household level because professors exhibit positive assortative matching.}

\item[] \hyperlink{sec:figssaldecompose}{Finding \ref{figs_sal_decompose}}: \Copy{clipboard:finding2}{Differences in institutions, faculty ranks, tasks, and sources of earnings can account for roughly half of the variation in earnings across professors.}

\item[] \hyperlink{sec:figsoutputsaldecompose}{Finding \ref{figs_outputsal_decompose}}: \Copy{clipboard:finding3}{There are significant differences across fields in the implied payoff to producing observable research output (e.g., earnings per publication). However, research output and payoff differences can only account for a small amount of the variation in earnings across professors.}

\item[] \hyperlink{sec:figstimecorr}{Finding \ref{figs_timecorr}}: \Copy{clipboard:finding4}{Professors with higher gross output (i.e., annual publications) are not always more productive on a per-research-hour basis because of substantial variation in professors' time allocations. This is especially true for non-tenure-track professors.}

\item[] \hyperlink{sec:figscorrrisk}{Finding \ref{figs_corr_risk}}: \Copy{clipboard:finding6}{Three of the strongest predictors of professors' beliefs about the riskiness of their research are: (1) the share of their time spent fundraising; (2) their risk-taking in their personal lives; and (3) their orientation towards generating new hypotheses with their research (as opposed to testing hypotheses).}

\item[] \hyperlink{sec:figsnrage}{Finding \ref{figs_nrage}}: \Copy{clipboard:finding5}{Older professors have different intended research outputs than younger professors (i.e., focusing on books as opposed to journal articles), but their intended audiences are the same. Administrative duties exhibit a rise and fall over professors' careers, often with a discontinuous increase after receiving tenure, which can explain a large fraction of the decline in research hours post-tenure.}

\item[] \hyperlink{sec:figscorrbescore}{Finding \ref{figs_corr_bescore}}: \Copy{clipboard:finding7}{Professors' position on the basic--applied spectrum can be proxied with the intended output and audience of their research; more applied ``Edison-like'' professors, whose output is more likely to be tools and products and whose audience is more likely to be businesses and policymakers, report a higher willingness to take risks in their personal lives.}
\end{itemize}

We do not report any causal effects in this paper, taking all equilibrium correlations as being representative of some combination of treatment and selection effects. Furthermore, the data currently exists only as a cross-section. Thus, variation across professors of different ages reflects both temporal dynamics as well as changes to the composition of this workforce.

In some cases, we report decompositions based on the $R^2$ and partial-$R^2$ statistics based on simple linear models.\footnote{The partial-$R^2$ for a given (independent) variable is the proportion of variation (in the dependent variable) explained by that variable that cannot be explained by the other (independent) variables.} Our goal with these exercises is to determine how much the variation in the focal outcome can be described by the covariates, which indicates the extent to which any treatment or selection effects are important along those dimensions. We also report the results of ``observational regressions'', a term which we will use to describe regressions of professors' features on a set of possibly endogenous covariates. Similarly, we sometimes employ ML-based covariate selection methods to identify variables with predictive power (e.g., \citealt{meinshausen2010stability}). The broad patterns that emerge provide useful views of equilibrium relationships and motivate new hypotheses about the incentives facing academic researchers.

An anonymous version of a subset of the individual-level data is publicly available here: \href{TBD}{[data preparation still in progress]}. We hope the results reported in this paper, and the public data, will spark further investigation into the academic research workforce. The rest of the paper is organized as follows: Section \ref{sec_methodology} describes the survey methodology and some summary statistics; Section \ref{sec_findings} walks through our key new findings; Section \ref{sec_discuss} concludes with a discussion.

%%%%%%%%%%%%%%%%%%%%%%%%%%%%%%%%%%%
%----------------------------------------------------------------------------------------%
\section{Methodology: Population, sampling, and survey}\label{sec_methodology}
%----------------------------------------------------------------------------------------%
%%%%%%%%%%%%%%%%%%%%%%%%%%%%%%%%%%%

%----------------------------------------------------------------------------------------%
\subsection{Population and sampling}
%----------------------------------------------------------------------------------------%
Our target population is US professors who conduct research at major institutions of higher education. We identify this population by selecting the 158 largest institutions in the U.S. per their total R\&D funding reported in the National Science Foundation's 2019 Higher Education R\&D (HERD) survey (\citealt{nsf2023herd}). We hired individuals to manually collect the emails of professors from these universities’ websites.\footnote{See Appendix \ref{sec_app_extramethods} for more.} We identified these individuals as people listed on institution's website with the word ``professor'' in their title, recording their title as well as information on each professor's name, program and/or department and/or college, and professorial rank. Our requirement of the word ``professor'' in the title was driven by the logistics of data collection.\footnote{It is a simple, observable feature to rule identifying individuals in and out of population. Anything beyond this proved too complicated in our data collection process. An important question is to what extent we miss relevant individuals without this moniker. We cannot arrive at a conclusive estimate here since there is no clear definition of relevance with which to benchmark. As one potential benchmark, data from the US National Center of Education Statistics, indicates that only a few percentage points of these institutions' full-time instructional staff are not considered ``faculty'' (\citealt{doe2021nces}).} Table \ref{tab_pop_sumstat} reports summary statistics for the institutions included in the population based on variables sourced from websites as well as the HERD survey. Appendix Figure \ref{fig_disinstrank_hists} illustrates some joint distributions of fields, institutions, and ranks showing a significant amount of heterogeneity in the organizational structures.

Our sampling process was as follows. Based on the information gathered, we classified these emails into one of twenty fields of study and one of four ranks (assistant, associate, full or emeritus, and adjunct or other). We then sent an e-mail invitation to a randomly-selected half of the e-mails within each field-rank cell. These e-mails were distributed from October 2022 to March 2023.

The population consisted of 262,343 unique e-mails. We e-mailed a total of 130,785 individuals and 4,357 (3.33\%) completed the survey.\footnote{This response rate is more than twice what has been obtained from sourcing academic researcher contacts from the corresponding author data contained within the publication record (e.g., \citealt{myers2020unequal}).} We then restrict the sample to the 4,357 individuals (100.0\% of respondents) who reported being a professor and spending a non-zero amount of time on research. Our final sample consists of professors from engineering, math and related sciences (733), the natural sciences (680; e.g., biology, chemistry, physics), social sciences (889; e.g., economics, political science, psychology, sociology), humanities and related science (818; e.g., art, history, education, linguistics), and health or medical sciences (1,237; e.g., schools of medicine or public health).
These five aggregate groupings of fields were chosen partly based on the results of a principal component analysis to identify fields where professors responded to the survey similarly. Appendix Table \ref{tab_aggregationpca} reports the results of a single-component PCA based on the entire survey, averaging at the field level. We then aggregated fields together by primarily relying on this score, with some minor adjustments to align with our understanding of these fields.\footnote{For instance, we group Medical School based professors with those in other Medicine- or Health-related fields, and we also assign sociology to the aggregate field of social science despite the market difference in average PCA score for this field.} We sometimes use these five aggregate groupings of fields in our discussion and empirical analyses given the small sample sizes within the narrower field definitions.

%----------------------------------------------------------------------------------------%
\subsection{Potential survey biases}
%----------------------------------------------------------------------------------------%
Ideally, our respondents would report all answers accurately and their responses would reflect the preferences and characteristics of the full population. We cannot formally test this, but we can take some steps to investigate the possibility of inattention and non-response bias and, in the case of non-response bias, possibly account for it. 

As a test of researchers' attention and their willingness to report truthfully, we can compare their self-reported salaries to their publicly-reported salaries for the subset of researchers at institutions that make such data public. To do so, we manually traced respondents at 89 institutions with public salary data to their records in these public sources. During this match, we used our data on researchers' e-mail addresses and institutional affiliations to maximize fidelity of the match. Still, there is likely still nonzero measurement error due to both (1) manual errors in the name merging process, and (2) our inability to perfectly confirm that the self- and publicly-reported salaries were referring to the same year of employment. Appendix Figure \ref{fig_salcompare_pub_slf} plots the relationship between these two sources of salary data. The correlation between the two is 0.75, and for roughly 75\% of observations the difference between the self- and publicly-reported salary is less than 30\%. This suggests the vast majority of respondents are responding truthfully along this dimension.

We use two auxiliary data sources to compare respondents to the population in terms of observable variables. First, we use our internally collected data on professors' fields and ranks to test for differences between the respondent sample and the population. In Appendix \ref{sec_app_extramethods_popsampcompare}, we show that our respondents are slightly more likely to be full professors and less likely to be adjunct, clinical or other professors compared to the population. We also see a significant under-response from the medical and health sciences relative to other fields that is sizeable. Overall, this suggests that the results reported here may be less generalizable to the full spectrum of professors across medical schools or those in adjunct or clinical professor tracks.

Second, in Appendix \ref{sec_app_extramethods_popsampcompare}, we also compare our respondents to the  population according to the HERD survey, which reports on the amounts of R\&D funding flowing to each of the institutions in our population. When we examine measures including total funding amounts, funding by source, and by type, we do find some evidence that professors who complete our survey are located at institutions with lower-than-average amounts of funding. However, graphical illustrations and statistical tests of these comparisons show that the difference is relatively small in economic magnitude. We have good representativeness over the full distribution, and the average difference in funding amounts between the institutions of respondents and non-respondents is generally in the range of 4--6\%. 

We are also able to merge roughly two-thirds of the population to their records in the Dimensions database (\citealt{dimension2018data}) using a fuzzy, name- and affiliation-based merging process. This allows us to compare individual-level grant input and publication output data across our population and respondents. This exercise is reported in Appendix \ref{sec_app_extramethods_popsampcompare}. Again, we find economically small differences between the full population and respondents, most of which are not statistically significant. Our respondents' average publication output rates, field-normalized citation rates, and grant receipt rates and amounts are all within a few percentages of what is observed in the full population. Likewise, graphical investigations show strong overlap in the support of these variables. Overall, our respondent sample appears very similar to the population along many observable dimensions.

%----------------------------------------------------------------------------------------%
\subsection{Sample summary statistics}
%----------------------------------------------------------------------------------------%
Table \ref{tab_samp_sumstat} reports summary statistics for the key questions in the survey, which are documented in further detail in Appendix \ref{sec_app_instrument}. The majority of respondents are full professors (40\%) followed by roughly equal proportions of assistant (25\%) and associate professors (25\%) with the remainder being adjunct, clinical, or other types of professors (10\%). The distribution across aggregate fields is relatively even, although we imposed these field groupings in a way that sought approximate balance in sub-sample sizes.\footnote{See Appendix Table \ref{tab_field_sumstat} for the groupings of the twenty narrower fields into these aggregate fields. The classification of fields into these broader groupings is chosen out of simplicity and with an understanding that it would be impossible to please all professors in terms of how the groups are constructed.} Most respondents are tenured (57\%), with another 21\% still on the tenure-track and 22\% not on the tenure track at all. The average number of years since tenure for those who are tenured is approximately 15 years. For those not on the tenure track, the average contract length is about 2.5 years, and for those pre-tenure, the average years until their tenure evaluation is about 2.6 years.

On average, individuals expect to work roughly 50 hours per week, which is nearly eight hours more than the national average for full-time workers (\citealt{bls2023atus}).\footnote{Specifically, respondents are asked to report their expected time use (i.e., total hours per week and allocations across different tasks) over the coming five years. This is done in order to solicit responses that approximate the short-run steady-state of professors' time use, and was motivated by pilot studies of our time-use questions.} The average annual income is approximately \$150,000 (s.d.=\$90,000), which is approximately the 90$^{th}$ percentile in the US (\citealt{census2023cps}).

When asked to report how much guaranteed research funding professors expect to have access to over the coming five years, professors report \$85,000 per year on average with a relatively large standard deviation (\$200,000 per year). The same is true of researchers' expected fundraising amounts, reporting funding amounts of same order of magnitude as their guarantees. 

The majority of the individuals report being White (79\%), followed by Asian (12\%), with these groups being slightly (Whites) and substantially (Asians) over-represented relative to their shares of the full US population (\citealt{census2023cps}; note: we allowed respondents to report multiple race/ethnicities). The representation of Black, Hispanic, and other ethnicities is relatively low at 3\%, 6\%, and 4\% respectively. Nearly 25\% of professors are non-US born. This is nearly double the rate of the full US population (\citealt{census2023cps}), which is yet another signal of the importance of immigration for the US research enterprise (e.g., \citealt{kerr2020gift}).

The survey also includes a battery of questions related to the nature of professors' research (see Appendix \ref{sec_app_instrument}). First, respondents are asked to rate on a scale of 0 to 10 whether their research is more about generating (0) or testing (10) hypotheses. In addition, respondents are asked to report the intended outputs and audiences for their research.\footnote{Options for intended output are: publications, books, tools (e.g. data, software, instruments), or practical applications (products, patents, policies). Options for intended audience are: other academics, policymakers, businesses, or the general public. Respondents whether each of these options is their intended audience or output ``Never or rarely'', ``Sometimes'', or ``Most or all of the time''.} Table \ref{tab_nat_sumstat} reports the summary statistics for these variables describing the nature of professors' research.

%%%%%%%%%%%%%%%%%%%%%%%%%%%%%%%%%%%
%----------------------------------------------------------------------------------------%
\section{Findings}\label{sec_findings}
%----------------------------------------------------------------------------------------%
%%%%%%%%%%%%%%%%%%%%%%%%%%%%%%%%%%%

\begin{figure}[htbp]\centering\footnotesize
\caption{Individual and household earnings across fields}
\label{figs_sal_field}
\subfloat[Own earnings]{
\includegraphics[width=0.475\linewidth, trim = 0mm 0mm 0mm 0mm , clip]{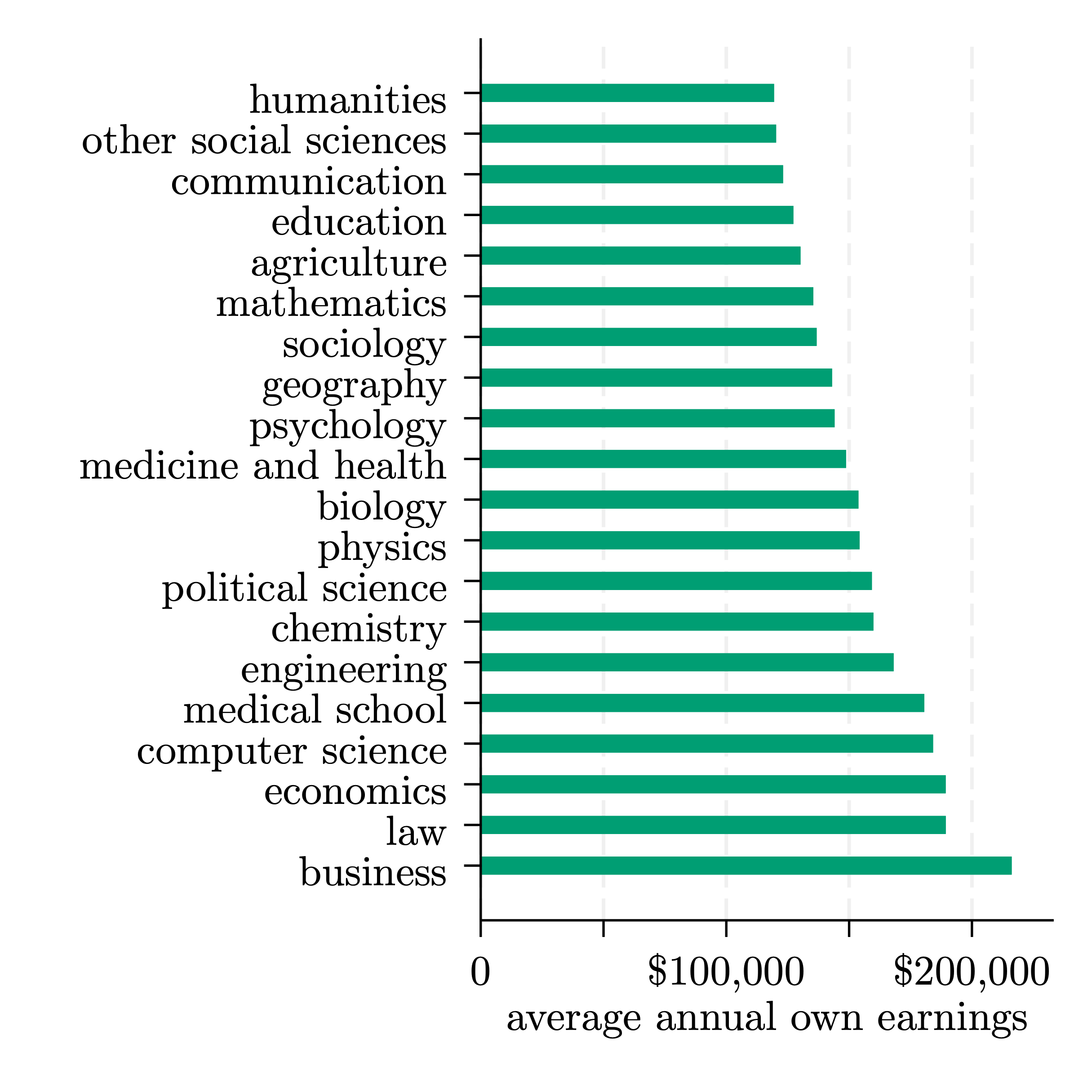}}
\subfloat[Inequality in own earnings]{
\includegraphics[width=0.475\linewidth, trim = 0mm 0mm 0mm 0mm , clip]{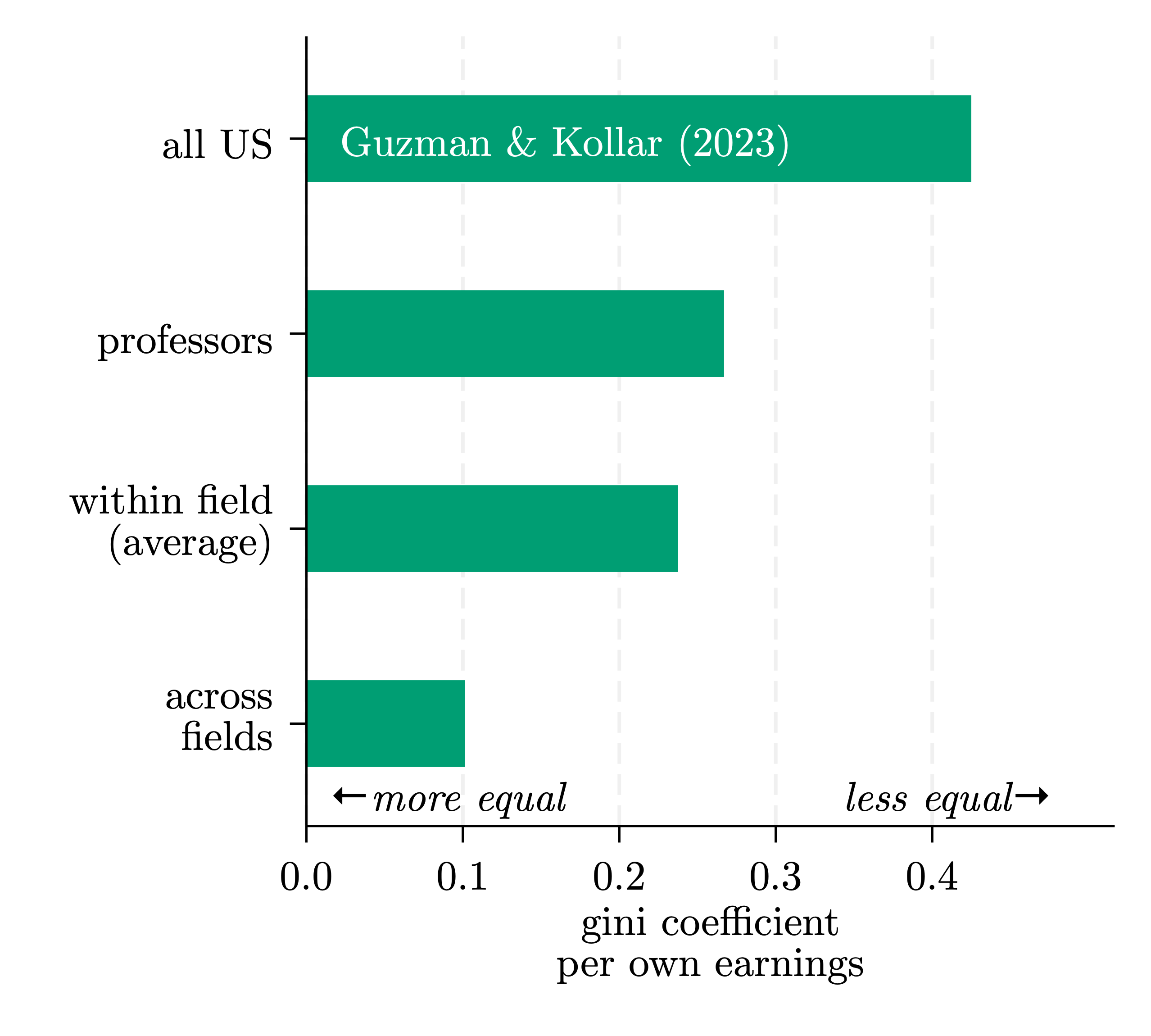}}\\
\-\\ \-\\
\subfloat[Own and rest-of-household earnings]{
\includegraphics[width=0.475\linewidth, trim = 0mm 0mm 0mm 0mm , clip]{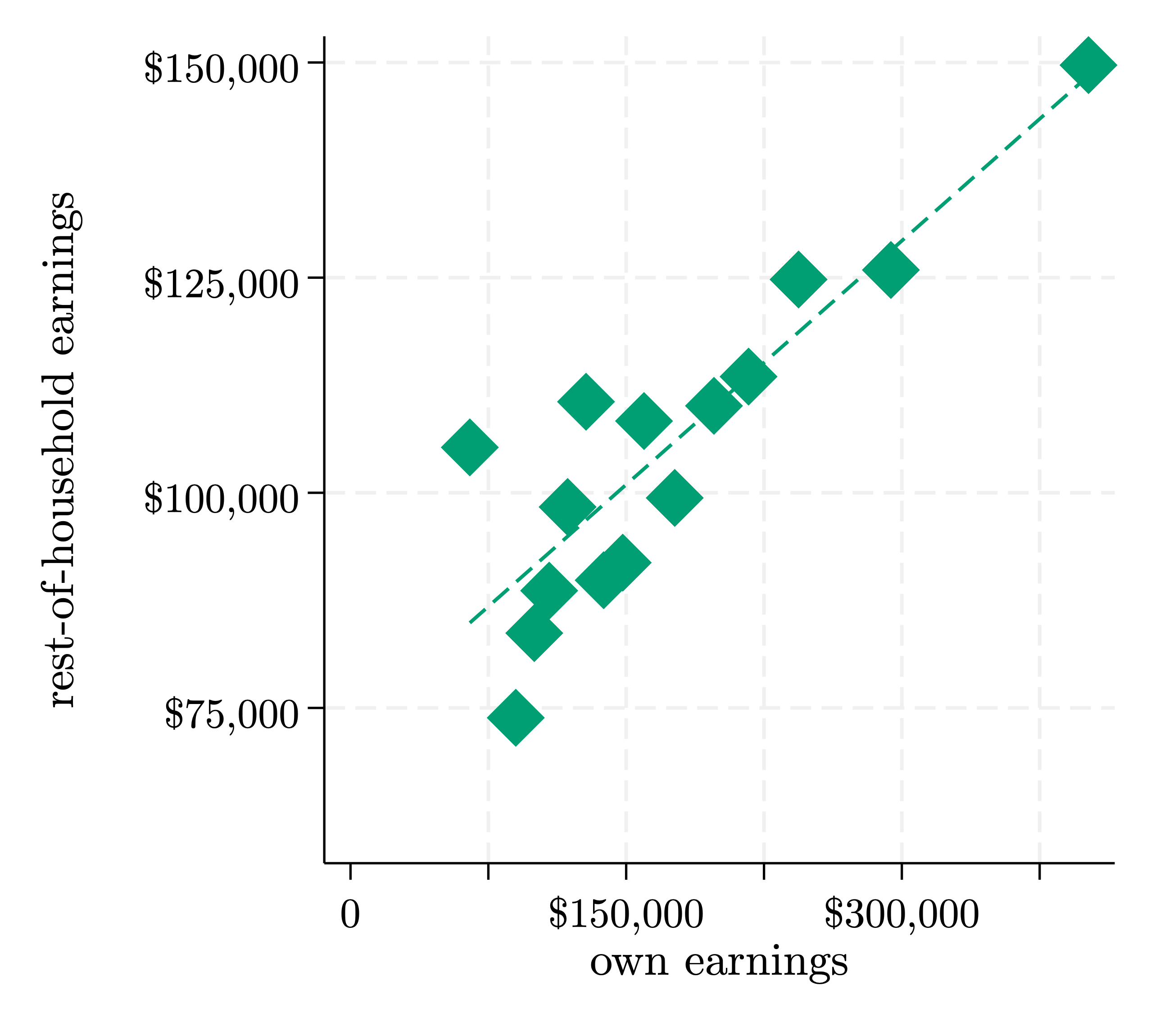}}
\subfloat[Inequality in household earnings]{
\includegraphics[width=0.475\linewidth, trim = 0mm 0mm 0mm 0mm , clip]{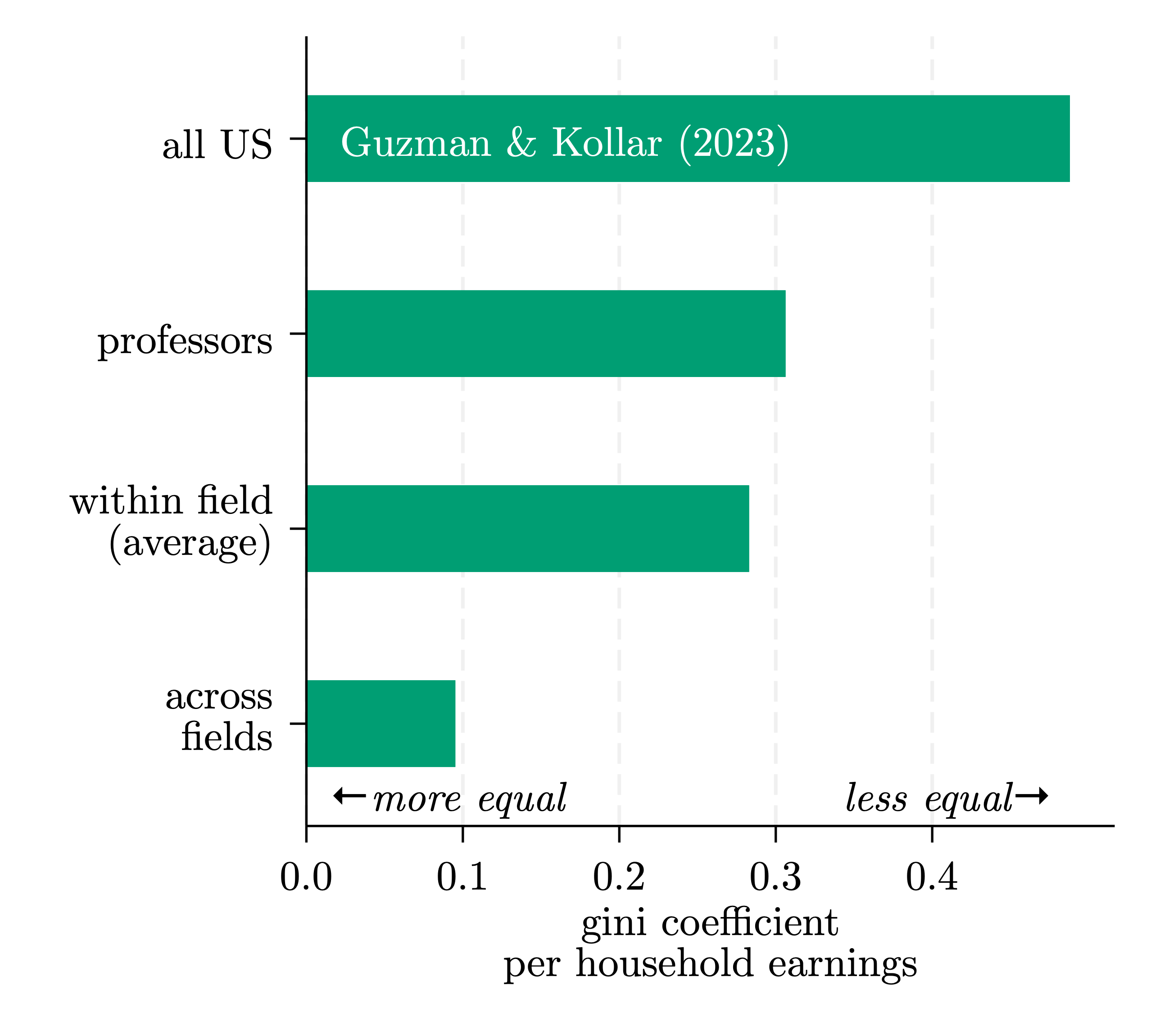}}\\
\begin{quote}
\emph{Note}: \input{figtab/stub_figs_sal_field_note1.tex}
\end{quote}
\end{figure}

%%%%%%%%%%%%%%%%%%%%%%%%%%%%%
\hypertarget{sec:figssalfield}{}
\begin{quote}
\textbf{Finding \ref{figs_sal_field}}: \emph{\Paste{clipboard:finding1}}
\end{quote}

Variation within and across fields in academic researchers' earnings is a regular conversation topic within the halls of most universities. But the difficulties of systematically matching researchers' fields to their earnings has limited investigations into this variation except in specific cases (e.g., \citealt{mohanty1986faculty,hamermesh1988salaries,bellas1997disciplinary,ehrenberg1998economics,baker2023pay}).\footnote{For some prior work on professors' earnings, with results that are relevant to this and other findings in this paper, see: \cite{tuckman1976structure,marsh1980academic,mohanty1986faculty,fairweather1993faculty,langton1994paying,hearn1999pay,perna2002sex,ehrenberg2006field,melguizo2007faculty}.} Furthermore, most existing data on researchers' earnings leave the remainder of the household (e.g., spouses' earnings) untouched despite the importance of the household as an economic unit (\citealt{chiappori2017static}).\footnote{We're referring to pre-tax annual earnings of any sources here, decomposing the sources of earnings below. For simplicity, the survey question that solicited earnings did not belabor the distinction between ``earned'' and ``non-earned'' incomes, which leaves open the possibility that some respondents reported non-earned income. Still, such income likely would've been reported via the ``other'' source category, which accounts for only 5\% of reported earnings on average.} 

Figure \ref{figs_sal_field} Panel (a) reports field-level average self-reported total annual earnings.\footnote{Figure \ref{figs_sal_field} reports post-shrinkage means using the empirical Bayes shrinkage methodology of  \cite{chandra2016productivity} in order to adjust for differences in sub-sample sizes across fields.} Professors' individual earnings span lower values of approximately \$130,000 per year in fields such as the humanities, communication, agriculture, and education. Earnings in the highest-paying fields, economics, medicine, law, and business are roughly \$200,000 per year.

How much variation in earnings do these field-level averages hide? To get a sense as to the variation in earnings both across and within fields, Panel (b) of Figure \ref{figs_sal_field} plots three alternative individual-level Gini coefficients for the sample and, for comparison, the Gini coefficient for the full US population of full-time workers per recent estimates from the US Census Bureau (\citealt{guzman2023income}). The overall Gini coefficient for professors' own earnings is roughly 0.27, which is significantly lower than the 0.43 observed across all US workers (\citealt{guzman2023income}) and is comparable to countries such as Belgium, the Netherlands, and Iceland (\citealt{worldbank2023gini}). Appendix Figure \ref{fig_incomegini_extra} reports additional views of earnings variation. Interestingly, there is very little variation in the within-field Gini coefficient across fields, and there is no significant relationship between field-level average earnings and within-field inequality.

To compare inequality in earnings within fields, we estimate a separate Gini coefficient for each field and report the average of these within-field estimates. To compare inequality in earnings across fields, we take the average salary within each field (held constant at associate professor rank) and calculate the Gini coefficient assuming there is one representative professor in each field who earns this (average) amount. Inequality in professors' earnings is much higher when focusing on the within-field variation (average Gini coef.=0.24, s.d.=0.02) compared to the variation in field-level averages (Gini coef.=0.10).

As evidenced by the strong positive relationship between individual and rest-of-household earnings for those with partners (Figure \ref{figs_sal_field} Panel c), there is clear evidence of positive assortative matching amongst research professors.\footnote{Amongst professors with partners, we estimate that each additional \$10 of a professor's own earnings is associated with approximately \$1.5 additional dollars in rest-of-household earnings.} To illustrate the role of this assortative matching (and the role of multi-earner households more generally) on household-level earnings variation, Figure \ref{figs_sal_field} Panel (d) reports the same Gini coefficients as Panel (b), this time based on household-level earnings. As expected given the positive matching, the pattern remains the same. Compared to the US population of households (Gini coef.=0.49), earnings across professors' households is more equal (Gini coef.=0.31) with most of the variation driven by earnings differences within fields.\footnote{The differences between household- and individual-level Gini coefficients in the full US population and professors is relatively similar on the scale of 5 p.p. and 15\%. This suggests the positive assortative matching observed amongst professors is similar to that observed in the full population (\cite{greenwood2014marry}).} 

There is much more inequality in earnings within fields than there is across fields, a fact that is of both inherent and policy relevance to the market for academic research labor. In many of the following analyses, we attempt to identify some of the sources of this variation based on how professors spend their time and what outputs they produce.

%%%%%%%%%%%%%%%%%%%%%%%%%%%%%
\hypertarget{sec:figssaldecompose}{}
\begin{quote}
\textbf{Finding \ref{figs_sal_decompose}}: \em{\Paste{clipboard:finding2}}
\end{quote}

Understanding the different incentives researchers face is key to understanding how they allocate their time. How much of the variation in professors' earnings can be explained by observable differences in their work? The rarity of jointly observing professors' time use across their many tasks alongside their earnings has limited our ability to investigate these issues.\footnote{There are numerous investigations into the determinants of professors' salaries (e.g., \citealt{siegfried1973teaching,delorme1979analysis,fairweather2005beyond,allgood2013economists,gibson2014journal}). However, most of these analyses are limited to a single scientific field and/or cannot specifically isolate inputs (e.g., time allocations) and outputs (e.g., articles published) of professors work. In Finding \#2 here, we focus on these inputs, and in the next finding we focus on their (research-oriented) outputs.} As evidenced in Appendix Figures \ref{fig_disinstrank_hists} and \ref{figs_app_extraresults_shrfields}, there is considerable heterogeneity in professors' ranks, in how they spend their time, and in the sources of their earnings. 

To better understand earnings differences, we first estimate rank- and institution-specific average earnings. The average annual earnings by rank are as follows: assistant, 126,485 (s.d.= 65,895); associate, 136,059 (s.d.= 67,665); full, 200,596 (s.d.= 102,001); other, 77,511 (s.d.= 82,356).
 Figure \ref{figs_sal_decompose} Panel (a) shows that there is considerable variation in average earnings across institutions. There is some right skew to this distribution, but it is much less skewed than the distribution of firm-level average earnings in the US economy. Here, the 75$^{th}$:25$^{th}$ percentile ratio is roughly 1.3 and the 90$^{th}$:10$^{th}$ percentile ratio is roughly 1.8. In the broader economy, these ratios tend to be closer to 2 and 5, respectively (\cite{song2019firming}). Notably, these are not adjusted for any observable differences across professors.

Next, we estimate observational regressions that relate professors' earnings to different features of their work. First, we regress their earnings on their time spent on six different tasks (research, fundraising, teaching, administration, clinical, or other). Next, we regress their total earnings on the source of their earnings (base salaries, grant-covered, supplemental earnings from their primary institution, clinical work, or other). These regressions are of the form: $Y_{i}=\alpha + \sum_j X^j_{i} \beta^j + \epsilon_i$, where $Y_i$ is professor $i$'s earnings and $X^j_{i}$ is the vector of either the professor's hours spent on each tasks indexed by $j$, or the percentage points of the professor's earnings due to each source indexed by $j$. The results of these two regressions (the estimates of $\beta^j$) represent the implied marginal wages of each task (or income source) holding the amount of time spent on the other tasks (or share of income from other sources) fixed. Figure \ref{figs_app_extraresults_shrfields} illustrates the variation in these metrics across fields.

There are substantial differences in the implied returns to different tasks (Figure \ref{figs_sal_decompose}, Panel b). Clinical work is associated with the most earnings, nearly \$75 per hour. This is to be expected since most ``clinical'' work performed by professors involves medical care delivery at academic medical systems. Research, fundraising, administration, and ``other'' tasks have implied marginal wages of roughly \$25--50 per hour. Additional time spent on teaching activities has no statistically significant association with earnings (and the point estimate is negative). Clearly, this partly reflects a selection effect whereby positions with larger teaching responsibilities also have lower salaries. Still, this observational regression makes the importance of this selection effect very clear.

Overall, Figure \ref{figs_sal_decompose} Panel (c) echoes the findings of Panel (b). In short, professors that have a larger fraction of their earnings coming from sources besides their base salaries have larger earnings levels. This is especially true for those undertaking clinical work.

We decompose the variation in earnings across professors more formally in Figure \ref{figs_sal_decompose} Panel (d).\footnote{Note that in a multivariate model, the sum of the partial-$R^2$ statistics for each variable need not sum to one.} Institution and rank averages explain a considerable amount of within-field variation, as do professors' tasks and earnings sources. Field-level averages alone describe roughly 8\% of earnings variation (Column 1), with the full set of covariates explaining roughly 58\% of earnings variation (Column 5). This suggests that nearly 50\%(=58\%-8\%) of earnings variation is due to these institutional and position specific features of professors' work and how compensation varies across these dimensions. It also implies that 42\%(=100\%-58\%) of the total variation is due to other factors or idiosyncratic differences. We focus on one of those possible factors, research output, next.

%%%%%%%%%%%%%%%%%%%%%%%%%%%%%
\hypertarget{sec:figsoutputsaldecompose}{}
\begin{quote}
\textbf{Finding \ref{figs_outputsal_decompose}}: \em{\Paste{clipboard:finding3}}
\end{quote}

How do different fields implicitly reward different types of research progress? And how much of the earnings variation across fields is due to differences in research output or the different way that output is rewarded?\footnote{For a narrow look at these questions within the field of marketing, see \cite{mittal2008publish}.} With approximately three quarters of our sample matched to their grant and publication histories, we can explore these questions further.\footnote{See Appendix \ref{sec_app_extramethods_dimensions} for more on this matching process and comparisons of the matched and un-matched sub-samples. All metrics are based on output from 2003--2023.} 

Figure \ref{figs_outputsal_decompose} Panel (a) reports the results of an observational regression of earnings on research inputs and outputs. To allow for some heterogeneous returns across fields, we interact the research metrics with indicators for each of the five aggregate fields in our sample. For each of the three metrics, we standardize the variation within each field so that a unit increase in each metric corresponds to a one s.d. increase per the distribution within the field.

We find significant differences across fields in terms of the implied earnings per research metric. Medical and natural sciences are the two fields that appear to implicitly reward grant funding, with a one s.d. increase being associated \$5,000 and \$20,000 in additional annual earnings, respectively. We estimate similarly-sized correlations for the humanities and social sciences, but there the relationships are not statistically significant.

In terms of the implied returns to research output, there appears to be two norms: rewards for publications or citations. In the humanities and social sciences, earnings are most clearly correlated with citation-based measures of output. In the other fields, earnings are much more closely connected to publication counts, with citation counts (conditional on publication counts) showing no clear relationship with earnings.

We again decompose earnings heterogeneity, this time focusing on these research metrics, in Figure \ref{figs_outputsal_decompose} Panel (b).\footnote{Recall, this is based only on observations matched to the Dimensions data, which can lead to some discrepancies if compared to Figure \ref{figs_sal_decompose} Panel (d).} Without accounting for any other covariates, differences in these research metrics appear roughly equally as important as field-level differences, with both explaining approximately 10\% of earnings variation. However, when we include the full set of institution, faculty rank, task, and earnings source covariates that we explored in Figure \ref{figs_sal_decompose} (Columns 6--7), we find that research output to be much less important. Conditional on these other covariates, research metrics can account for only 3\%(=60\%-57\%) of earnings variation.

When paired, with the prior finding, it appears that the traditional metrics of research output often used in the science of science are much less related to professors' earnings than other attributes of their job. This doesn't necessarily imply anything about the validity of these metrics as indicators of scientific progress. Nor does this imply anything about the optimality of researchers' payoffs from conducting research. However, it illustrates that there is potentially a large gap between the way in which professors are financially compensated and the way in which scholars in the science of science field might characterize their performance. We turn more specifically to the notion of research productivity next.

\begin{figure}[htbp]\centering\footnotesize
\caption{Research productivity and time allocations}
\label{figs_timecorr}
\subfloat[Correlation between hourly and annual output]{
\includegraphics[width=0.475\linewidth, trim = 0mm 0mm 0mm 0mm , clip]{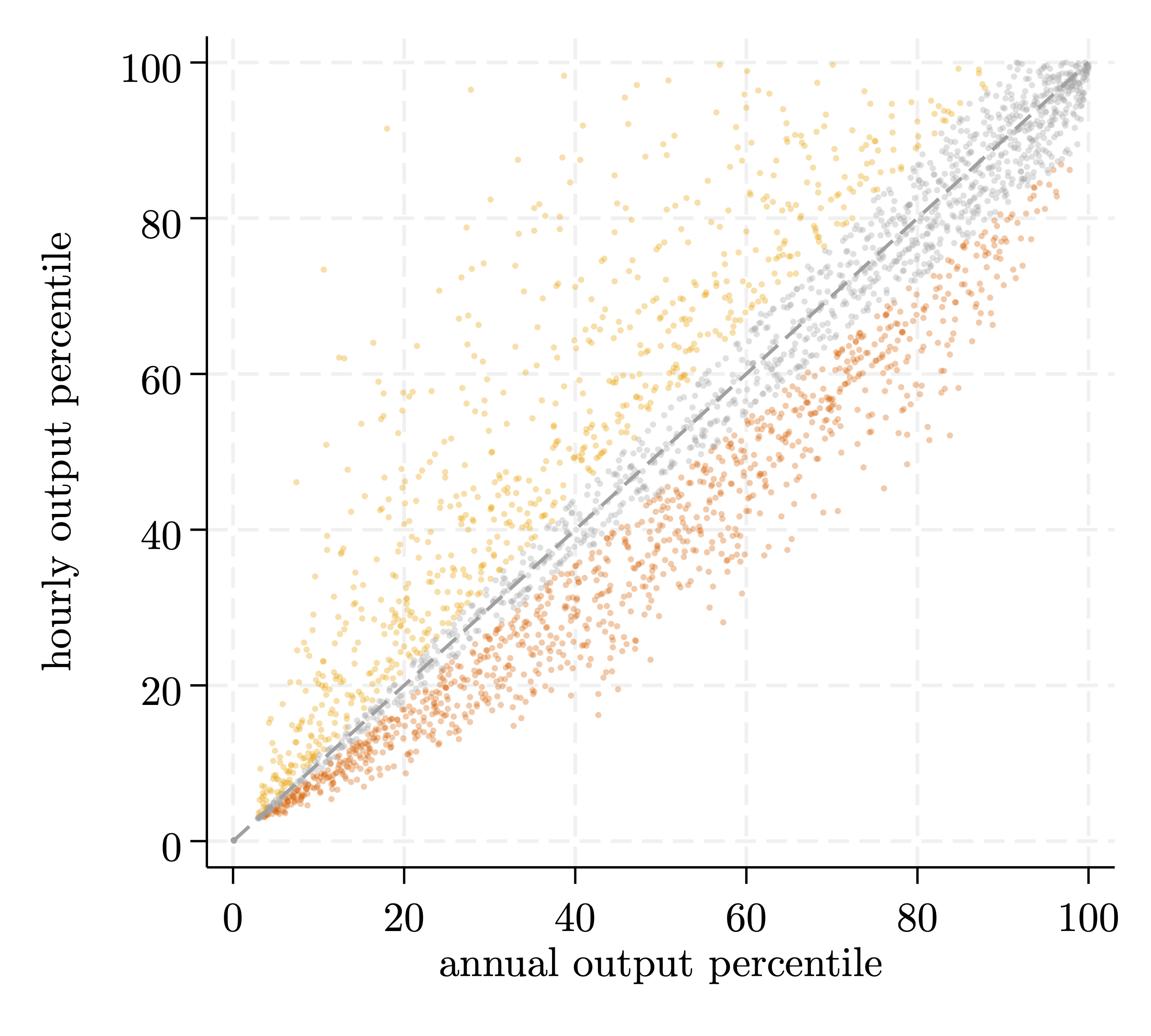}}
\subfloat[Ratio of hourly and annual output]{
\includegraphics[width=0.475\linewidth, trim = 0mm 0mm 0mm 0mm , clip]{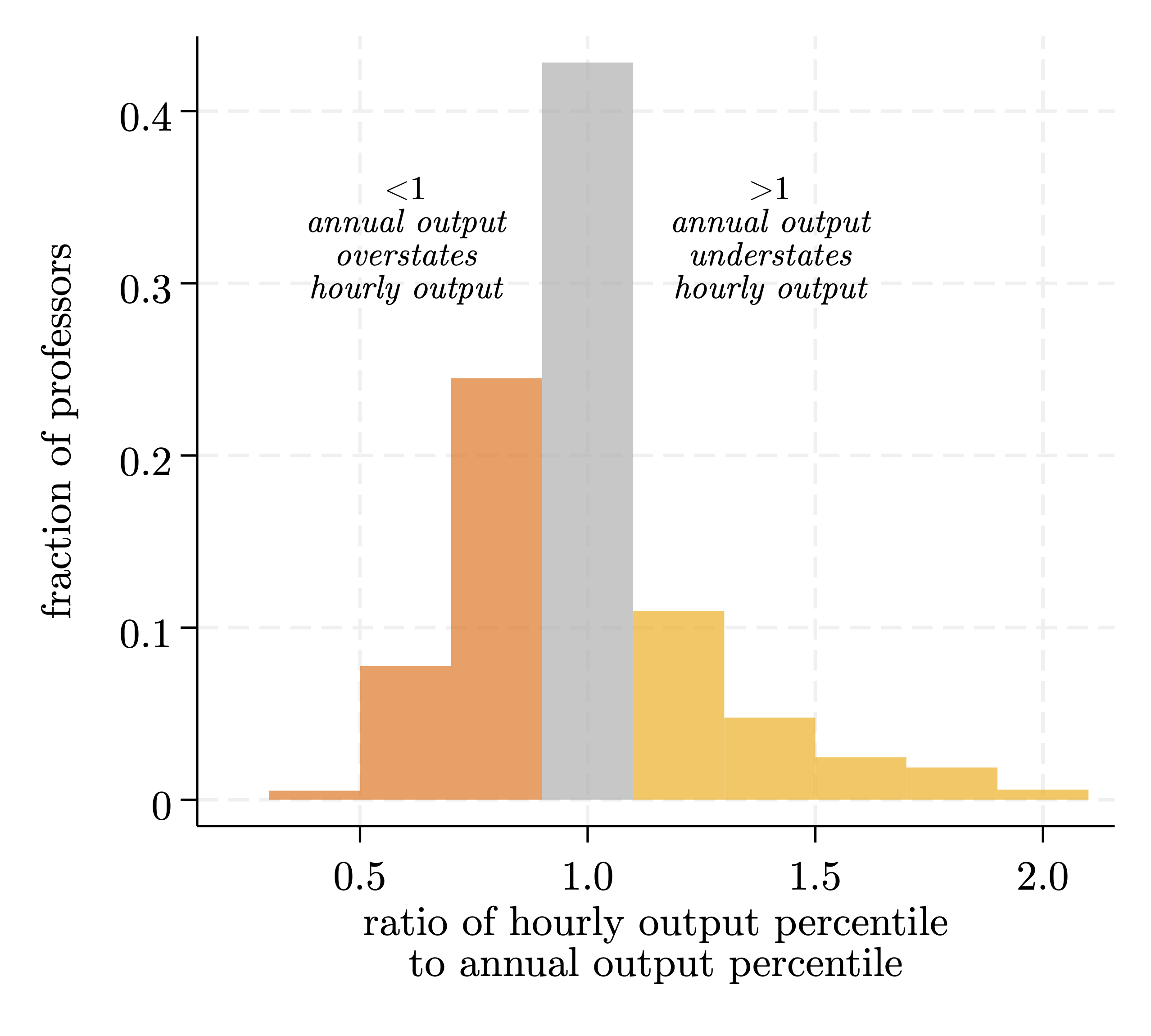}}
\\ \-\\ \-\\ 
\subfloat[Lasso-selected predictors \protect\\ of output ratio]{\small\begin{tabular}{lcc}

\hline\hline
&\multicolumn{1}{c}{stability}&\\
&\multicolumn{1}{c}{selection}&\multicolumn{1}{c}{univar.}\\
 &\multicolumn{1}{c}{share}&\multicolumn{1}{c}{corr.} \\
\hline
share of hrs., research&\phantom{-}1.00&--0.59*** \\
share of hrs., clinical&\phantom{-}1.00&\phantom{-}0.39*** \\
non-tenure-track&\phantom{-}1.00&\phantom{-}0.25*** \\
total work hrs.&\phantom{-}1.00&--0.23*** \\
adj. or clin. prof.&\phantom{-}1.00&\phantom{-}0.19*** \\
share of hrs., fundrais.&\phantom{-}1.00&--0.13*** \\
output: journal articles&\phantom{-}0.88&--0.17*** \\
audience: academics&\phantom{-}0.51&--0.14*** \\
earnings, other sources&\phantom{-}0.51&\phantom{-}0.04**\phantom{*} \\
test (vs. generate) hypoth.&\phantom{-}0.45&\phantom{-}0.08*** \\
\hline\hline
\end{tabular}

}\\ \-\\ 
\begin{quote}
\emph{Note}: \input{figtab/stub_figs_timecorr_note1.tex}
\end{quote}
\end{figure}

%%%%%%%%%%%%%%%%%%%%%%%%%%%%%
\hypertarget{sec:figstimecorr}{}
\begin{quote}
\textbf{Finding \ref{figs_timecorr}}: \em{\Paste{clipboard:finding4}}
\end{quote}

Despite time being a key scientific input, the difficulty of observing even a proxy of researchers' time allocations has severely limited our understanding of the labor component of the scientific production function.\footnote{There has long been studies of higher education focused on professors time allocations, with particular focus on the ``research-teaching nexus'' as it is often referred (e.g., \citealt{jauch1976relationships,halse2007research,horta2012output,garcia2015research,duff2017teaching,guarino2017faculty,reymert2023task}). But this work has focused less on analysing professors scientific productivity per se.} This has led many meta-science analyses to assume that researchers all have access to the same amount of time per year and either explicitly or implicitly use researchers' gross output per year as a measure of productivity. 

But of course, professors generally balance multiple roles in a university, only one of them being a researcher. Teaching and advising responsibilities, administrative duties, and grant-writing tasks all can loom large.\footnote{For example, in grant-intensive fields, many have raised concerns that scientists devote too much time to unproductive activities in order to win grants (\citealt{gross2019contest,collison2021fastgrants}); however, it is difficult to estimate the social value of these efforts since not all time and effort devoted to fundraising is necessarily wasteful (\citealt{ayoubi2019important,myers2022potential}).} Table \ref{tab_pwcorr_hrstot_uncond_fullsamp} reports the pairwise correlations for the main categories of time allocation we focus on: research (including supervising others); fundraising for research; teaching or advising (not as a part of their own research); clinical or medical practice; all other activities.\footnote{We ask respondents to forecast their weekly hours they will spend on each of these activities over the coming 5-year horizon in hopes of them estimating something close to their steady-state time allocations that are not driven by year-to-year idiosyncrasies.} Except in the case of fundraising time, which appears to be a partial complement to research time, time spent on all other tasks is associated with a decline in time spent on research.

The key question is whether these other time constraints are allocated in a way that is correlated with researchers' actual scientific productivity. If time constraints (e.g., administrative duties) are often allocated to researchers with lower hourly scientific productivity (i.e., the two are negatively correlated), then researchers' annual and hourly output will be very closely aligned. However, to the extent certain researchers with high hourly productivity have fewer opportunities to conduct research because they face additional time constraints, then the alignment between annual and hourly productivity will begin to deteriorate.\footnote{If the positive correlation between hourly productivity and time constraints was large enough, there could feasibly be no correlation between annual and hourly research output.}

In order to understand how well traditional measures of gross output per year correlate with measures that account for differences in input levels (i.e., output per hour worked), we again focus on the Dimensions-matched sample where we can see professors' publication output. We calculate their publication output on both an annual and hourly basis (per their field-normalized publication counts). The correlation between the two measures is $\rho$=0.69, which suggests that annual output is indeed informative of hourly productivity, but it may be misleading for some. To get a better view, Figure \ref{figs_timecorr} Panel (a) plots each individual's percentile of annual output on the $x$-axis and their percentile of hourly output on the $y$-axis, noting that this compression into percentiles hides the skewed nature of these measures. Researchers below the 45$^o$ line have an annual output that overstates their hourly output, and vice versa. 

To further investigate this issue, Figure \ref{figs_timecorr} Panel (b) plots a histogram of the ratio of each professor's percentile on an hourly basis compared to the same on an annual basis. We find about 40\% of professors have an annual output percentile within 10\% their hourly output percentile (see: the grey shaded bar in the histogram). Another 35\% have output percentiles within 20\% of each other. There are a significant number of researchers for whom annual output is not a strong proxy for their hourly output. Understanding who these researchers are and whether more resources ought to be allocated to them is a key policy question. Figure \ref{figs_output_f} shows that the same general pattern holds when focusing on professors' fundraising productivity (i.e., grant dollars per year or per fundraising-hour).

As a first look, we use the stability selection method of \cite{meinshausen2010stability} to identify ``important'' predictors of the gap between researchers' annual and hourly output (per percentiles).\footnote{This method proceeds as follows: (1) a random 50\% sub-sample is drawn; (2) a standard cross-validation Lasso is used to select the relevant covariates (we use the standard $k$-fold cross-validation \texttt{lasso} program in \texttt{Stata} with all default options); (3) Steps (1--2) are repeated 100 times, recording the share of samples each covariate is selected by the Lasso (i.e., the stability selection share).} Figure \ref{figs_timecorr} Panel (b) reports the stability selection share for the top ten covariates along with the univariate correlation between each covariate and the ratio of researchers' annual and hourly output. Unsurprisingly, the strongest predictors are variables related to researchers' work hours. Individuals with the most understated hourly productivity are in non-tenure-track, adjunct, or other position, presumably, because they have the largest constraints on their time. Interestingly, there is some evidence that researchers pursuing non-traditional research outputs (that is, not journal articles intended for academics) also appear to have understated hourly productivities. This again may be due to their time allocations focusing more on non-research-specific tasks. Overall, this new view of researchers' time indicates that gross output measures like publications per year may not provide an unbiased view into researchers' true underlying productivity in terms of their ability to convert their actual \emph{research} time into scientific output.

\begin{figure}[htbp]\centering\footnotesize
\captionsetup[subfloat]{format=hang,singlelinecheck=false,justification=centering}
\caption{Correlates of research risk-taking}
\label{figs_corr_risk}
\subfloat[Lasso-selected predictors \protect\\ of research risk-taking]{\small
\begin{tabular}{lcc}

\hline\hline
&\multicolumn{1}{c}{stability}&\\
&\multicolumn{1}{c}{selection}&\multicolumn{1}{c}{univar.}\\
 &\multicolumn{1}{c}{share}&\multicolumn{1}{c}{corr.} \\
\hline
share of hrs., fundrais.&\phantom{-}1.00&\phantom{-}0.22*** \\
risk-taking in personal life&\phantom{-}1.00&\phantom{-}0.21*** \\
test (vs. generate) hypoth.&\phantom{-}1.00&--0.20*** \\
share of hrs., research&\phantom{-}1.00&\phantom{-}0.18*** \\
non-tenure-track&\phantom{-}1.00&--0.15*** \\
output: materials, methods&\phantom{-}1.00&\phantom{-}0.13*** \\
female&\phantom{-}1.00&--0.13*** \\
total work hrs.&\phantom{-}1.00&\phantom{-}0.10*** \\
household earnings&\phantom{-}1.00&\phantom{-}0.09*** \\
fundraising expectations&\phantom{-}0.99&\phantom{-}0.18*** \\
\hline\hline
\end{tabular}

}\hfill
\subfloat[Research risk and fundraising]{
\includegraphics[width=0.475\linewidth, trim = 0mm 0mm 0mm 0mm , clip]{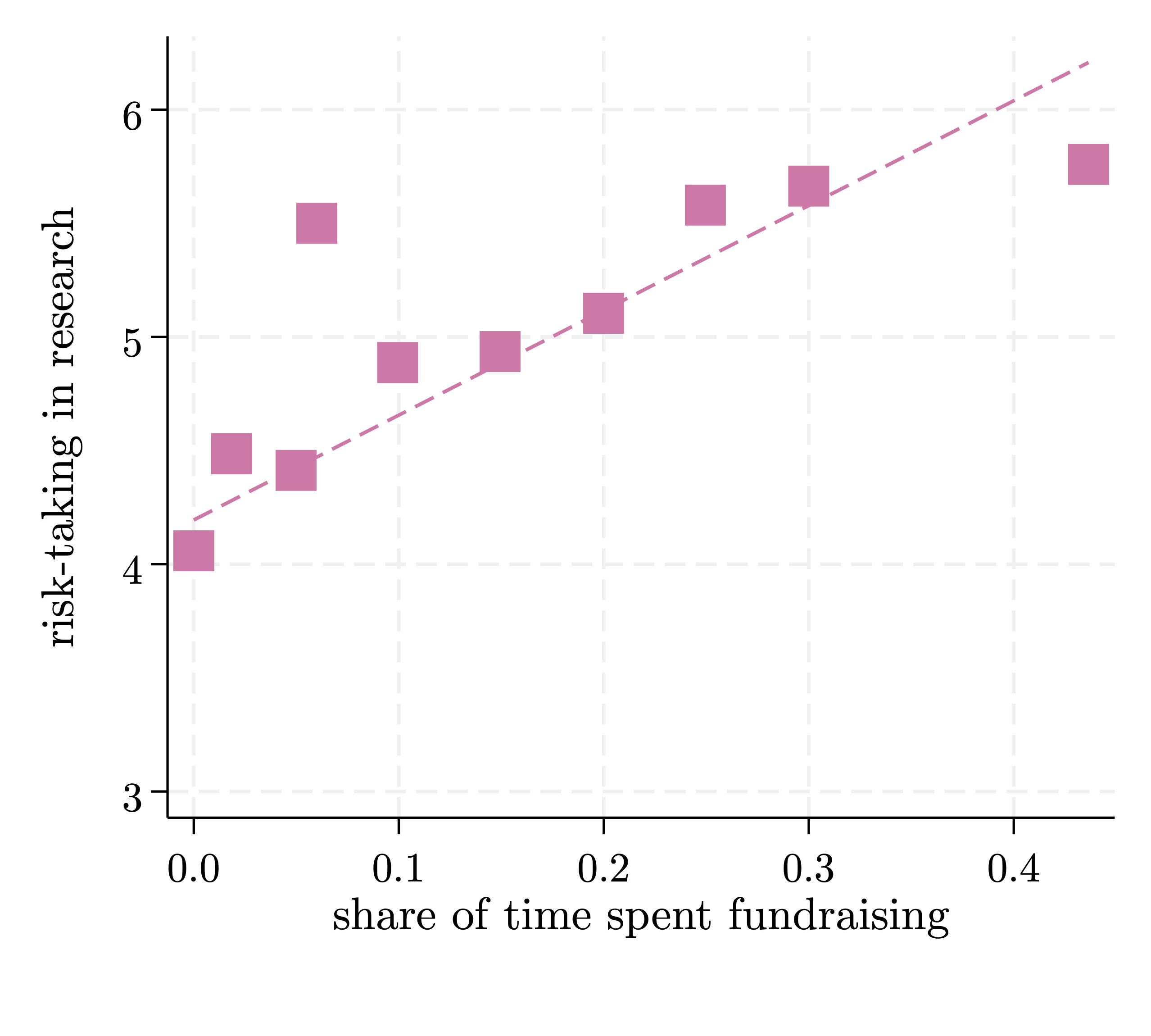}}\\
\-\\ \-\\
\subfloat[Research risk and personal risk-taking]{
\includegraphics[width=0.475\linewidth, trim = 0mm 0mm 0mm 0mm , clip]{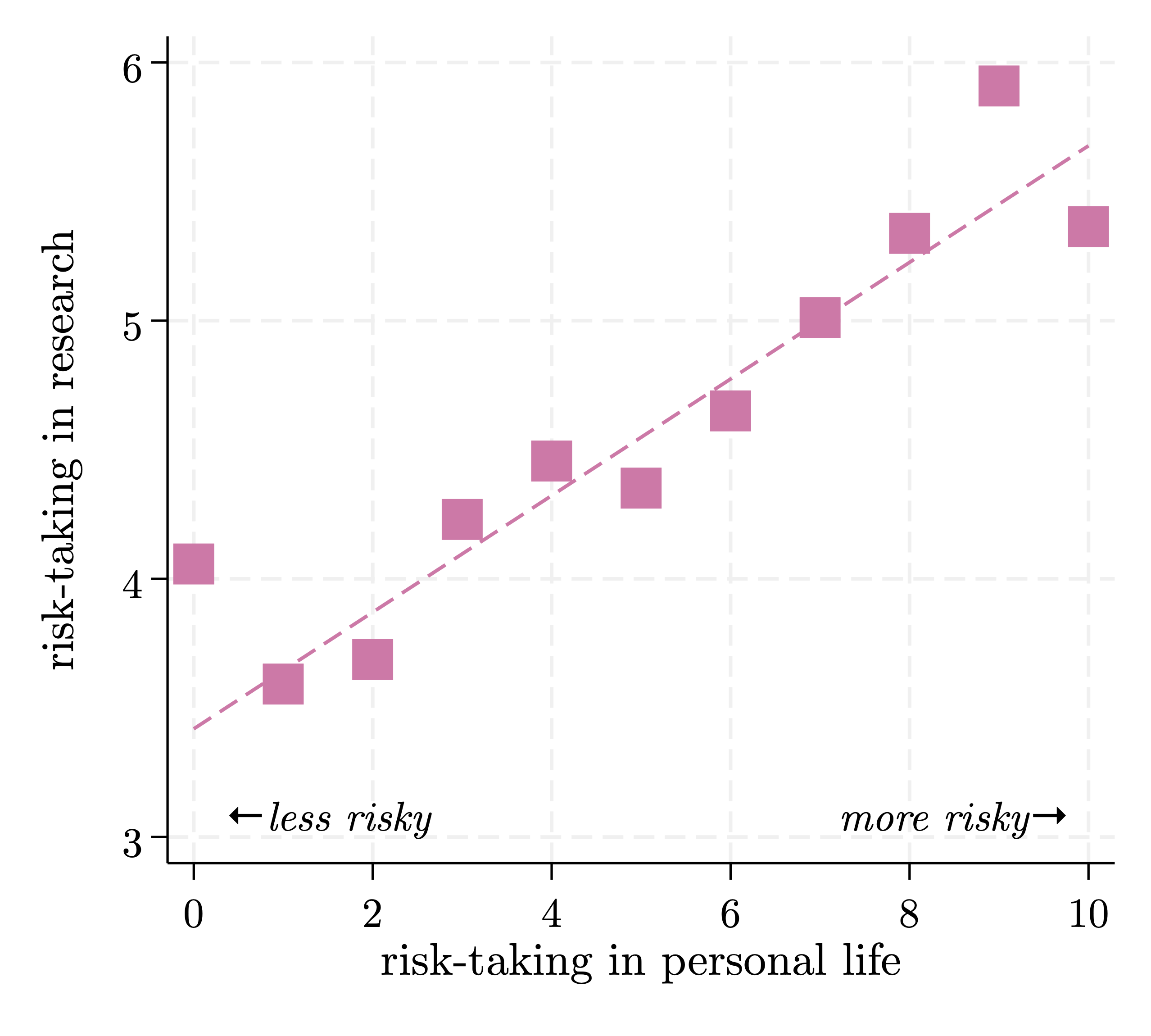}}\hfill
\subfloat[Research risk and strategy]{
\includegraphics[width=0.475\linewidth, trim = 0mm 0mm 0mm 0mm , clip]{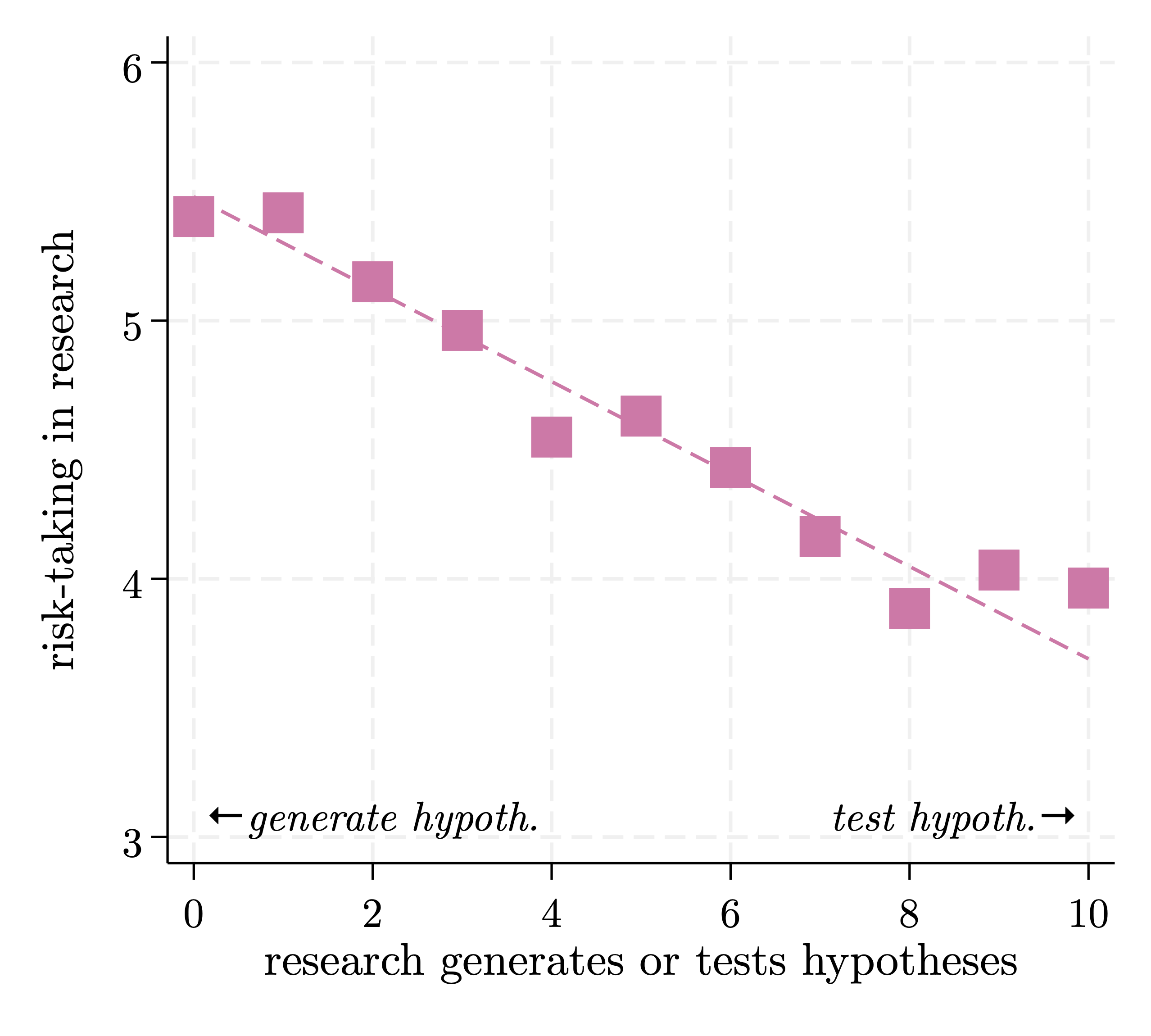}}
\begin{quote}
\emph{Note}: \input{figtab/stub_figs_corrrisk_note1.tex}
\end{quote}
\end{figure}

%%%%%%%%%%%%%%%%%%%%%%%%%%%%%
\hypertarget{sec:figscorrrisk}{}
\begin{quote}
\textbf{Finding \ref{figs_corr_risk}}: \em{\Paste{clipboard:finding6}}
\end{quote}

Discourse about innovation and science policy often asserts that the system overly discourages scientists from taking risks, causing society to miss out on high-impact scientific discoveries and inventions (e.g. \citealt{alberts2014rescuing}). Most empirical work has focused on this issue relies on ex-post measures of risk-taking based on bibliometric measures.\footnote{For example, \cite{zoller2014assessing,franzoni2017academic,brogaard2018economists}. See \cite{greenblatt2022does} for an effort that includes a wide range of proxies for concepts underlying risk and novelty, and see \cite{figueira2018unveiling}, \cite{carson2022risk}, or \cite{carson2023choose} alternative approaches.} These ex-post measures of publications are clearly limited in their ability to proxy for ex-ante risk-taking by researchers.

To provide a new, alternative view of risk in science, the survey solicits professors' subjective beliefs of their own risk-taking behaviors.\footnote{Constructing any sort of field-agnostic (relatively) objective measure of risk-taking that mirrored those commonly used in lab experiments (e.g., gambles over outcomes) proved extremely difficult in pilot tests due to the heterogeneity in relevant outcomes. Hence, our more subjective, but much simpler measure.} Using questions structured in the same format as the more general risk preference questions of \cite{dohmen2011individual}, researchers report how risky they think their own research is, as well as how risky they think their peers think their research is (on a scale from 0 to 10, with larger values indicating more risk).\footnote{Pilot interviews with scientists suggested that both approaches would prove useful avenues for soliciting scientists beliefs.} Our preferred metric reported throughout this paper is the average of these two responses, which hopefully serves to reduce some of the measurement error inherent to either phrasing of the question. Figure \ref{fig_hist_qriskresavg} reports the distribution of this (averaged) risk score, which illustrates significant support across the full range of possible values except for the uppermost tail of risk-taking. 

We use the ML-based approach of stability selection to identify the covariates that best predict researchers' perceptions about the riskiness of their research. Figure \ref{figs_corr_risk} Panel (a) reports the results of this exercise, with Panels (b--d) showing binned scatterplots of researchers' risk perceptions based on three of the top predictors we identify via stability selection: (Panel b) the share of time researchers spend on fundraising; (Panel c) researchers' willingness to take risks in their personal lives; and (Panel d) researchers orientation towards generating (as opposed to testing) hypotheses.

Figure \ref{figs_corr_risk} Panel (b) suggests that professors who undertake more fundraising may inherently perceive more risk in their research. This is interesting because ``fundraising risk'' has not traditionally received much attention due to the difficulties of observing professors' specific funding streams. This may be driven by the fact that professors who spend a large fraction of their time fundraising tend to be in ``soft-money'' positions where a portion of their salary is derived from their fundraising. \footnote{Every percentage point increase in the share of researchers' time spent on fundraising is associated with an additional \$1,095 (s.e.=\$58) in annual earnings.} Alternatively, it may be that, projects that are more ambitious and risky tend to be projects that also require more resources and hence more fundraising. This points to the need for more work on understanding what risk means to researchers (e.g., \citealt{holzmeister2020drives}): What are the outcomes they care about? How do they perceive risk?

Risk-taking in personal life is another strong predictor of researchers' perceptions of the riskiness in their science. Of course, the simplest explanation here is a survey response bias whereby individuals inflate their risk-taking in both questions and this generates the correlation. However, the pattern is also consistent with researchers' latent risk-aversion being a key determinant of how they pursue their science. To the extent this is true, it suggests that understanding the extensive margin of selection into science and how it may screen individuals with higher or lower levels of latent risk-aversion would be a fruitful avenue for future research.

Lastly, we see that researchers who report focusing on generating new hypotheses also tend to report higher perceived risk in their research. This aligns with the idea that it is inherently more difficult to capture the value of a good question compared to a good answer; in other words, hypotheses themselves have stronger public-good attributes than tests of hypotheses. Undertaking projects where your ability to capture the value of your efforts is less certain would likely be perceived as more risky. We dig further into researchers' strategies next.

\begin{figure}[htbp]\centering\footnotesize
\caption{Nature of research, lifecycle, and tenure}
\label{figs_nrage}
\subfloat[Intended outputs, by age]{
\includegraphics[width=0.475\linewidth, trim = 0mm 0mm 0mm 0mm , clip]{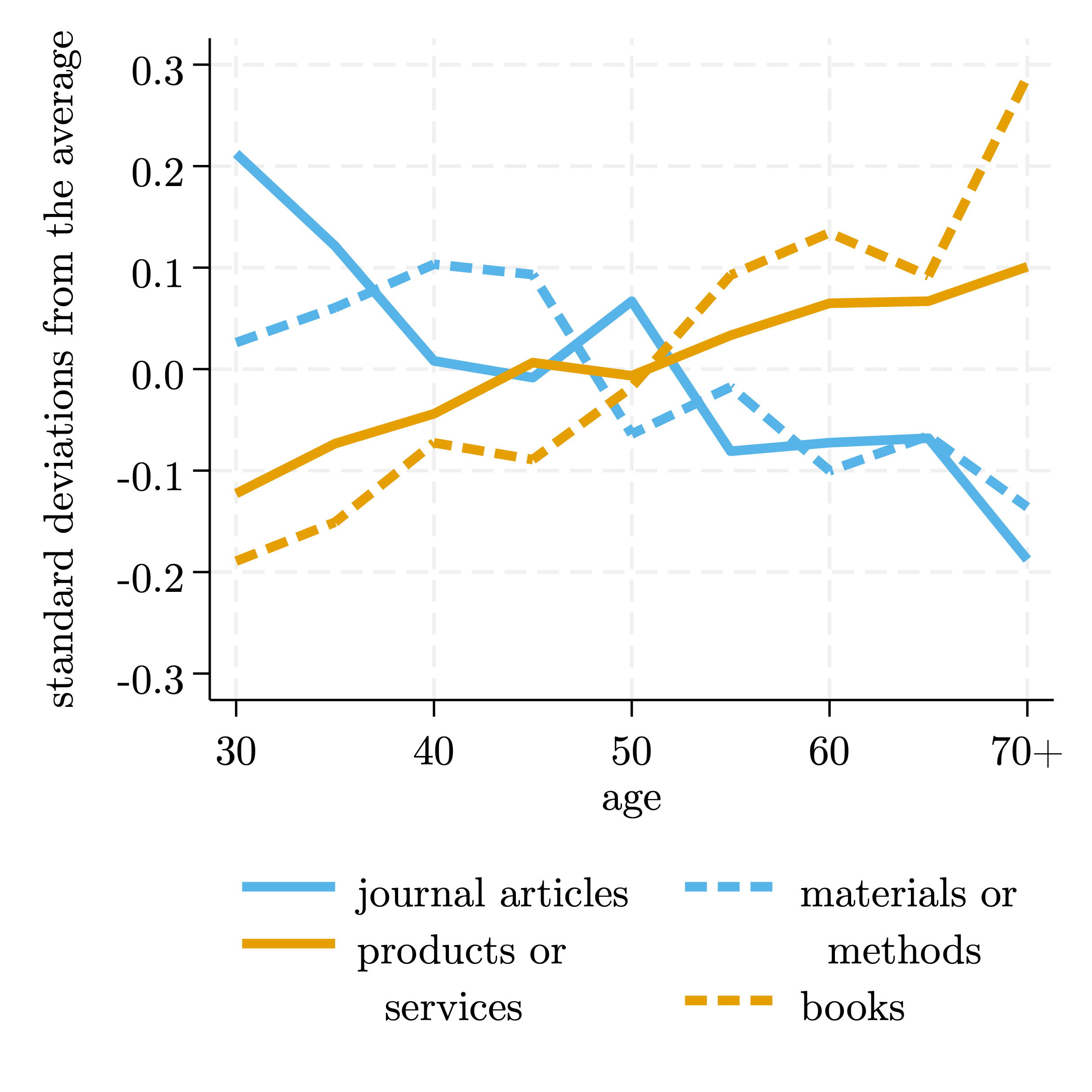}}
\subfloat[Intended audience, by age]{
\includegraphics[width=0.475\linewidth, trim = 0mm 0mm 0mm 0mm , clip]{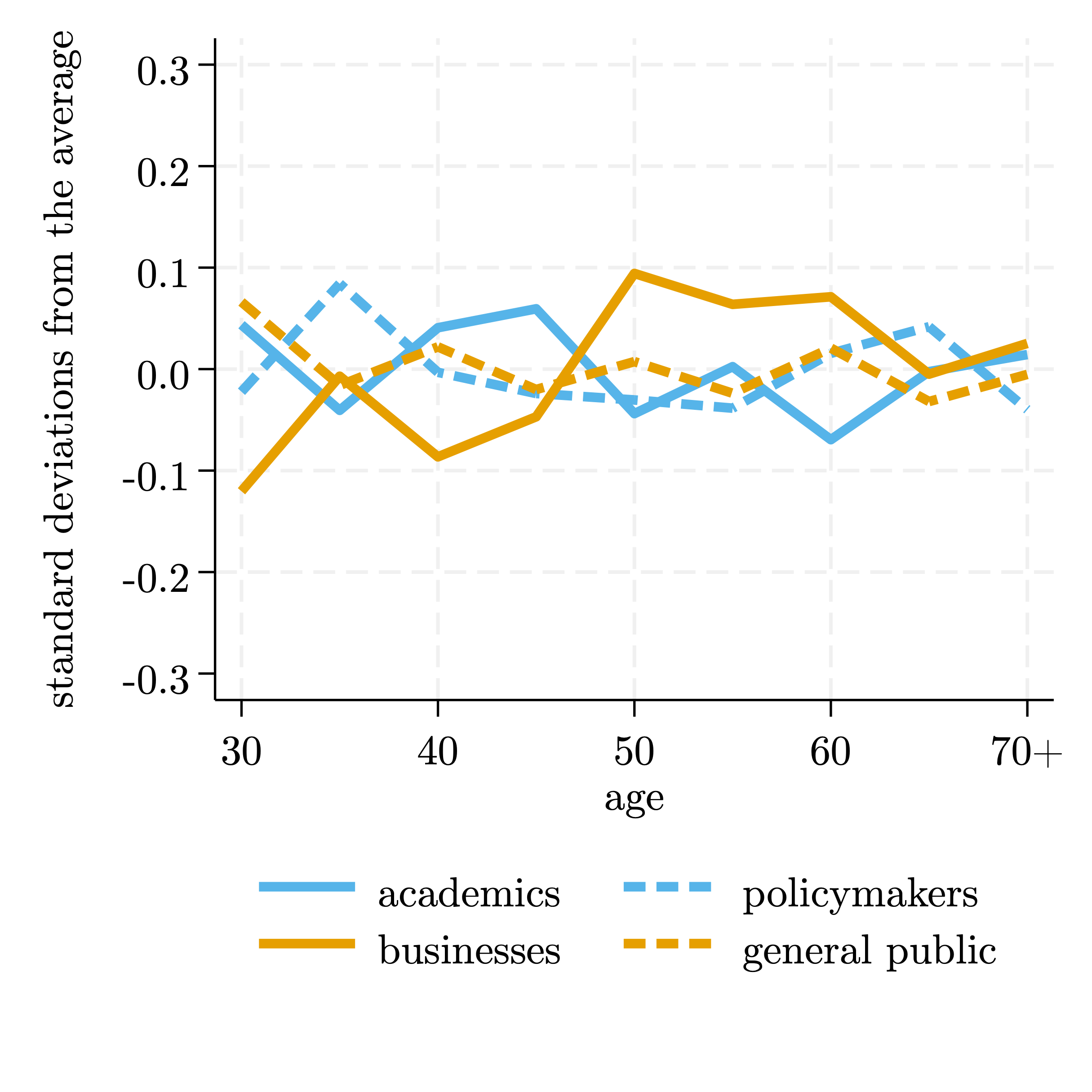}} \\
\-\\ \-\\
\subfloat[Work hours, by age]{
\includegraphics[width=0.475\linewidth, trim = 0mm 0mm 0mm 0mm , clip]{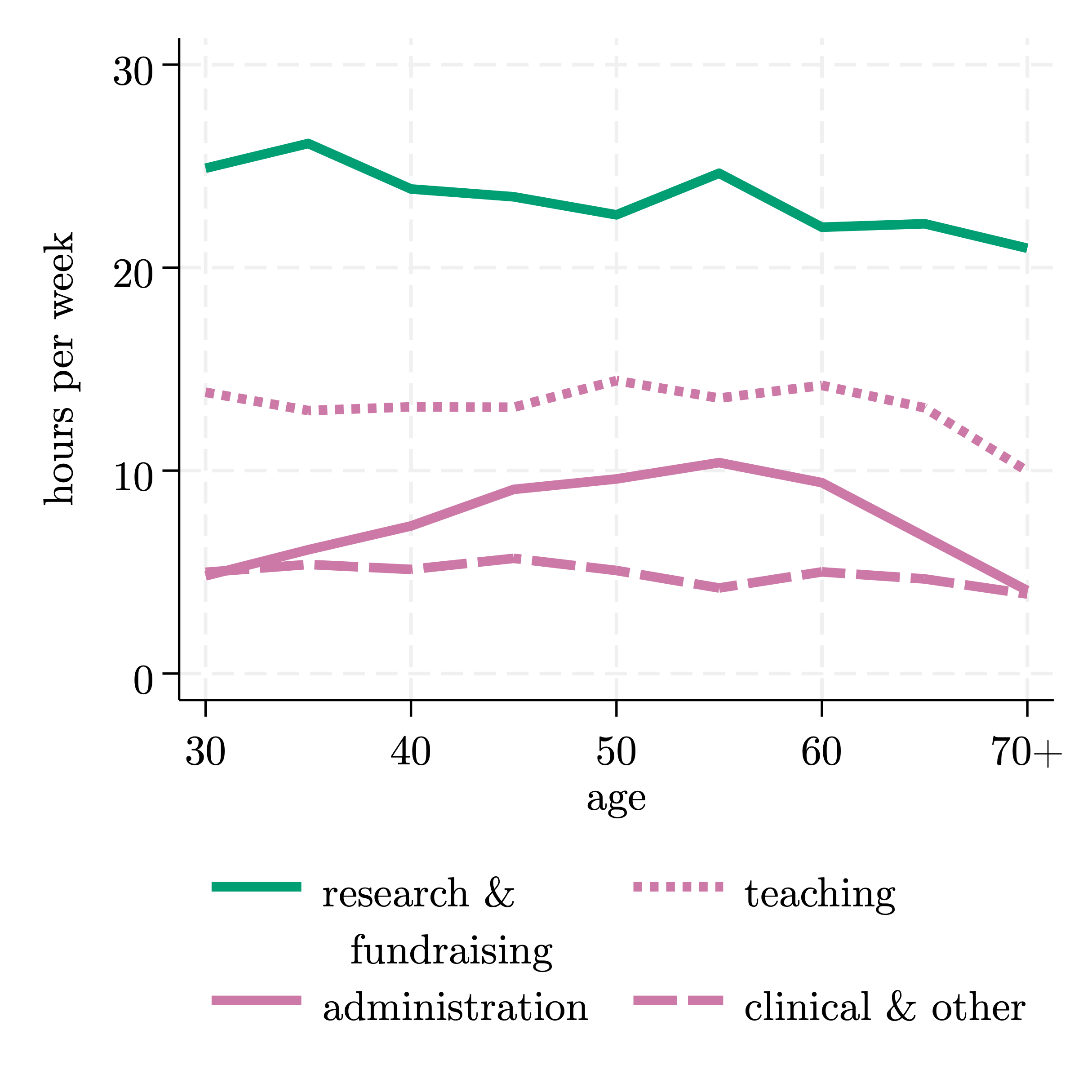}}
\subfloat[Work hours, relative to tenure evaluation]{
\includegraphics[width=0.475\linewidth, trim = 0mm 0mm 0mm 0mm , clip]{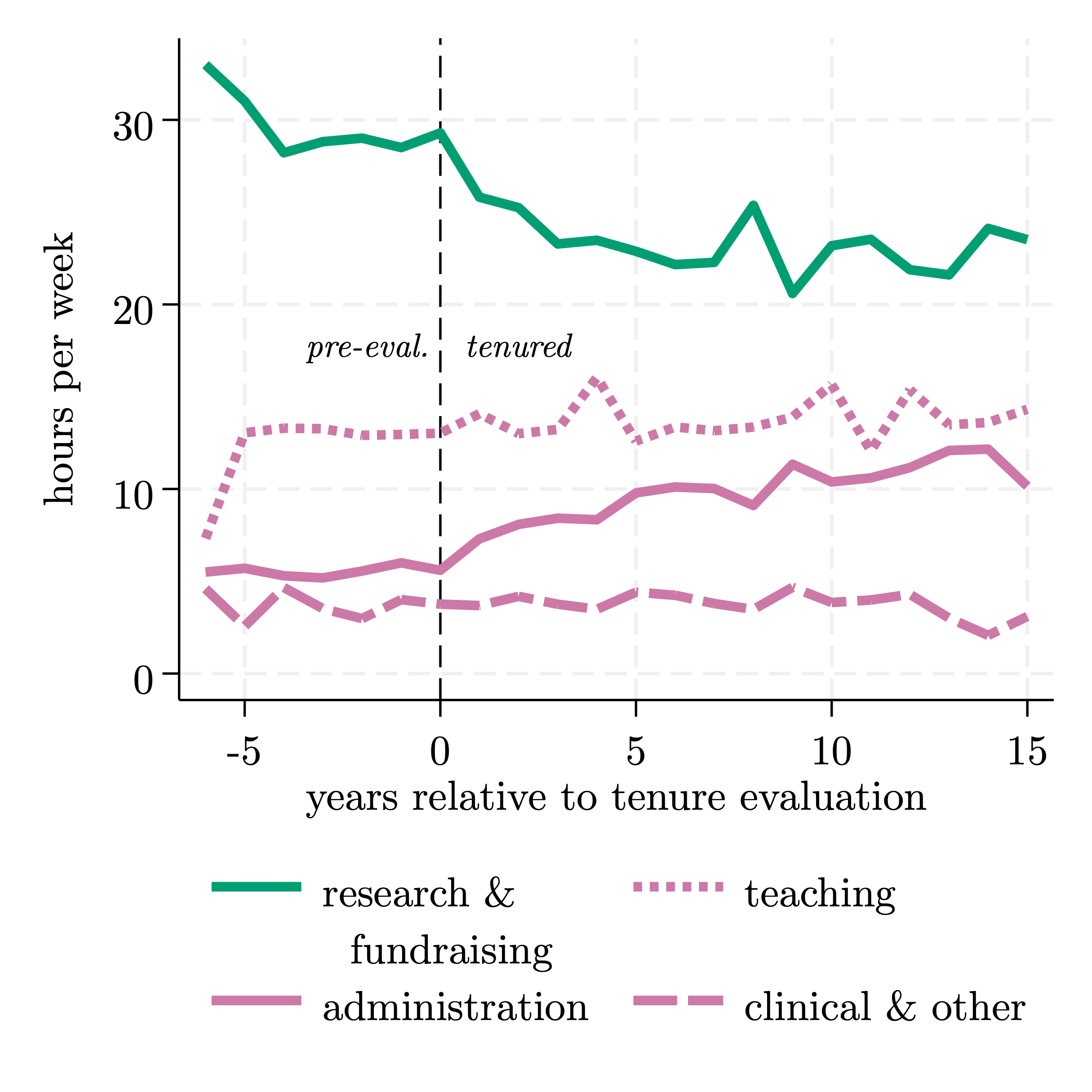}}
\\ 
\begin{quote}
\emph{Note}: \input{figtab/stub_figs_lifecycle_note1.tex}
\end{quote}
\end{figure}

%%%%%%%%%%%%%%%%%%%%%%%%%%%%%
\hypertarget{sec:figsnrage}{}
\begin{quote}
\textbf{Finding \ref{figs_nrage}}: \em{\Paste{clipboard:finding5}}
\end{quote}

Besides taking (or avoiding) risks, what exactly are research professors intending to do with their research? Here, we dig into our questions related to the intended outputs and audience of professors' research (see Table \ref{tab_nat_sumstat}).\footnote{All of the questions related to professors' intended output and audience are solicited on a likert scale with three values of frequency (``Rarely'', ``Sometimes'', ``Most of the time''), which we convert into a variable valued \{0,1,2\} (see Table \ref{tab_nat_sumstat}). These measures are intended to reflect the underlying \emph{share} of professors' scientific production that is destined for a particular output or audience type. Thus, for all of the following analyses, we assume that professors have a fixed level of intentions in these two dimensions and re-scale each their responses into fractions that the sum to one for each dimension. For example, if a professor reports that all four output types are their intentions ``Most of the time'', then we assume that the share of their intended output of each type is 1/4.} Specifically, motivated by early work in the economics of science related to life-cycle effects (\citealt{levin1991research}), we focus on temporal changes across ages and professional experience. An important caveat to reiterate here is that the cross-sectional nature of the survey means that any temporal dynamics reflect both age- or experience-related effects in addition to any selection effects that occur over the life cycle and/or career cycle.

Figure \ref{figs_nrage} Panels (a--b) plot professors' intended outputs and audiences across the forty years of ages in our sample. Notably, there is a marked evolution in \emph{what} professors are focusing on producing, shifting from a focus on journal articles, materials, or methods in their early decades into a focus on books, products, or services in their later decades (Panel a). However, there is no significant change in \emph{who} professors are focusing their efforts towards (Panel b). This pattern suggests that professors' preferences over their audience are quite stable, but the optimal way of reaching this audience is not. Books as a scientific output have not received much attention by the science of science community, likely because of data limitations.\footnote{Exceptions include \cite{wanner1981research,sabharwal2013comparing,gimenez2016taking}, and in a more general sense \cite{giorcelli2022does}.}

Figure \ref{figs_nrage} Panels (c--d) revisit professors' time use, now looking over the life cycle (Panel c) and more narrowly around the tenure evaluation process for those on the tenure track (Panel d).\footnote{Here, we group research and fundraising time given their positive correlation as shown in the pairwise time-use correlations of Table \ref{tab_pwcorr_hrstot_uncond_fullsamp}.} As most experienced professors can attest, age and experience are associated with a clear increase in administrative duties. However, this provides one of the first views of this shift in task composition that allows us to quantify the relative increase in administrative duties relative to the change in professors' time spent on their research. A common result publicized by studies that can observe only publication output across researchers' careers is the marked decline in their publication output after their early years of work, and especially after receiving tenure (\citealt{brogaard2018economists}). Focusing specifically on Panel (d), we can see that the receipt of tenure is associated with a marked increase in administrative duties and, to a much lesser degree, some increases in teaching and other duties. Aggregating these changes together indicates that, in the first ten years post tenure, approximately 80\% of the decline in research and fundraising efforts can be explained by the increase in teaching, administrative, and other effort. Here again, much like our prior finding on the differences between annual and hourly output, our ability to observe researchers' time allocations indicates that the ``post-tenure glut'' in publication may not be any change in productivity on a publication-per-research-hour basis but, to a large degree, may more simply be a decline in input levels. This distinction is important because it speaks to the trade-offs of the institutions of academic science and professors' responsibilities therein.\footnote{For instance, the ideal distribution of professors' administrative duties over the course of their career will depend on, among other things, how their productivity evolves over their life-cycle, which may follow field-specific patterns (\citealt{levin1991research,galenson2001creating,weinberg2019creative}).}

\begin{figure}[htbp]\centering\footnotesize
\captionsetup[subfloat]{format=hang,singlelinecheck=false,justification=centering}
\caption{Correlates of the Bohr--Edison (``basic--applied'') spectrum}
\label{figs_corr_bescore}
\subfloat[Principal component analysis \protect\\ of nature-of-research questions]{\small
\begin{tabular}{l*{1}{r}}
\hline\hline
                    &     loading\\
\hline
objective is to generate (0) versus&            \\
test hypotheses (10)&        0.17\\
\\ \underline{\emph{intended output}}&            \\
journal articles    &       --0.25\\
books               &       --0.07\\
research materials, tools, etc.&        0.24\\
products            &        0.51\\
\\ \underline{\emph{intended audience}}&            \\
academics           &       --0.37\\
policymakers        &        0.38\\
businesses and organizations&        0.43\\
general public      &        0.35\\
\hline\hline
\end{tabular}
}\hfill
\subfloat[Bohr-Edison scores, by field]{
\includegraphics[width=0.475\linewidth, trim = 0mm 0mm 0mm 0mm , clip]{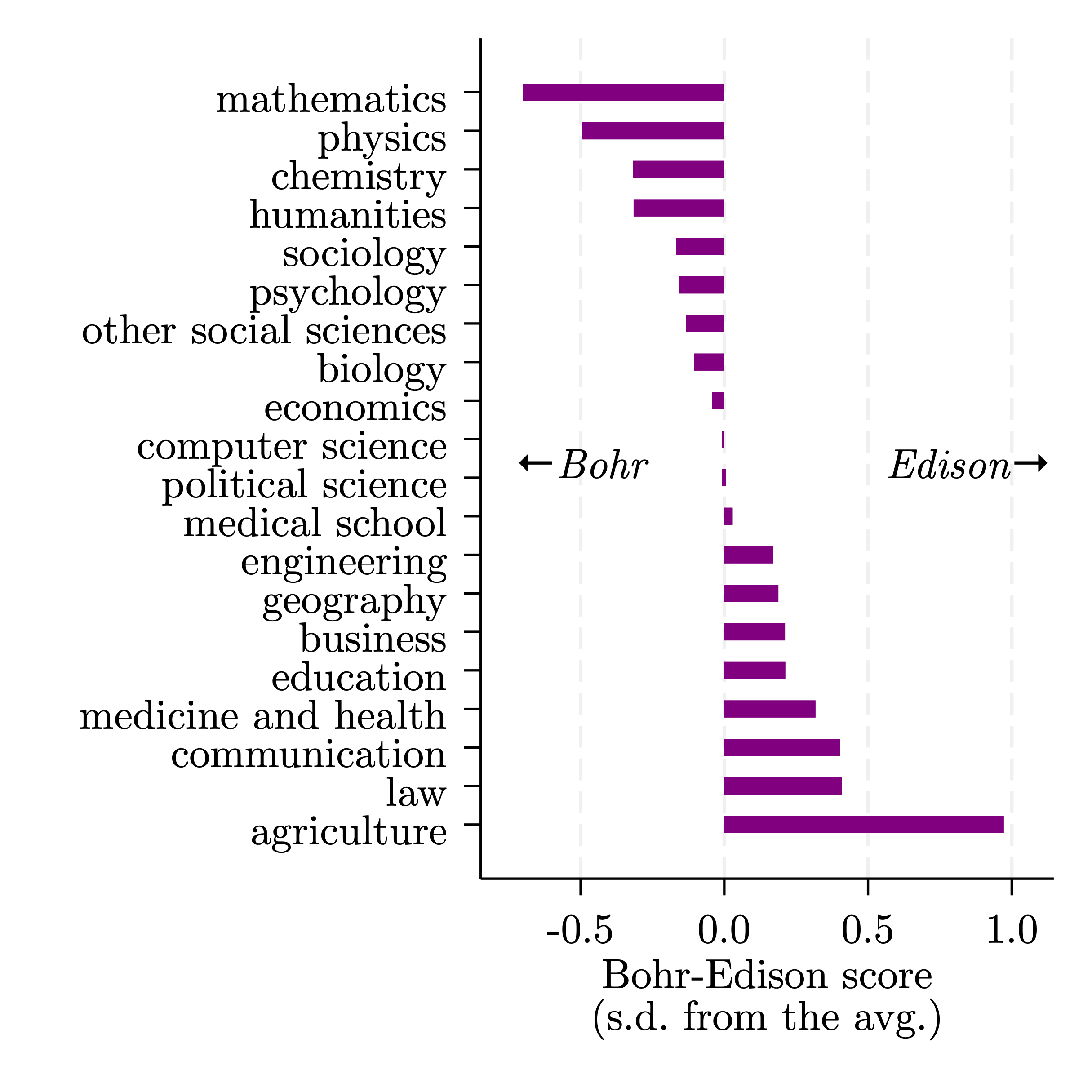}}\\
\-\\ \-\\
\subfloat[Bohr-Edison scores and earnings]{
\includegraphics[width=0.475\linewidth, trim = 0mm 0mm 0mm 0mm , clip]{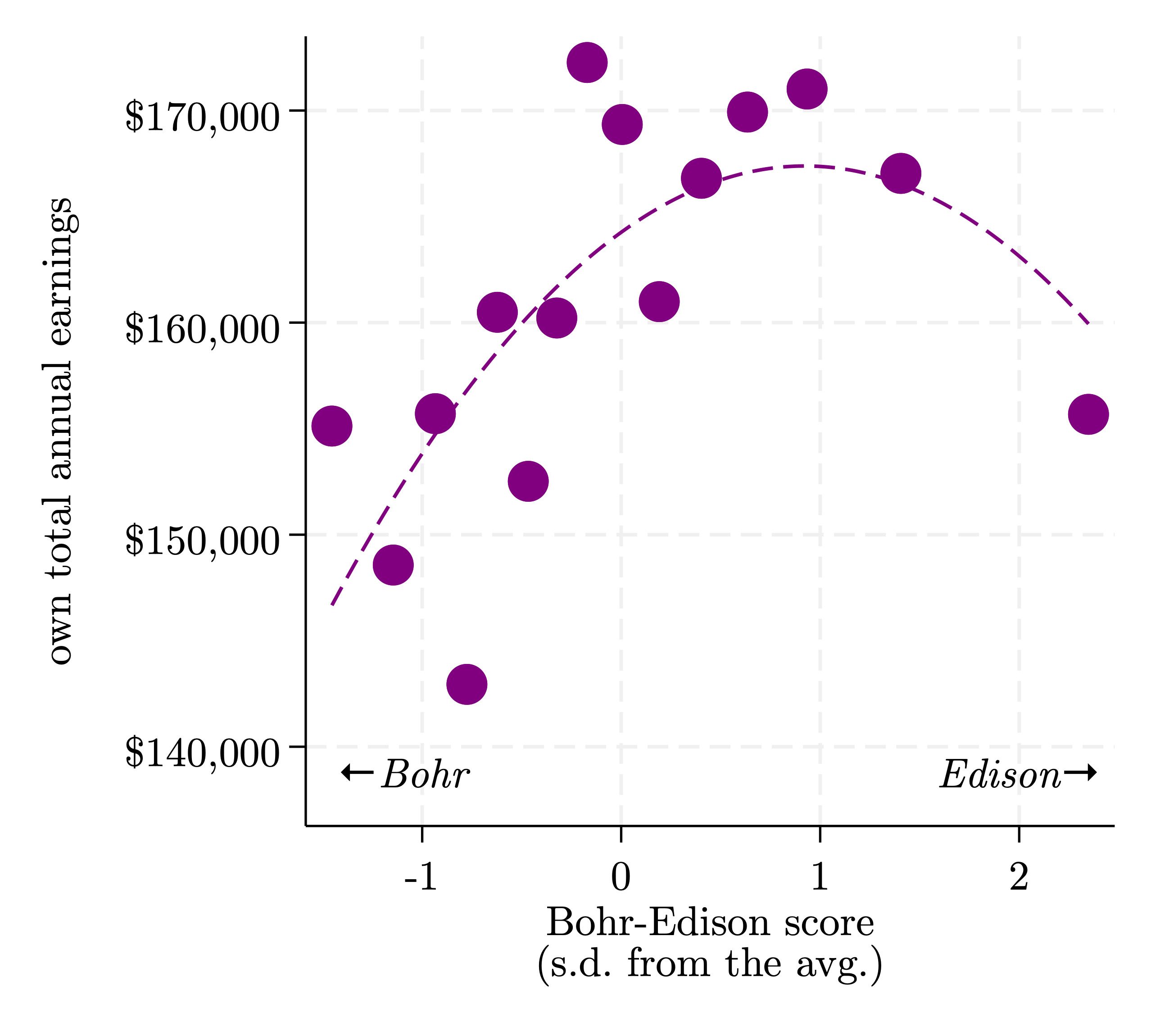}}\hfill
\subfloat[Lasso-selected predictors \protect\\ of Bohr-Edison score]{\small
\begin{tabular}{lcc}

\hline\hline
&\multicolumn{1}{c}{stability}&\\
&\multicolumn{1}{c}{selection}&\multicolumn{1}{c}{univar.}\\
 &\multicolumn{1}{c}{share}&\multicolumn{1}{c}{corr.} \\
\hline
share of hrs., other&\phantom{-}1.00&\phantom{-}0.13*** \\
personal risk-taking&\phantom{-}1.00&\phantom{-}0.13*** \\
tenured&\phantom{-}1.00&--0.11*** \\
earnings, other sources&\phantom{-}1.00&\phantom{-}0.11*** \\
share of hrs., research&\phantom{-}1.00&--0.08*** \\
has dependents&\phantom{-}1.00&\phantom{-}0.07*** \\
share of hrs., fundrais.&\phantom{-}0.99&\phantom{-}0.08*** \\
age&\phantom{-}0.99&--0.00\phantom{***} \\
non-tenure-track&\phantom{-}0.98&\phantom{-}0.12*** \\
non-recent-generation&\phantom{-}0.97&--0.05*** \\
\hline\hline
\end{tabular}

} \\
\begin{quote}
\emph{Note}: \input{figtab/stub_figs_corrbescore_note1.tex}
\end{quote}
\end{figure}

%%%%%%%%%%%%%%%%%%%%%%%%%%%%%
%%%%%%%%%%%%%%%%%%%%%%%%%%%%%
\hypertarget{sec:figscorrbescore}{}
\begin{quote}
\textbf{Finding \ref{figs_corr_bescore}}: \em{\Paste{clipboard:finding7}}
\end{quote}

Understanding the selection of research topics by professors and the rewards for these choices is crucial to understanding the direction of science.\footnote{See \cite{neumann1977standards} for an early investigation into the differences in research outputs across fields of science. More specifically related to the finding here, see \cite{stern2004scientists} and \cite{roach2010taste} for work on scientists' ``taste'' for commercially-oriented science.} One of the most common approaches to characterizing research is on a spectrum of ``basic'' to ``applied'' (e.g.,\citealt{cockburn1999balancing,aghion2008academic,cohen2020not}). However, creating a quantitative measure solely based on existing data is challenging. For example, bibliometric measures such as patent citations may have limited validity in some fields where patenting is rare. To provide an alternative view, the survey includes multiple questions related to the nature of professors' research (see Table \ref{tab_nat_sumstat}). Each of these questions were designed to both capture different types of scientific outputs and audiences, but also to reflect different dimensions of the basic--applied spectrum as often described.\footnote{See \cite{bentley2015relationship} for another survey-based approach that more directly attempts to solicit researchers' position on this spectrum, which yields findings complementary to ours here.}

To combine the information contained in all of these nature-of-research questions, we use Principal Components Analysis (PCA) to estimate a single-dimension, standardized index.  Figure \ref{figs_corr_bescore} Panel (a) reports the results from the PCA. On one end of this spectrum are professors focused on generating hypotheses and writing journal articles for academics, and on the other end of this spectrum are professors focused on testing hypotheses and making tools and products for policymakers, businesses, and the general public. Hence, in the spirit of \citeauthor{stokes1997pasteur}'s (\citeyear{stokes1997pasteur}) quadrants, we term this uni-dimensional index the ``Bohr-Edison'' score. More negative values indicate more ``Bohr''-like basic science and more positive values indicate more ``Edison''-like applied science. 

Figure \ref{figs_corr_bescore} Panel (b) reports the field-level average Bohr-Edison scores. The ranking is intuitive, with traditionally ``theoretical'' fields like mathematics and physics scoring more towards the Bohr end of the spectrum, and more applied and technically-oriented fields like agriculture, law, and medicine scoring more towards the Edison end of the spectrum.

One way to conceptualize the basic-applied spectrum through the lens of economics is the ability of the researcher to appropriate the value of their outputs. The application of more basic research is fundamental in nature and may be harder to appropriate, and vice versa as the research becomes more applied. In this vein, Figure \ref{figs_corr_bescore} Panel (c) plots professors' earnings as a function of their Bohr-Edison score. As expected, we find a strong positive correlation. The correlation reverses at the extreme Edison-end of the spectrum, which may be driven by the fact that some of the most applied professors in the sample come from fields such as education, communication, and agriculture (Figure \ref{figs_corr_bescore} Panel b) which are fields with some of the lowest average earnings (Figure \ref{figs_sal_field} Panel b). 

To understand what variables are most predictive of the Bohr-Edison score, we again use the stability selection approach. Figure \ref{figs_corr_bescore} (d) reports the top 10 predictors per their stability selection share along with their univariate correlations with the Bohr-Edison score. A noteworthy finding here is that personal risk-taking is one of the best predictors of doing more applied, Edison-like work. This is an example of a pattern that would be difficult to find using ex-post measures of risk aversion (e.g., based on citations or text of publications and patents) because it is difficult to know whether cross-field differences in such measures are due to underlying professor characteristics or simply reflect field differences (e.g., different citation norms). To the extent the Bohr-Edison score reflects Edison-like entrepreneurship, this finding echoes other work showing that entrepreneurs tend to be more risk tolerant (\citealt{van2001roots,cramer2002low,herranz2015entrepreneurs}). More generally, this approach to transforming researchers' intentions in terms of the outputs and audience of their science could prove useful in generating observable, ex-ante variation in researchers' positions along the basic-applied spectrum.

%%%%%%%%%%%%%%%%%%%%%%%%%%%%%%%%%%%
%----------------------------------------------------------------------------------------%
\section{Discussion}\label{sec_discuss}
%----------------------------------------------------------------------------------------%
%%%%%%%%%%%%%%%%%%%%%%%%%%%%%%%%%%%
The emergence of large, curated datasets derived from publication and grant records has facilitated a surge of new empirical studies of science. However, there are many important variables that even high-quality administrative datasets do not capture (\citealt{stantcheva2022run}). In this paper, we document our survey efforts to solicit and codify some of these important but hard-to-observe variables for the academic research workforce: professors' time allocations, their earnings sources, the nature of their research, and their risk aversion. This new survey data, combined with existing datasets, yields new insights into variation amidst US academic researchers both \emph{within} and \emph{across} fields at a national scale.

We are certainly not the first to use survey methods to learn about the academic research workforce. But our approach provides one of the first broad views across the full spectrum of science in modern research universities. We do not report any causal effects here, and we are limited in our ability to precisely disentangle sources of heterogeneity across researchers in many dimensions. But we have highlighted a number of novel features of this workforce. 

We take a narrow view of the broader academic research workforce here, focusing only on professors. The dramatic rise in contingent and part-time faculty (\citealt{colby2023data}) and prevalence of ``staff scientists'' (\citealt{carpenter2012hidden}) at US universities suggest that we clearly are missing some important workers. Targeting these researchers would likely require alternative outreach techniques and new survey instruments, but would certainly be worthwhile. Future survey work can build on our efforts more broadly by investing more resources into recruiting respondents, eliciting preferences with more precise methods, or eliciting a wider spectrum of preferences. Such efforts can continue to provide a complementary view of the science of science.

%%%%%%%%%%%%%%%%%%%%%%%%%%%%%%%%%%%
%----------------------------------------------------------------------------------------%
%BIB
%----------------------------------------------------------------------------------------%
%%%%%%%%%%%%%%%%%%%%%%%%%%%%%%%%%%%
\clearpage
\singlespacing
\bibliographystyle{apalike}
\bibliography{bibliography.bib}

%%%%%%%%%%%%%%%%%%%%%%%%%%%%%%%%%%%
%----------------------------------------------------------------------------------------%
\clearpage
\section*{Tables and Figures}
%----------------------------------------------------------------------------------------%
%%%%%%%%%%%%%%%%%%%%%%%%%%%%%%%%%%%

%%%%%%%%%%%%%%%%%%%%%%%%%%%%%%%%%%%%%%%%%%%%%%%%%%%%%%%%%%%%%%%
%%%%%%%%%%%%%%%%%%%%%%%%%%%%%%%%%%%%%%%%%%%%%%%%%%%%%%%%%%%%%%%
\begin{table}[htbp]\centering\small
\caption{Institution-level population summary statistics}
\label{tab_pop_sumstat}
{
\def\sym#1{\ifmmode^{#1}\else\(^{#1}\)\fi}
\begin{tabular}{lc*{1}{rrrrr}}
\hline\hline
                    &    type    &       count&        mean&          sd\\
\hline
\underline{\emph{prof. share, by rank}}&            &            &            &            \\
assistant           &  $ [0,1]$  &         158&        0.25&        0.10\\
associate           &  $ [0,1]$  &         158&        0.23&        0.06\\
full                &  $ [0,1]$  &         158&        0.37&        0.11\\
adjunct, clinical, other&  $ [0,1]$  &         158&        0.15&        0.11\\
\\ \underline{\emph{prof. share, by field}}&            &            &            &            \\
engineering, math \& related sciences&  $ [0,1]$  &         158&        0.17&        0.13\\
humanities \& related sciences&  $ [0,1]$  &         158&        0.17&        0.10\\
medicine \& health sciences&  $ [0,1]$  &         158&        0.36&        0.30\\
social sciences     &  $ [0,1]$  &         158&        0.13&        0.09\\
natural sciences    &  $ [0,1]$  &         158&        0.17&        0.11\\
                    &            &            &            &            \\
num. prof., total   &$ [0,\infty)$&         158&      843.60&      555.92\\
num. schools        &$ [0,\infty)$&         158&       11.73&        5.76\\
num. departments    &$ [0,\infty)$&         158&       80.28&       40.55\\
\\ \underline{\emph{HERD data}}&            &            &            &            \\
total R\&D funding, M-\$&$ [0,\infty]$&         158&      469.55&      414.93\\
federal R\&D funding, M-\$&$ [0,\infty)$&         158&      249.60&      273.42\\
non-federal R\&D funding, M-\$&$ [0,\infty)$&         158&      219.96&      182.89\\
total wage bill, M-\$&$ [0,\infty)$&         158&      205.98&      183.85\\
num. principal investigators&$ [0,\infty)$&         158&      862.89&      729.41\\
\hline\hline
\end{tabular}
}

\begin{quote}\footnotesize
\emph{Note}: All variables are from the NSAR population, except for the last rows, which are sourced from HERD.
\end{quote}
\end{table}

%%%%%%%%%%%%%%%%%%%%%%%%%%%%%%%%%%%%%%%%%%%%%%%%%%%%%%%%%%%%%%%
%%%%%%%%%%%%%%%%%%%%%%%%%%%%%%%%%%%%%%%%%%%%%%%%%%%%%%%%%%%%%%%
\begin{table}[htbp]\centering\small
\caption{Professor-level sample summary statistics --- Professional and socio-demographic}
\label{tab_samp_sumstat}
{
\def\sym#1{\ifmmode^{#1}\else\(^{#1}\)\fi}
\begin{tabular}{lc*{1}{rrrrr}}
\hline\hline
                    &    type    &       count&        mean&          sd\\
\hline
\underline{\emph{rank}}&            &            &            &            \\
assistant           & $ \{0,1\}$ &       4,357&        0.25&        0.43\\
associate           & $ \{0,1\}$ &       4,357&        0.25&        0.43\\
full                & $ \{0,1\}$ &       4,357&        0.40&        0.49\\
adjunct, clinical, other& $ \{0,1\}$ &       4,357&        0.10&        0.30\\
\\ \underline{\emph{aggregate field}}&            &            &            &            \\
engineering, math \& related sciences& $ \{0,1\}$ &       4,357&        0.17&        0.37\\
humanities \& related sciences& $ \{0,1\}$ &       4,357&        0.19&        0.39\\
medicine \& health sciences& $ \{0,1\}$ &       4,357&        0.28&        0.45\\
social sciences     & $ \{0,1\}$ &       4,357&        0.16&        0.36\\
natural sciences    & $ \{0,1\}$ &       4,357&        0.20&        0.40\\
\\ \underline{\emph{tenure status}}&            &            &            &            \\
not on tenure track & $ \{0,1\}$ &       4,357&        0.22&        0.42\\
contract length (non tenure track)&$ [1,\infty)$&         932&        2.47&        1.98\\
pre-tenure          & $ \{0,1\}$ &       4,357&        0.21&        0.40\\
years until tenure eval. &$ [1,\infty)$&         880&        2.64&        1.94\\
tenured             & $ \{0,1\}$ &       4,357&        0.57&        0.49\\
years since tenure  &$ (-\infty,0)$&       2,414&       15.05&       11.57\\
\\ \underline{\emph{work hrs., earnings, research funding}}&            &            &            &            \\
work hours per week &$ [1,\infty)$&       4,357&       49.13&       13.54\\
work-hrs. share, research&  $ [0,1]$  &       4,357&        0.37&        0.22\\
work-hrs. share, fundraising&  $ [0,1]$  &       4,357&        0.09&        0.11\\
work-hrs. share, teaching&  $ [0,1]$  &       4,357&        0.27&        0.19\\
work-hrs. share, administration&  $ [0,1]$  &       4,357&        0.16&        0.15\\
work-hrs. share, clinical&  $ [0,1]$  &       4,357&        0.04&        0.15\\
work-hrs. share, other&  $ [0,1]$  &       4,357&        0.07&        0.12\\
own annual earnings &$ [0,\infty)$&       4,357&  153,184.53&   93,733.51\\
earnings share, base salary&  $ [0,1]$  &       4,357&        0.70&        0.34\\
earnings share, grant-sponsored&  $ [0,1]$  &       4,357&        0.14&        0.24\\
earnings share, supplemental&  $ [0,1]$  &       4,357&        0.03&        0.07\\
earnings share, other&  $ [0,1]$  &       4,357&        0.04&        0.12\\
earnings share, clinical&  $ [0,1]$  &       4,357&        0.02&        0.10\\
5-year guaranteed research funding&$ [0,\infty)$&       4,357&  398,931.03&  989,173.36\\
5-year fundraising expectations&$ [0,\infty)$&       4,357&  511,043.15&  974,479.03\\
\\ \underline{\emph{socio-demographics}}&            &            &            &            \\
asian               & $ \{0,1\}$ &       4,285&        0.12&        0.33\\
black               & $ \{0,1\}$ &       4,285&        0.03&        0.18\\
hispanic            & $ \{0,1\}$ &       4,285&        0.06&        0.23\\
other               & $ \{0,1\}$ &       4,285&        0.04&        0.19\\
white               & $ \{0,1\}$ &       4,285&        0.79&        0.41\\
US born             & $ \{0,1\}$ &       4,259&        0.73&        0.44\\
non-US born         & $ \{0,1\}$ &       4,259&        0.24&        0.43\\
married or in partnership& $ \{0,1\}$ &       4,214&        0.84&        0.36\\
num. dependents     & $ \{0,1\}$ &       4,276&        0.97&        1.12\\
household annual earnings&$ [0,\infty)$&       4,206&  254,698.05&  157,539.20\\
\hline\hline
\end{tabular}
}

\begin{quote}\footnotesize
\emph{Note}: All variables are from the NSAR sample. Any counts that do not equal the full sample size are due to the question either being not applicable (i.e., the tenure status questions) or were socio-demographic questions where respondents were not forced to answer.
\end{quote}
\end{table}

%%%%%%%%%%%%%%%%%%%%%%%%%%%%%%%%%%%%%%%%%%%%%%%%%%%%%%%%%%%%%%%
%%%%%%%%%%%%%%%%%%%%%%%%%%%%%%%%%%%%%%%%%%%%%%%%%%%%%%%%%%%%%%%
\begin{table}[htbp]\centering\small
\caption{Professor-level sample summary statistics --- Nature of research}
\label{tab_nat_sumstat}
{
\def\sym#1{\ifmmode^{#1}\else\(^{#1}\)\fi}
\begin{tabular}{lc*{1}{rrrrr}}
\hline\hline
                    &    type    &       count&        mean&          sd\\
\hline
\\ \underline{\emph{intended output}}&            &            &            &            \\
journal articles    &$ \{0,1,2\}$&       4,357&        1.78&        0.54\\
books               &$ \{0,1,2\}$&       4,357&        0.50&        0.68\\
research materials, tools, etc.&$ \{0,1,2\}$&       4,357&        0.67&        0.68\\
products            &$ \{0,1,2\}$&       4,357&        0.45&        0.64\\
\\ \underline{\emph{intended audience}}&            &            &            &            \\
academics           &$ \{0,1,2\}$&       4,357&        1.79&        0.53\\
policymakers        &$ \{0,1,2\}$&       4,357&        0.80&        0.69\\
businesses and organizations&$ \{0,1,2\}$&       4,357&        0.51&        0.62\\
general public      &$ \{0,1,2\}$&       4,357&        0.79&        0.63\\
\\ \underline{\emph{other}} \\ objective is to test&            &            &            &            \\
\hspace{3mm} (vs. generate) hypotheses& $ [0,10]$  &       4,357&        4.66&        2.68\\
risk of research    & $ [0,10]$  &       4,357&        4.39&        2.39\\
\hline\hline
\end{tabular}
}

\begin{quote}\footnotesize
\emph{Note}: All variables are from the NSAR sample. The variables of type ``\{0,1,2\}'' are conversions of a three-point likert scale increasing in intensity. The variables of type ``$[0,10]$'' were solicited directly as real numbers on a scale from 0 to 10.
\end{quote}
\end{table}

%%%%%%%%%%%%%%%%%%%%%%%%%%%%%%%%%%%%%%%%%%%%%%%%%%%%%%%%%%%%%%%
%%%%%%%%%%%%%%%%%%%%%%%%%%%%%%%%%%%%%%%%%%%%%%%%%%%%%%%%%%%%%%%
%%%%%%%%%%%%
\clearpage %FIGURES
%%%%%%%%%%%%

%%%%%%%%%%%%%%%%%%%%%%%%%%%%%%%%%%%%%%%%%%%%%%%%%%%%%%%%%%%%%%%
%%%%%%%%%%%%%%%%%%%%%%%%%%%%%%%%%%%%%%%%%%%%%%%%%%%%%%%%%%%%%%%

%%%%%%%%%%%%%%%%%%%%%%%%%%%%%%%%%%%%%%%%%%%%%%%%%%%%%%%%%%%%%%%
%%%%%%%%%%%%%%%%%%%%%%%%%%%%%%%%%%%%%%%%%%%%%%%%%%%%%%%%%%%%%%%
\begin{figure}[htbp]\centering\footnotesize
\caption{Earnings per institutions, ranks, tasks, and sources}
\label{figs_sal_decompose}
\subfloat[Earnings across institutions]{
\includegraphics[width=0.475\linewidth, trim = 0mm 0mm 0mm 0mm , clip]{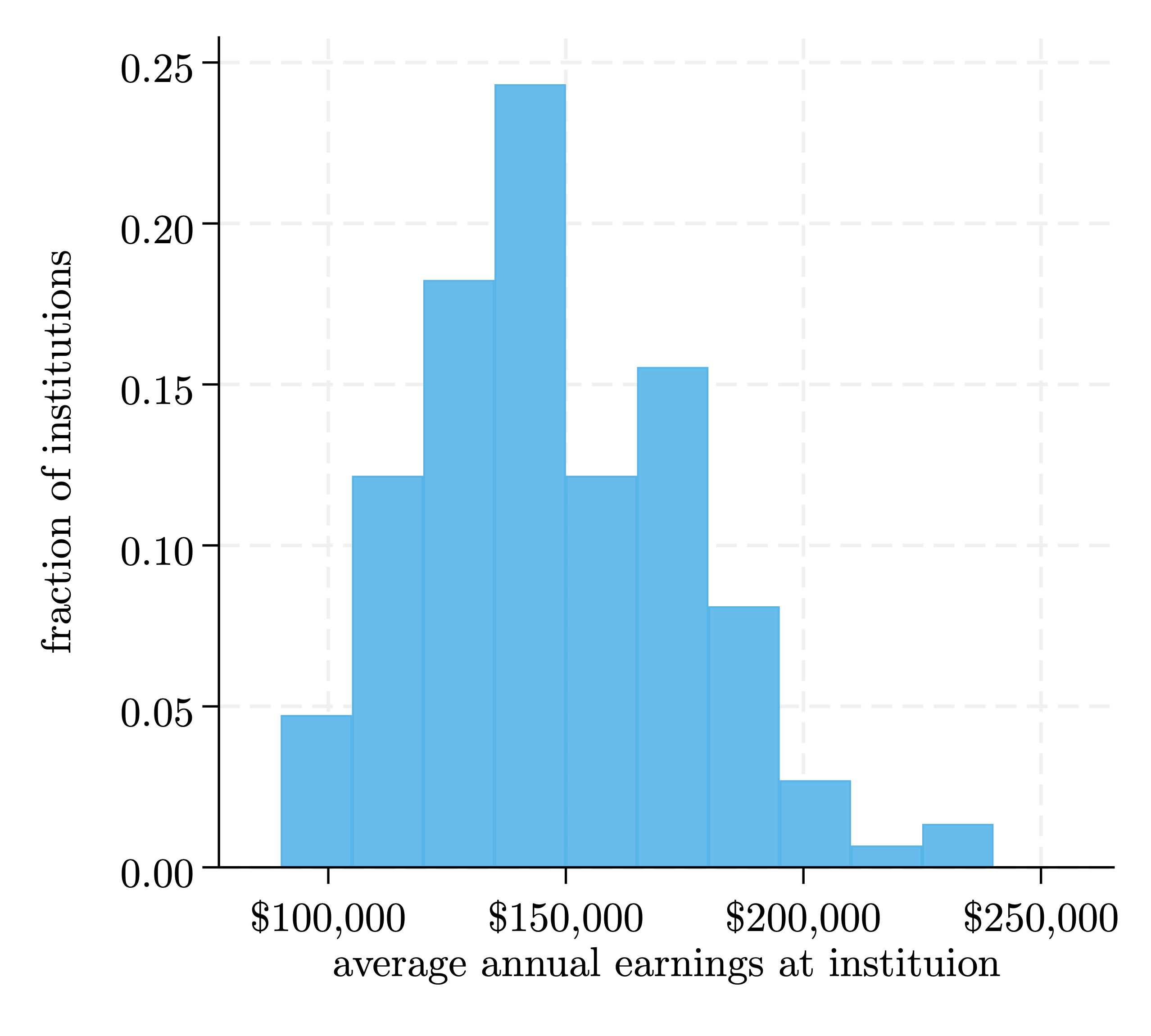}}
\subfloat[Observational earnings regression, by task]{
\includegraphics[width=0.475\linewidth, trim = 0mm 0mm 0mm 0mm , clip]{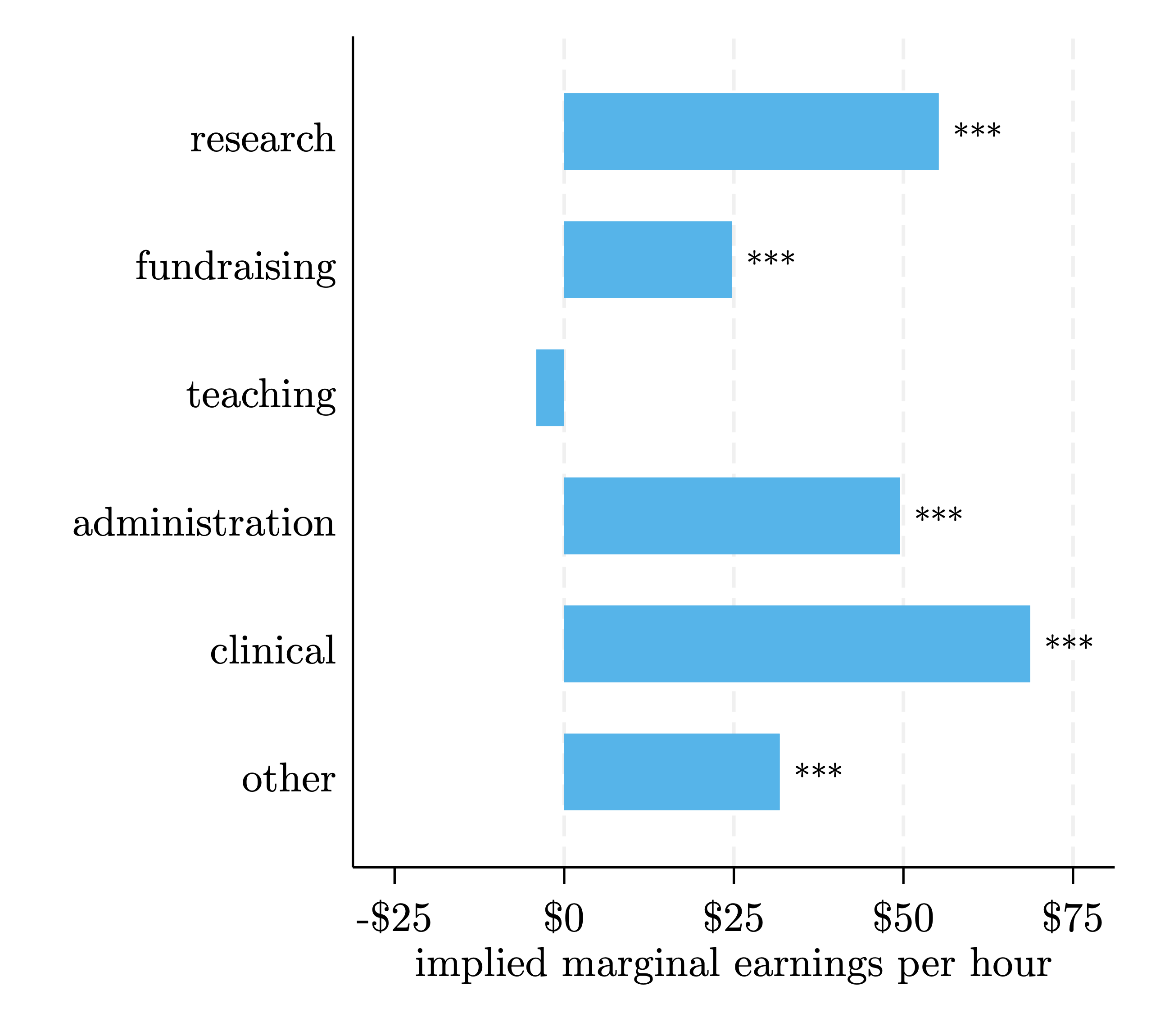}}\\
\-\\ \-\\
\subfloat[Observational earnings regression, by source]{
\includegraphics[width=0.475\linewidth, trim = 0mm 0mm 0mm 0mm , clip]{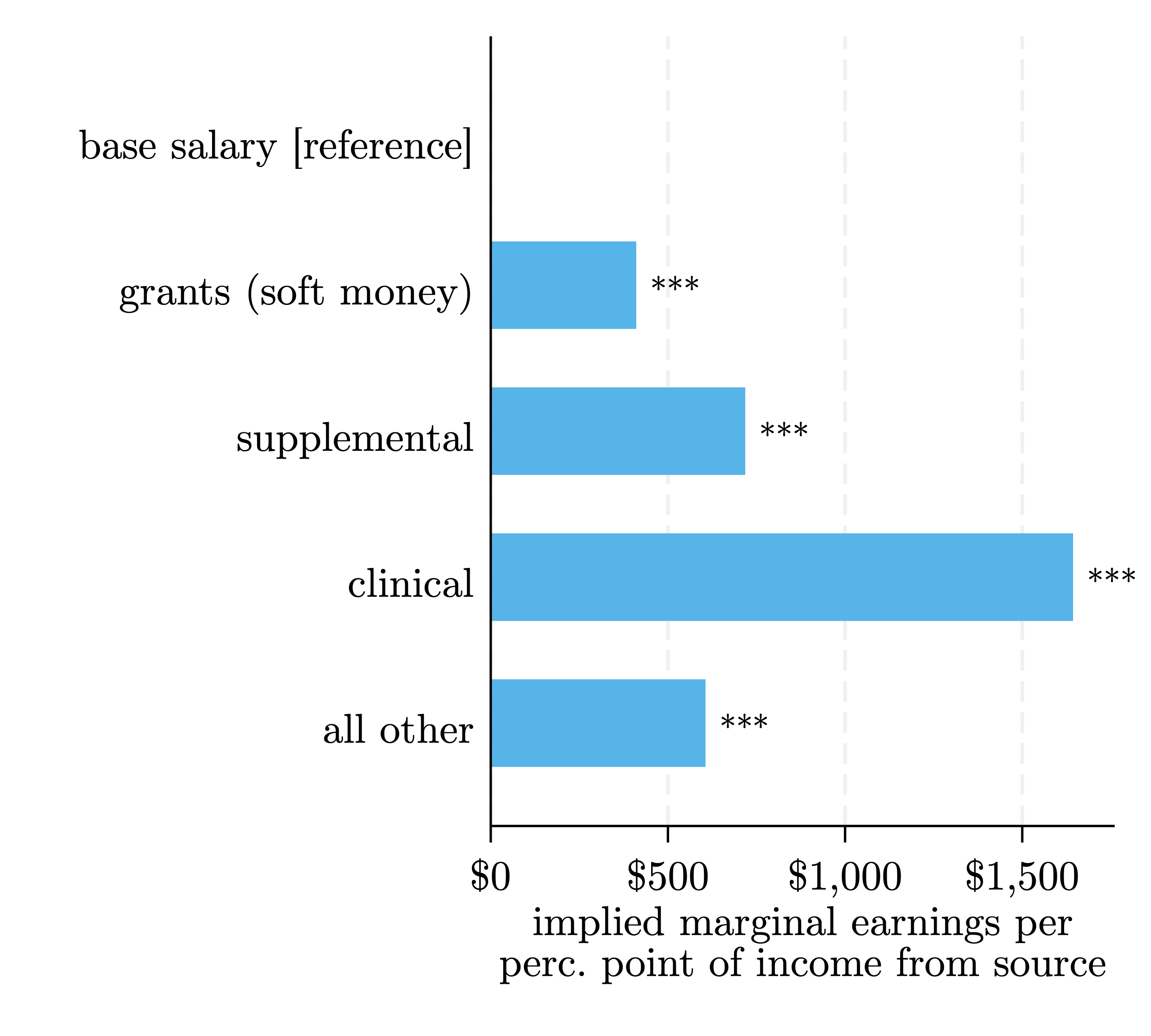}}
\subfloat[Decomposing heterogeneity]{\small{
\def\sym#1{\ifmmode^{#1}\else\(^{#1}\)\fi}
\begin{tabular}{l*{5}{c}}
\hline\hline
            &\multicolumn{1}{c}{(1)}         &\multicolumn{1}{c}{(2)}         &\multicolumn{1}{c}{(3)}         &\multicolumn{1}{c}{(4)}         &\multicolumn{1}{c}{(5)}         \\
\hline

partial $ R^2$: \\ \hspace{2mm} field&        0.08         &        0.11         &        0.07         &        0.11         &        0.14         \\
\hspace{2mm} inst. \& rank&                     &        0.45         &                     &                     &        0.43         \\
\hspace{2mm} tasks&                     &                     &        0.16         &                     &        0.08         \\
\hspace{2mm} sources&                     &                     &                     &        0.38         &        0.37         \\
\\ total $ R^2$&        0.08         &        0.36         &        0.20         &        0.33         &        0.58         \\
\hline\hline
\end{tabular}
}
}\\
\begin{quote}
\emph{Note}: \input{figtab/stub_figs_sal_decompose_note1.tex}
\end{quote}
\end{figure}

%%%%%%%%%%%%%%%%%%%%%%%%%%%%%%%%%%%%%%%%%%%%%%%%%%%%%%%%%%%%%%%
%%%%%%%%%%%%%%%%%%%%%%%%%%%%%%%%%%%%%%%%%%%%%%%%%%%%%%%%%%%%%%%
\begin{figure}[htbp]\centering\footnotesize
\caption{Earnings per research metrics}
\label{figs_outputsal_decompose}
\subfloat[Observational earnings regression, by research metric]{
\includegraphics[width=0.75\linewidth, trim = 0mm 0mm 0mm 0mm , clip]{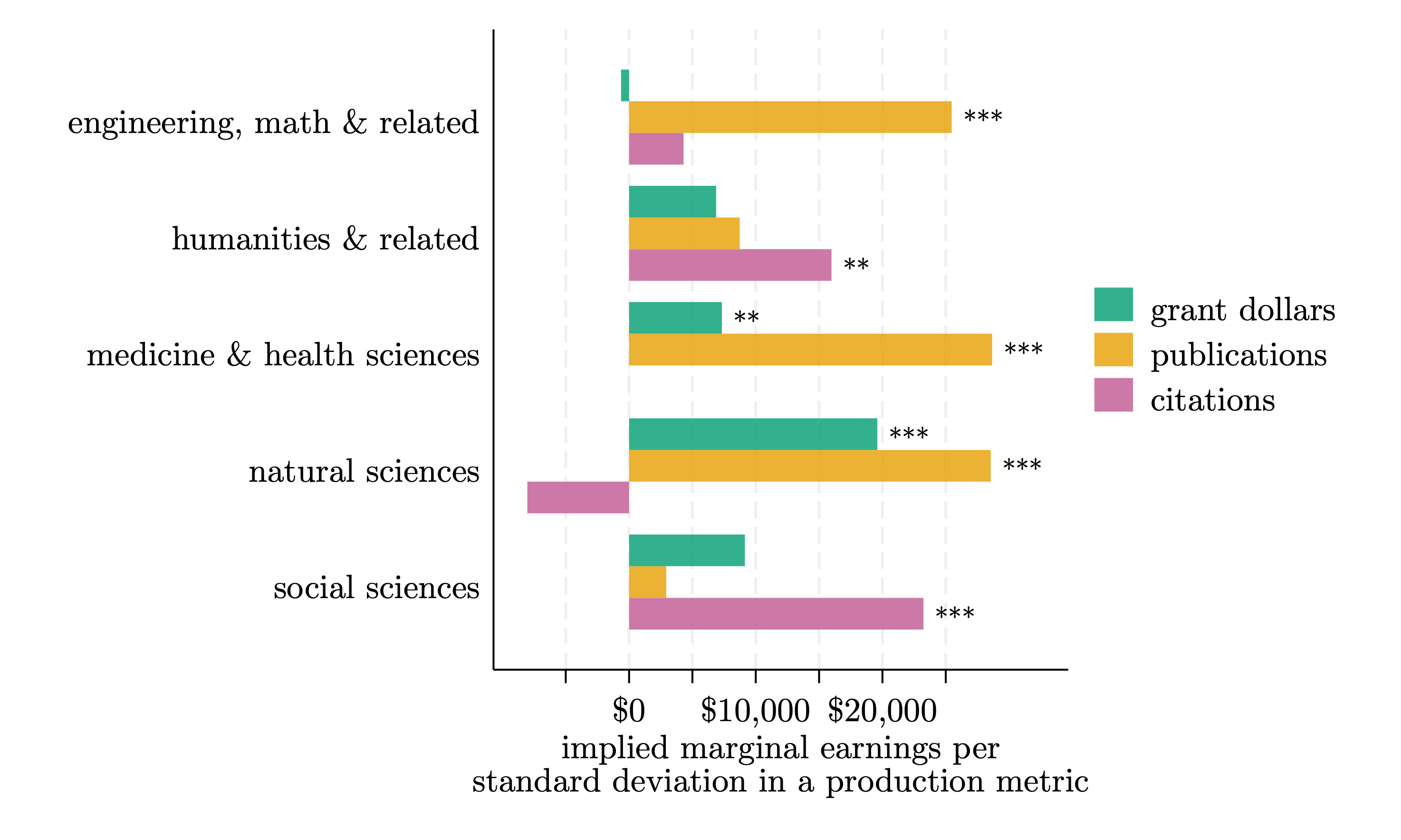}}\\ \-\\ \-\\
\subfloat[Decomposing heterogeneity]{\small{
\def\sym#1{\ifmmode^{#1}\else\(^{#1}\)\fi}
\begin{tabular}{l*{7}{c}}
\hline\hline
            &\multicolumn{1}{c}{(1)}         &\multicolumn{1}{c}{(2)}         &\multicolumn{1}{c}{(3)}         &\multicolumn{1}{c}{(4)}         &\multicolumn{1}{c}{(5)}         &\multicolumn{1}{c}{(6)}         &\multicolumn{1}{c}{(7)}         \\
\hline

partial $ R^2$: \\ \hspace{2mm} field&        0.07         &        0.08         &        0.09         &        0.08         &        0.09         &        0.13         &        0.14         \\
\hspace{2mm} grant dollars&                     &        0.03         &                     &                     &        0.01         &                     &        0.02         \\
\hspace{2mm} publications&                     &                     &        0.10         &                     &        0.03         &                     &        0.02         \\
\hspace{2mm} citations&                     &                     &                     &        0.07         &        0.01         &                     &        0.01         \\
\hspace{2mm} inst. \& rank&                     &                     &                     &                     &                     &        0.47         &        0.35         \\
\hspace{2mm} tasks&                     &                     &                     &                     &                     &        0.09         &        0.09         \\
\hspace{2mm} sources&                     &                     &                     &                     &                     &        0.35         &        0.38         \\
\\ total $ R^2$&        0.07         &        0.10         &        0.16         &        0.14         &        0.18         &        0.57         &        0.60         \\
\hline\hline
\end{tabular}
}
}\\  \-\\
\begin{quote}
\emph{Note}: \input{figtab/stub_figs_corrs_outputsal_note1.tex}
\end{quote}
\end{figure}

%%%%%%%%%%%%%%%%%%%%%%%%%%%%%%%%%%%%%%%%%%%%%%%%%%%%%%%%%%%%%%%
%%%%%%%%%%%%%%%%%%%%%%%%%%%%%%%%%%%%%%%%%%%%%%%%%%%%%%%%%%%%%%%

%%%%%%%%%%%%%%%%%%%%%%%%%%%%%%%%%%%%%%%%%%%%%%%%%%%%%%%%%%%%%%%
%%%%%%%%%%%%%%%%%%%%%%%%%%%%%%%%%%%%%%%%%%%%%%%%%%%%%%%%%%%%%%%

%%%%%%%%%%%%%%%%%%%%%%%%%%%%%%%%%%%%%%%%%%%%%%%%%%%%%%%%%%%%%%%
%%%%%%%%%%%%%%%%%%%%%%%%%%%%%%%%%%%%%%%%%%%%%%%%%%%%%%%%%%%%%%%

%%%%%%%%%%%%%%%%%%%%%%%%%%%%%%%%%%%%%%%%%%%%%%%%%%%%%%%%%%%%%%%
%%%%%%%%%%%%%%%%%%%%%%%%%%%%%%%%%%%%%%%%%%%%%%%%%%%%%%%%%%%%%%%

%%%%%%%%%%%%%%%%%%%%%%%%%%%%%%%%%%%
%----------------------------------------------------------------------------------------%
%APPENDIX
%----------------------------------------------------------------------------------------%
%%%%%%%%%%%%%%%%%%%%%%%%%%%%%%%%%%%
\appendix
\onehalfspacing
\clearpage
\fancypagestyle{mystyle}{%
    \fancyhead{}
    \fancyhead[C]{Appendices}\fancyfoot{}
}%
\setlength{\headheight}{15pt}
\addtolength{\topmargin}{-2.5pt}
\thispagestyle{mystyle}

%----------------------------------------------------------------------------------------%
\section{Additional information on methods}\label{sec_app_extramethods}
\setcounter{figure}{0}
\renewcommand{\thefigure}{A\arabic{figure}}
\setcounter{table}{0}
\renewcommand{\thetable}{A\arabic{table}}
\setcounter{equation}{0}
\renewcommand{\theequation}{A\arabic{equation}}
%%%%%%%%%%%%%%%%%%%%%%%%%%%%%%%%%%%

%----------------------------------------------------------------------------------------%
\subsection{Population data collection}\label{sec_app_extramethods_data}
%----------------------------------------------------------------------------------------%
Our process for the population data collection was as follows. We hired workers to collect publicly available contact information. The workers received detailed instructions on the layout of institutions' websites, the required fields, and data formatting guidelines. After a first round of data collection, the research team manually checked each submitted file by navigating to the relevant institutions' webpage and using random spot-checking to ensure the collected information was complete, accurate, and correctly formatted. If discrepancies were identified, workers were then instructed to add any missing information to ensure a complete record. This process repeated as many times as necessary to complete each file. Finally, the research team cleaned the files to ensure consistent formatting.

%----------------------------------------------------------------------------------------%
\subsection{Survey instrument}\label{sec_app_instrument}
%----------------------------------------------------------------------------------------%
Below is an abbreviated version of the survey instrument, including the wording of questions and response formats. Some components of the survey not referenced in this paper, as well as other guidance, conditional logic requirements, and notes provided to respondents, are not shown here. Thus, not all respondents were shown all of the following questions.

\begin{enumerate}
    \item Which of the following best describes your current tenure status or applicability?
        \begin{itemize}
\item On tenure track, not tenured
\item Tenured
\item Not on tenure track or tenure does not apply
        \end{itemize}
    \item In what year do you expect to be evaluated for tenure at your primary institution?
        \begin{itemize}
\item 2022 (this year)
\item 2023 (in about 1 year)
\item 2024 (in about 2 years)
\item ... (continue in this format)
\item 2033 or after
        \end{itemize}
    \item In what year did you first receive tenure at any institution?
        \begin{itemize}
\item 2022
\item 2021
\item 2020
\item ... (continue in this format)
\item 1965 or earlier
        \end{itemize}
    \item In what year do you expect your current contract or appointment with your primary institution to be reviewed or evaluated?
        \begin{itemize}
\item 2022 (this year)
\item 2023 (in about 1 year)
\item 2024 (in about 2 years)
\item ... (continue in this format)
\item 2033 or after
        \end{itemize}
    \item Approximately how often are your contracts or appointments with your primary institution reviewed or evaluated?
        \begin{itemize}
\item Every year
\item Every 2 years
\item Every 3 years
\item ... (continue in this format)
\item Every 10 or more years
        \end{itemize}
    \item Over the next 5 years, what is your best guess as to how many hours per week you will work on average? (Please try to include any time spent actively engaged in your research, teaching, administration, or effort towards any other positions you hold.)
        \begin{itemize}
\item 10 hours or less per week
\item 11-15 hours per week
\item 16-20 hours per week
\item ... (continue in this format)
\item More than 90 hours per week
        \end{itemize}
    \item Over the next 5 years, what is your best guess as to how your work time will be spent across the following categories in percentage terms? (Please enter a number between 0 and 100 for each category. Choose the category that best describes the work. Your answers must sum to 100.)
        \begin{itemize}
\item Research (including the supervision of others) : \_\_\_\_\_\_\_
\item Fundraising for your research : \_\_\_\_\_\_\_
\item Teaching or advising (not as a part of your own research) : \_\_\_\_\_\_\_
\item Administration or committee service at your institution : \_\_\_\_\_\_\_
\item Clinical and/or medical practice : \_\_\_\_\_\_\_
\item All other (e.g., professional service, consulting, other) : \_\_\_\_\_\_\_
\item Total : \_\_\_\_\_\_\_
        \end{itemize}
    \item Approximately, what do you expect your earnings to be from each of the following sources this year? (Note: Do not count earnings in more than one option; choose the option that best describes the source. Please report a pre-tax estimate. If you are in a "soft money" position or you must fund all or a portion of your salary, please report the amount of your salary you must cover in the "Salary covered by research grants or awards" category.)
        \begin{itemize}
\item Primary institution: Guaranteed base salary (e.g., “hard money”): \_\_\_\_\_\_\_
\item Primary institution: Salary covered bt research grants or awards (e.g., “soft money”) : \_\_\_\_\_\_\_
\item Primary institution: Supplemental teaching or other duties: \_\_\_\_\_\_\_
\item Primary institution: Clinical practice and/or medical practice: \_\_\_\_\_\_\_
\item All other wages or salaries from positions outside your primary institution: \_\_\_\_\_\_\_
\item {Each category includes a dropdown menu of dollar amounts in \$5,000 increments up to \$100,000, then \$10,000 increments to \$500,000, then a single option for \$500,000 and above.}
        \end{itemize}
    \item Over the next 5 years, approximately how much total research funding are you guaranteed to have from any previous or active funding streams? (Note: Please ignore any indirect or overhead costs and report only funding amounts that you can directly spend. Include funding remaining from any "start-up" packages, internal awards, external grants, or any other funding that you have already obtained and/or are guaranteed to receive. If in doubt, please take "guaranteed" to imply there is more than a 95% chance you will receive the funding.)
        \begin{itemize}
\item No guaranteed funding
\item \$5,000 total (avg. \$1.0k per year) or less
\item \$12,500 total (avg. \$2.5k per year)
\item \$25,000 total (avg. \$5.0k per year)
\item \$37,500 total (avg.\$7.5k per year)
\item \$50,000 total (avg. \$10k per year)
\item ... (continue in this format)
\item \$250,000,000 total (avg. \$50M per year) or more
        \end{itemize}
    \item Over the next 5 years, approximately how much funding do you expect to obtain from new research awards given the amount of fundraising you plan to do? (Note: Please ignore any indirect or overhead costs and do not count any guaranteed funding or awards.)
        \begin{itemize}
\item No fundraising expected
\item \$5,000 total (avg. \$1.0k per year) or less
\item \$12,500 total (avg. \$2.5k per year)
\item \$25,000 total (avg. \$5.0k per year)
\item \$37,500 total (avg. \$7.5k per year)
\item \$50,000 total (avg. \$10k per year)
\item ... (continue in this format)
\item \$250,000,000 total (avg. \$50M per year) or more
        \end{itemize}
    \item Use the scale below to rate whether the overall objective of your current research is more (A) to generate new theories and hypotheses, or is more (B) to test existing theories and hypotheses, or is somewhere in between.
        \begin{itemize}
\item {Slider scale spanning 0[generate theories] – 10[test theories]}
        \end{itemize}
    \item In general, how risky do you think your current research projects are? (Note: Use this scale from 0 to 10, where 0 means “very safe” and 10 means “very risky”)
        \begin{itemize}
\item {Slider scale spanning 0[very safe] – 10[very risky]}
        \end{itemize}
    \item In general, how risky do you think your peers think your current research projects are? (Note: Use this scale from 0 to 10, where 0 means your peers think your projects are “very safe” and 10 means “very risky”)
        \begin{itemize}
\item {Slider scale spanning 0[very safe] – 10[very risky]}
        \end{itemize}
    \item How often are the following items the intended outputs of your work? {Never or barely (0); Sometimes (1); Most or all of the time (2)}
        \begin{itemize}
\item Academic publications in journals or proceedings
\item Books
\item Data, instruments, materials, methods, software, or tools for other researchers
\item Any kind of consumer-oriented or practical application (e.g., products, patents, policies, etc.)
        \end{itemize}
    \item How often are the following groups the intended audience of your work? {Never or barely (0); Sometimes (1); Most or all of the time (2)}
        \begin{itemize}
\item Other academic researchers
\item Policymakers, governments, or other public organizations
\item Businesses or other private organizations
\item General public
        \end{itemize}
    \item In general, how willing are you to take risks in your personal life? (Note: Use this scale from 0 to 10, where 0 means “completely unwilling” and 10 means “very willing”)
        \begin{itemize}
\item {Slider scale spanning 0[completely unwilling] – 10[very willing]}
        \end{itemize}
    \item What is your household’s total annual income from all sources including salaries, bonuses, investments, etc.? (Note: Include all income received by any persons, including yourself, residing in your home. Please report a pre-tax estimate)
        \begin{itemize}
\item {Dropdown menu of dollar amounts in \$5,000 increments up to \$100,000, then \$10,000 increments to \$200,000, then \$25,000 increments up to \$500,000, then \$50,000 increments up to \$1,000,000, then \$250,000 increments up to \$3,000,000, then a single option for \$3,000,000 and above.}
        \end{itemize}
    \item What is your age?
        \begin{itemize}
\item 19 or younger
\item 20-24  
\item ...
\item 75-79  
\item 80 or older  
\item Prefer not to say  
        \end{itemize}
    \item What is your gender identity?
        \begin{itemize}
\item Female 
\item Male  
\item Non-binary, genderqueer, or other not listed  
\item Prefer not to say  
        \end{itemize}
    \item What best describes your race and/or ethnicity? (Note: You may choose more than one. These categories are based on the U.S. Census Bureau's definitions.)
        \begin{itemize}
\item American Indian or Alaska Native 
\item Asian 
\item Black or African American  
\item Hispanic, Latino, or Spanish  
\item Native Hawaiian or Other Pacific Islander  
\item White or Caucasian  
\item Other, not listed  
        \end{itemize}
    \item What best describes your U.S. citizenship status?
        \begin{itemize}
\item Domestic-born, U.S. citizen  
\item Foreign-born, Naturalized U.S. citizen or Legal Permanent Resident  
\item Foreign-born, non-citizen and non-permanent resident
\item Prefer not to say            
        \end{itemize}
    \item How many generations have you and your direct ancestors lived in the U.S.?
        \begin{itemize}
\item 1
\item 2
\item 3
\item 4 or more
\item Prefer not to say
        \end{itemize}
    \item What is your relationship or marital status?
        \begin{itemize}
\item Single  
\item Married or in a domestic partnership  
\item Other  
\item Prefer not to say  
        \end{itemize}
\end{enumerate}

%----------------------------------------------------------------------------------------%
\subsection{Population and sample comparisons}\label{sec_app_extramethods_popsampcompare}
%----------------------------------------------------------------------------------------%
A key challenge for all surveys is ensuring representativeness. Here we report three tests of representativeness using data that is observable for both respondents and non-respondents. First, we test for differences in means of the few observable variables extracted from professors' online information: their aggregate field and their rank. Table \ref{tab_ttest_pop_samp} reports $t$ tests showing both that our respondents includes a slightly larger share of full professors with a relatively equal smaller share of adjunct, clinical or other professors. We also see under-responding from the medical and health sciences relative to all other fields that is sizeable; the recruited share in this field is nearly 45\% while the respondent share accounts for just under 30\%. Our best hypothesis as to this difference is that is driven in part by the unique structure of medical schools, which account for the vast number of professors in this field. The title of ``professor'' can often imply notably different duties and positions at medical schools relative to the rest of academia, with professors at medical schools often carrying a significantly larger load of clinical duties. Thus, to some individuals we recruit from these schools, a survey soliciting responses about one's ``research'' may seem disproportionately irrelevant. This, combined with the relatively less flexible job arrangements given their clinical duties, may be the cause for this discrepancy. Altogether, this implies that caution should be had when interpreting results for this particular field.

\begin{table}[htbp]\centering\small
\caption{Test of differences between recruited and respondent samples}
\label{tab_ttest_pop_samp}
{
\def\sym#1{\ifmmode^{#1}\else\(^{#1}\)\fi}
\begin{tabular}{l*{1}{ccc}}
\hline\hline
                    &\multicolumn{2}{c}{\underline{mean}}&                     \\
                    & respondents&non-respondents&       diff.   \\
\hline
assistant           &        0.26&        0.25&        0.01   \\
associate           &        0.23&        0.22&        0.01   \\
full                &        0.40&        0.35&        0.05***\\
adjunct, clinical, other&        0.11&        0.18&       --0.07***\\
engineering, math \& related sciences&        0.17&        0.15&        0.02***\\
humanities \& related sciences&        0.19&        0.15&        0.04***\\
medicine \& health sciences&        0.28&        0.44&       --0.15***\\
social sciences     &        0.16&        0.11&        0.04***\\
natural sciences    &        0.20&        0.15&        0.05***\\
\hline
$N$ obs. &        4,357 &      126,428 & \\
\hline\hline
\end{tabular}
}

\begin{quote}\footnotesize
\emph{Note}: Based on 130,785 e-mail-based observations in the 50\% of the population sampled for recruitment. $t$ test of differences in means; * $p<0.1$, ** $p<0.05$, *** $p<0.01$.

\end{quote}
\end{table}

\begin{figure}[htbp] \centering
\caption{Recruitment versus completion sample comparison per HERD metrics}\label{bigfig_nonresponse_herd}
\subfloat[Distributions per HERD R\&D measures]{\label{fig_nonresponse_herd}\includegraphics[width=0.95\textwidth, trim=0mm 0mm 0mm 0mm, clip]{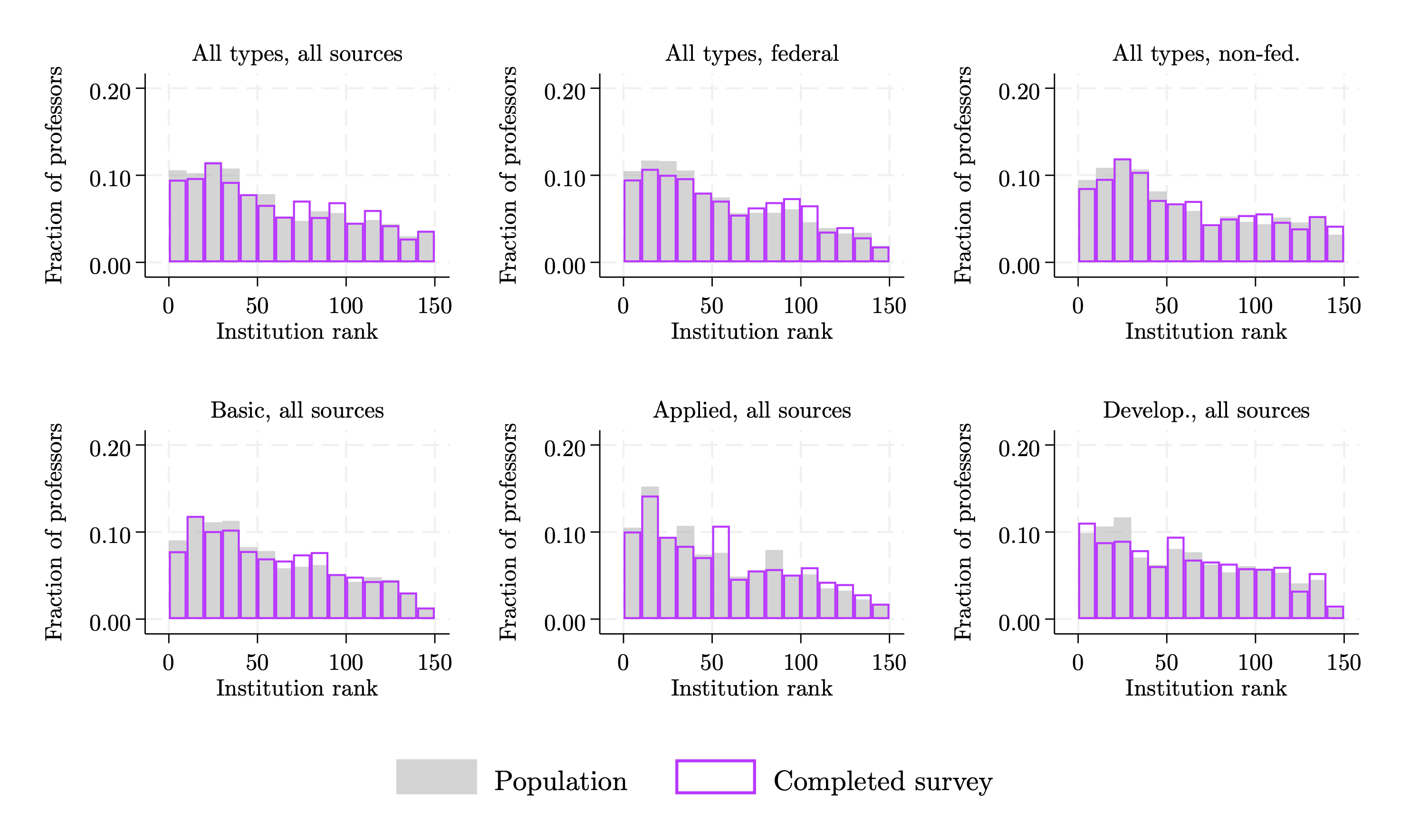}}\-\\
\subfloat[Regression estimates of mean differences]{\label{tab_nonresponse_herd}\centering \footnotesize {
\def\sym#1{\ifmmode^{#1}\else\(^{#1}\)\fi}
\begin{tabular}{l*{7}{c}}
\hline\hline
& \multicolumn{3}{c}{All sources, by type} & &\multicolumn{3}{c}{All types, by source} \\ 
& All & Federal & Non--fed. & & Basic & Applied & Develop. \\
            &\multicolumn{1}{c}{(1)}         &\multicolumn{1}{c}{(2)}         &\multicolumn{1}{c}{(3)}         &            &\multicolumn{1}{c}{(4)}         &\multicolumn{1}{c}{(5)}         &\multicolumn{1}{c}{(6)}         \\
\hline
Completed survey&      --35.01\sym{***}&      --21.30\sym{***}&      --13.71\sym{***}&            &      --24.31\sym{***}&      --7.170\sym{***}&      --3.812\sym{***}\\
            &     (6.953)         &     (4.511)         &     (3.062)         &            &     (4.867)         &     (2.492)         &     (1.328)         \\
[1em]
Constant    &       649.1\sym{***}&       351.7\sym{***}&       297.4\sym{***}&            &       413.2\sym{***}&       176.7\sym{***}&       60.46\sym{***}\\
            &     (1.308)         &     (0.863)         &     (0.578)         &            &     (0.917)         &     (0.459)         &     (0.254)         \\
[1em]
\hline
\% diff.    &      --5.4\%         &      --6.1\%         &      --4.6\%         &            &      --5.9\%         &      --4.1\%         &      --6.3\%         \\
$ N$ obs.   &     130,785         &     130,785         &     130,785         &            &     130,785         &     130,735         &     128,169         \\
\hline\hline
\end{tabular}
}
}
\begin{quote} \footnotesize
\emph{Note}: Panel (a) compares the distributions of e-mailed professors and respondents per the rank of their institution along each dimension of R\&D funding. Panel (b) reports estimates from a regression of each sampled professor's institutional R\&D funding (in 2019 \$-M) on a dummy for whether the sampled individual completed the survey; the ``\% diff.'' row reports the mean difference in the measure as a percentage of the non-respondent average (i.e., it is the ratio of the two coefficients); robust standard errors in parentheses; $^{*}$ \(p<0.1\), $^{**}$ \(p<0.05\), $^{***}$ \(p<0.01\). All institutional data is from the 2019 NSF HERD (\citealt{nsf2023herd}).
\end{quote}
\end{figure}

Figure \ref{bigfig_nonresponse_herd} shows this comparison based on data from the National Science Foundation's HERD survey (\citealt{nsf2023herd}), which reports institutional-level data on the total amount of funding flows into all of the institutions in our population (recall, our population was constructed using the HERD, which is why this data is available for the full population).

Figure \ref{bigfig_nonresponse_herd} reports the distributions and regression tests for mean differences in six metrics of institution-level research funding that compare respondents to the full set of researchers invited to participate (``Sample e-mailed''). The distributions overlap to a large degree (see Panel a). We do estimate statistically significant differences in means (see Panel b), but the magnitudes of these differences are all in the range of approximately 4--6\%. 

\begin{figure}[htbp] \centering
\caption{Recruitment versus completion sample comparison per Dimensions metrics}\label{bigfig_nonresponse_dim}
\subfloat[Distributions per Dimensions publications and grants]{\label{fig_nonresponse_dim}\includegraphics[width=0.8\textwidth, trim=0mm 0mm 0mm 0mm, clip]{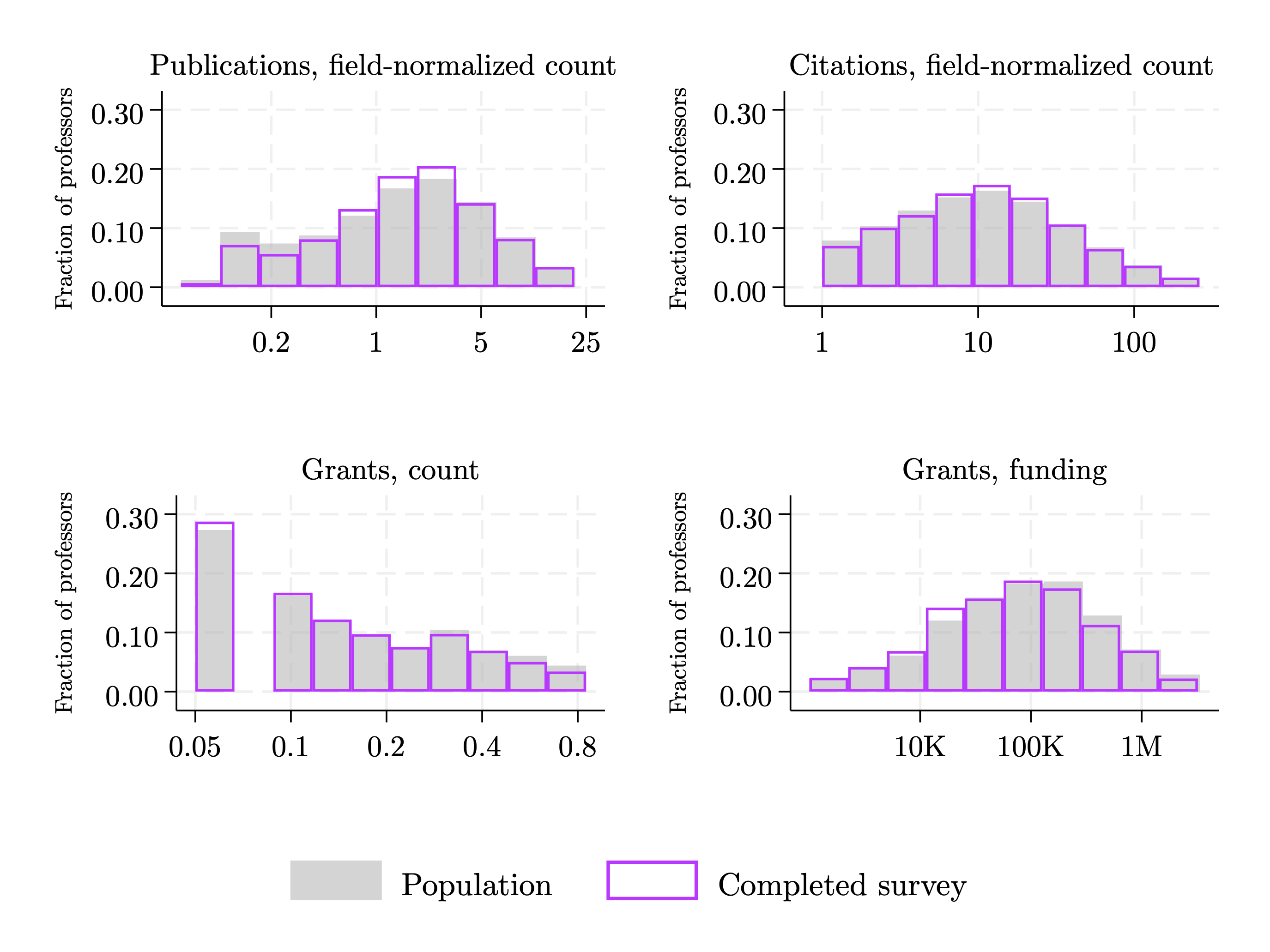}}\-\\
\subfloat[Regression estimates of mean differences]{\label{tab_nonresponse_dim}\centering \footnotesize {
\def\sym#1{\ifmmode^{#1}\else\(^{#1}\)\fi}
\begin{tabular}{l*{4}{c}}
\hline\hline
& Pub., & Pub. cites, & Grant, & Grant, \\
& count & normalized & count & total \$ \\
            &\multicolumn{1}{c}{(1)}         &\multicolumn{1}{c}{(2)}         &\multicolumn{1}{c}{(3)}         &\multicolumn{1}{c}{(4)}         \\
\hline
Completed survey&      --0.116\sym{*}  &      --0.573         &     0.00857\sym{**} &     --5250.7         \\
            &    (0.0689)         &     (0.810)         &   (0.00344)         &    (8556.2)         \\
[1em]
Constant    &       2.902\sym{***}&       20.56\sym{***}&       0.117\sym{***}&    121115.1\sym{***}\\
            &    (0.0158)         &     (0.186)         &  (0.000717)         &    (2100.2)         \\
[1em]
\hline
\% diff.    &      --4.0\%         &      --2.8\%         &       7.3\%         &      --4.3\%         \\
$ N$ obs.   &      86,504         &      86,504         &      86,504         &      86,504         \\
\hline\hline
\end{tabular}
}
}
\begin{quote} \footnotesize
\emph{Note}: Panel (a) compares the distributions of e-mailed professors and respondents per each dimension of individual-level publication output and grant funding per year (2003--2022). Panel (b) reports estimates from a regression of each sampled professor's publication or grant metric on a dummy for whether the sampled individual completed the survey; the ``\% diff.'' row reports the mean difference in the measure as a percentage of the non-respondent average (i.e., it is the ratio of the two coefficients); robust standard errors in parentheses; $^{*}$ \(p<0.1\), $^{**}$ \(p<0.05\), $^{***}$ \(p<0.01\). All publication and grant data is from the Dimensions database (\citealt{dimension2018data}).
\end{quote}
\end{figure}

Figure \ref{bigfig_nonresponse_dim} reports the results of a similar exercise, instead using individual-level data on researchers' publication output and grant receipts. This data was obtained by performing a fuzzy match of our population (i.e., using names and institution data) to the Dimensions database (\citealt{dimension2018data}), which includes disambiguated researcher-level records. We focus on researchers' publications and grants during the twenty years prior (2003-2022) and see very little differences between our respondents and the full set of individuals invited to the survey. The distributions have strong overlap over the full support (see Panel a), and the mean differences are all insignificant and/or smaller than 7.5\%.

%----------------------------------------------------------------------------------------%
\subsection{Self- and publicly-reported salary comparisons}\label{sec_app_extramethods_salcompare}
%----------------------------------------------------------------------------------------%
One limitation of our survey-based methodology is that respondents do not report truthfully due to inattention or some bias such social desirability or an experimenter demand effect. Testing for such issues is notoriously challenging. But one of the survey questions, which solicits respondents' annual salary, provides one test of the degree to which respondents are reporting truthfully since a sub-sample of respondents' salaries are publicly observable. 

A number of, primarily publicly owned, institutions in our population publicly post the salaries of their workforce. Using this data we can estimate the extent to which researchers' self- and publicly-reported salaries align. 

We used a manual approach to match respondents to their public records (when possible), relying on researchers' names and affiliations to make as high-fidelity of a match as possible. Across and within the 89 institutions that had public salary data, there was a significant amount of heterogeneity in how salaries were reported. In the matching process we made note of any irregularities or confusion for each record matched. Below, we focus only on matches where both  we had a high degree of confidence that both (1) the respondent was in fact the individual listed in the public record, and (2) the salary listed represented the full annual salary of that individual.

\begin{figure}[htbp] \centering
\caption{Self- and publicly-reported salaries}\label{fig_salcompare_pub_slf}
\includegraphics[width=0.6\textwidth, trim=0mm 0mm 0mm 0mm, clip]{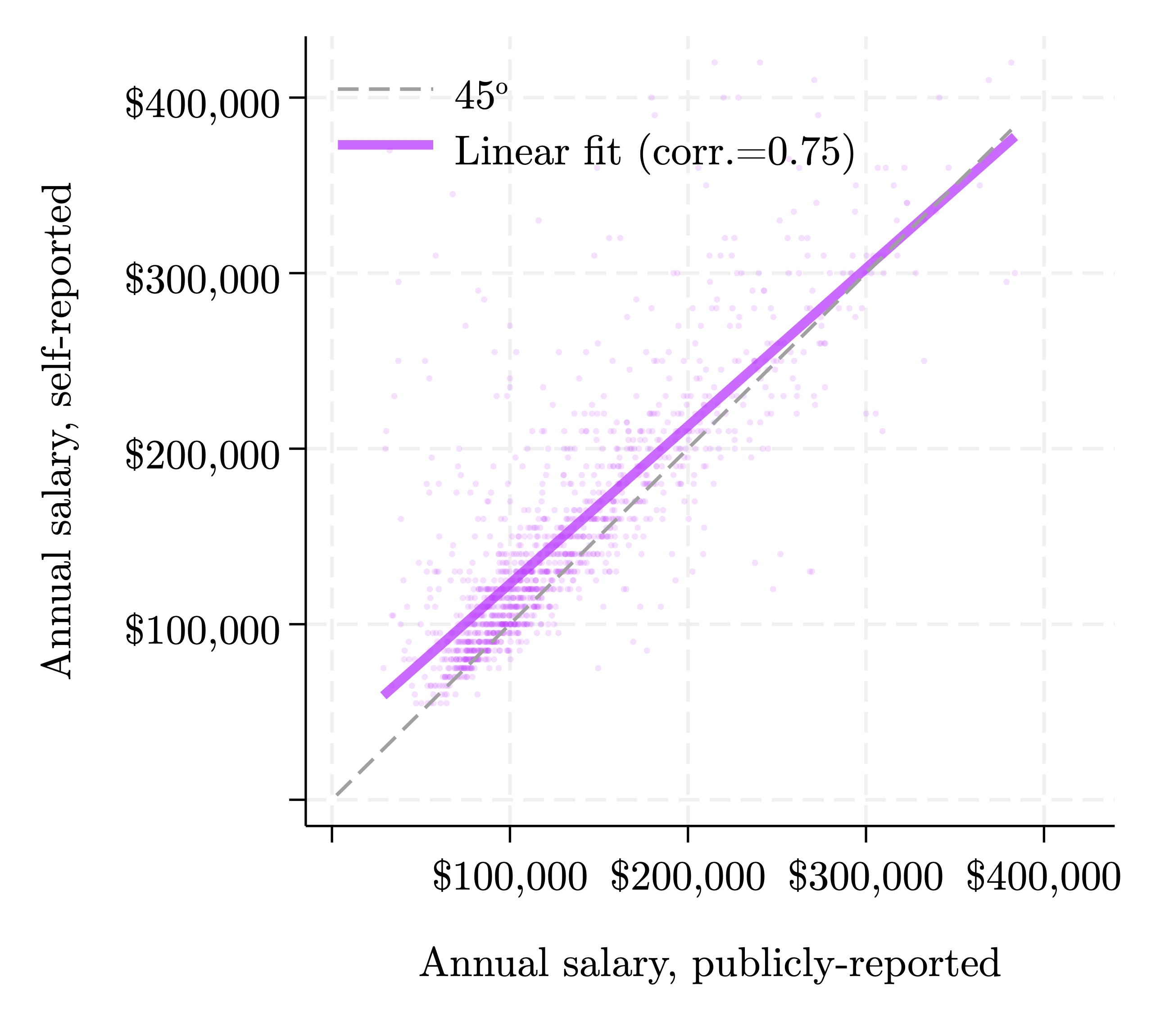}
\begin{quote} \footnotesize
\emph{Note}: Shows a scatterplot of salaries by source along with a linear fit and 45$^o$ line for comparison.
\end{quote}
\end{figure}

We expect a high degree of alignment between these two measures, but not perfect alignment. Beyond the traditional concern of misreporting by the respondents, misalignment may occur for three other possible reasons: (1) human error in the matching process; (2) the publicly reported salary was for a different fiscal- or calendar-year than what the respondent was reporting or did not in fact reflect the individuals' full annual salary; or (3) we solicited salaries in the survey using a pre-populated menu of discrete values to minimize effort, which in turn creates ``lumpiness'' in the self-reported values.

Figure \ref{fig_salcompare_pub_slf} reports a scatterplot of respondents self- and publicly-reported salaries, which have a correlation of approximately 0.75. Overall, the figure indicates a high degree of alignment between the two measures, suggesting that respondents are answering this question truthfully. Given the traditional sensitivity surrounding salaries amongst this population, especially relative to the questions asked in our thought experiments, this alignment suggests that most respondents were in fact attentive and truthful.

%----------------------------------------------------------------------------------------%
\subsection{Narrow and broad fields}\label{sec_app_extramethods_fields}
%----------------------------------------------------------------------------------------%
\begin{table}[htbp]\centering\small
\caption{Field composition of respondents}
\label{tab_field_sumstat}
{
\def\sym#1{\ifmmode^{#1}\else\(^{#1}\)\fi}
\begin{tabular}{lc*{1}{rcccc}}
\hline\hline
                    &    type    &       count&        mean&          sd\\
\hline
\\ \underline{\emph{humanities \& related}}&            &            &            &            \\
communication       &  $ [0,1]$  &       4,357&        0.03&        0.17\\
education           &  $ [0,1]$  &       4,357&        0.04&        0.20\\
humanities          &  $ [0,1]$  &       4,357&        0.11&        0.31\\
law                 &  $ [0,1]$  &       4,357&        0.02&        0.13\\
\\ \underline{\emph{medicine \& health}}&            &            &            &            \\
medical school      &  $ [0,1]$  &       4,357&        0.21&        0.41\\
medicine and health &  $ [0,1]$  &       4,357&        0.07&        0.26\\
\\ \underline{\emph{natural}}&            &            &            &            \\
agriculture         &  $ [0,1]$  &       4,357&        0.02&        0.15\\
biology             &  $ [0,1]$  &       4,357&        0.05&        0.22\\
chemistry           &  $ [0,1]$  &       4,357&        0.02&        0.14\\
engineering         &  $ [0,1]$  &       4,357&        0.06&        0.24\\
geography           &  $ [0,1]$  &       4,357&        0.05&        0.21\\
physics             &  $ [0,1]$  &       4,357&        0.04&        0.19\\
\\ \underline{\emph{social \& math}}&            &            &            &            \\
business            &  $ [0,1]$  &       4,357&        0.05&        0.21\\
computer science    &  $ [0,1]$  &       4,357&        0.02&        0.13\\
economics           &  $ [0,1]$  &       4,357&        0.02&        0.15\\
mathematics         &  $ [0,1]$  &       4,357&        0.03&        0.18\\
other social sciences&  $ [0,1]$  &       4,357&        0.04&        0.19\\
political science   &  $ [0,1]$  &       4,357&        0.04&        0.19\\
psychology          &  $ [0,1]$  &       4,357&        0.04&        0.20\\
sociology           &  $ [0,1]$  &       4,357&        0.04&        0.19\\
\hline\hline
\end{tabular}
}

\end{table}

%----------------------------------------------------------------------------------------%
\subsection{Match to grant and publication database}\label{sec_app_extramethods_dimensions}
%----------------------------------------------------------------------------------------%
Professors in the full population data were matched to their corresponding records in the Dimensions grant and publication database (\citealt{dimension2018data}) using a iterative, fuzzy-merge process based on their name (e.g., transformations and abbreviations thereof) and institution. To minimize false positives, only matches where one record in our population data were matched to one record in the Dimensions data were used. Overall, roughly 78\% of observations are successfully matched.

Table \ref{tab_ttest_samp_indim} reports $t$ tests using a select set of covariates observable for the full sample to test for differences between the full sample and those professors merged to Dimensions. Clearly, it is not a random sub-sample. Overall, matched professors seem to be more experienced, perform more research, and have higher earnings. This is consistent with the observation (based on our exploration of the Dimensions database) that Dimensions' dismabiguation and record-making processes appear to be positively correlated with research output.

\begin{table}[htbp]\centering\small
\caption{Comparison of sub-sample matched to grants and publication data}
\label{tab_ttest_samp_indim}
{
\def\sym#1{\ifmmode^{#1}\else\(^{#1}\)\fi}
\begin{tabular}{l*{1}{ccc}}
\hline\hline
                    &\multicolumn{2}{c}{\underline{mean}}&                     \\
                    &     matched&  un-matched&       diff.   \\
\hline
assistant           &        0.25&        0.24&        0.00   \\
associate           &        0.24&        0.30&       --0.06***\\
full                &        0.43&        0.29&        0.14***\\
adjunct, clinical, other&        0.08&        0.16&       --0.08***\\
engineering, math \& related sciences&        0.18&        0.14&        0.03***\\
humanities \& related sciences&        0.16&        0.28&       --0.13***\\
medicine \& health sciences&        0.28&        0.28&        0.00   \\
social sciences     &        0.17&        0.10&        0.07***\\
natural sciences    &        0.21&        0.19&        0.02   \\
not on tenure track &        0.19&        0.32&       --0.12***\\
pre-tenure          &        0.20&        0.22&       --0.03*  \\
tenured             &        0.61&        0.46&        0.15***\\
work hours per week &       49.10&       49.25&       --0.15   \\
work-hrs. share, research&        0.39&        0.31&        0.09***\\
work-hrs. share, fundraising&        0.09&        0.07&        0.02***\\
work-hrs. share, teaching&        0.26&        0.33&       --0.08***\\
work-hrs. share, administration&        0.15&        0.17&       --0.01** \\
work-hrs. share, clinical&        0.04&        0.05&       --0.01** \\
work-hrs. share, other&        0.07&        0.07&       --0.00   \\
own annual earnings &  159,882.10&  132,063.87&   27,818.23***\\
earnings share, base salary&        0.70&        0.71&       --0.01   \\
earnings share, grant-sponsored&        0.16&        0.09&        0.07***\\
earnings share, supplemental&        0.03&        0.03&       --0.01** \\
earnings share, other&        0.04&        0.04&       --0.00   \\
earnings share, clinical&        0.02&        0.01&        0.00   \\
5-year guaranteed research funding&  450,408.10&  236,599.14&  213,808.96***\\
5-year fundraising expectations&  564,527.66&  342,380.84&  222,146.82***\\
\hline
$N$ obs. &        3,308 &        1,049 & \\
\hline\hline
\end{tabular}
}

% \begin{quote}\footnotesize
% \emph{Note}: XXX
% \end{quote}
\end{table}

\clearpage

%%%%%%%%%%%%%%%%%%%%%%%%%%%%%%%%%%%
%----------------------------------------------------------------------------------------%
\clearpage
\fancypagestyle{mystyle}{%
    \fancyhead{}
    \fancyhead[C]{Appendices}\fancyfoot{}
}%
\thispagestyle{mystyle}

\section{Additional summary statistics and results}\label{sec_app_extraresults}
%----------------------------------------------------------------------------------------%
\setcounter{figure}{0}
\renewcommand{\thefigure}{B\arabic{figure}}
\setcounter{table}{0}
\renewcommand{\thetable}{B\arabic{table}}
\setcounter{equation}{0}
\renewcommand{\theequation}{B\arabic{equation}}
%%%%%%%%%%%%%%%%%%%%%%%%%%%%%%%%%%%

%%%%%%%%%%%%%%%%%%%%%%%%%%%%%%%%%%%
%%%%%%%%%%%%%%%%%%%%%%%%%%%%%%%%%%%
\begin{table}[htbp]\centering\small
\caption{Full-survey one-dimensional PCA, by field}
\label{tab_aggregationpca}
{
\def\sym#1{\ifmmode^{#1}\else\(^{#1}\)\fi}
\begin{tabular}{lc*{1}{rcccc}}
\hline\hline
                   field &aggregate field&       count&        mean&          sd\\
\hline
law                 &social sciences&          71&        0.64&        2.87\\
political science   &social sciences&         163&        0.56&        1.94\\
business            &social sciences&         198&        0.29&        2.43\\
economics           &social sciences&          99&        0.27&        2.20\\
psychology          &social sciences&         177&        0.26&        1.83\\
medical school      &natural sciences&         858&        0.20&        2.39\\
chemistry           &medicine \& health sciences&          86&        0.19&        2.17\\
physics             &natural sciences&         159&        0.16&        2.08\\
biology             &natural sciences&         211&        0.12&        2.36\\
geography           &natural sciences&         202&        0.07&        2.02\\
sociology           &social sciences&         155&       -0.00&        2.03\\
medicine and health &medicine \& health sciences&         292&       -0.03&        2.12\\
other social sciences&humanities \& related sciences&         157&       -0.11&        1.88\\
engineering         &humanities \& related sciences&         261&       -0.31&        2.56\\
education           &humanities \& related sciences&         170&       -0.31&        2.07\\
agriculture         &engineering, math \& related sciences&         103&       -0.32&        1.86\\
humanities          &engineering, math \& related sciences&         456&       -0.33&        2.13\\
communication       &engineering, math \& related sciences&         117&       -0.34&        2.16\\
computer science    &engineering, math \& related sciences&          76&       -0.54&        2.74\\
mathematics         &engineering, math \& related sciences&         145&       -0.69&        2.41\\
\hline\hline
\end{tabular}
}

\begin{quote}\footnotesize
\emph{Note}: Reports one-dimensional PCA score using all relevant survey questions, averaged at the field level, alongside the aggregate field assigned.
\end{quote}
\end{table}

%%%%%%%%%%%%%%%%%%%%%%%%%%%%%%%%%%%
%%%%%%%%%%%%%%%%%%%%%%%%%%%%%%%%%%%
\begin{figure}[htbp!]\centering\footnotesize
\caption{Distributions of fields, institutions, and ranks}
\label{fig_disinstrank_hists}
\subfloat[Field composition of institutions]{
\includegraphics[width=0.65\linewidth, trim = 0mm 0mm 0mm 15mm , clip]{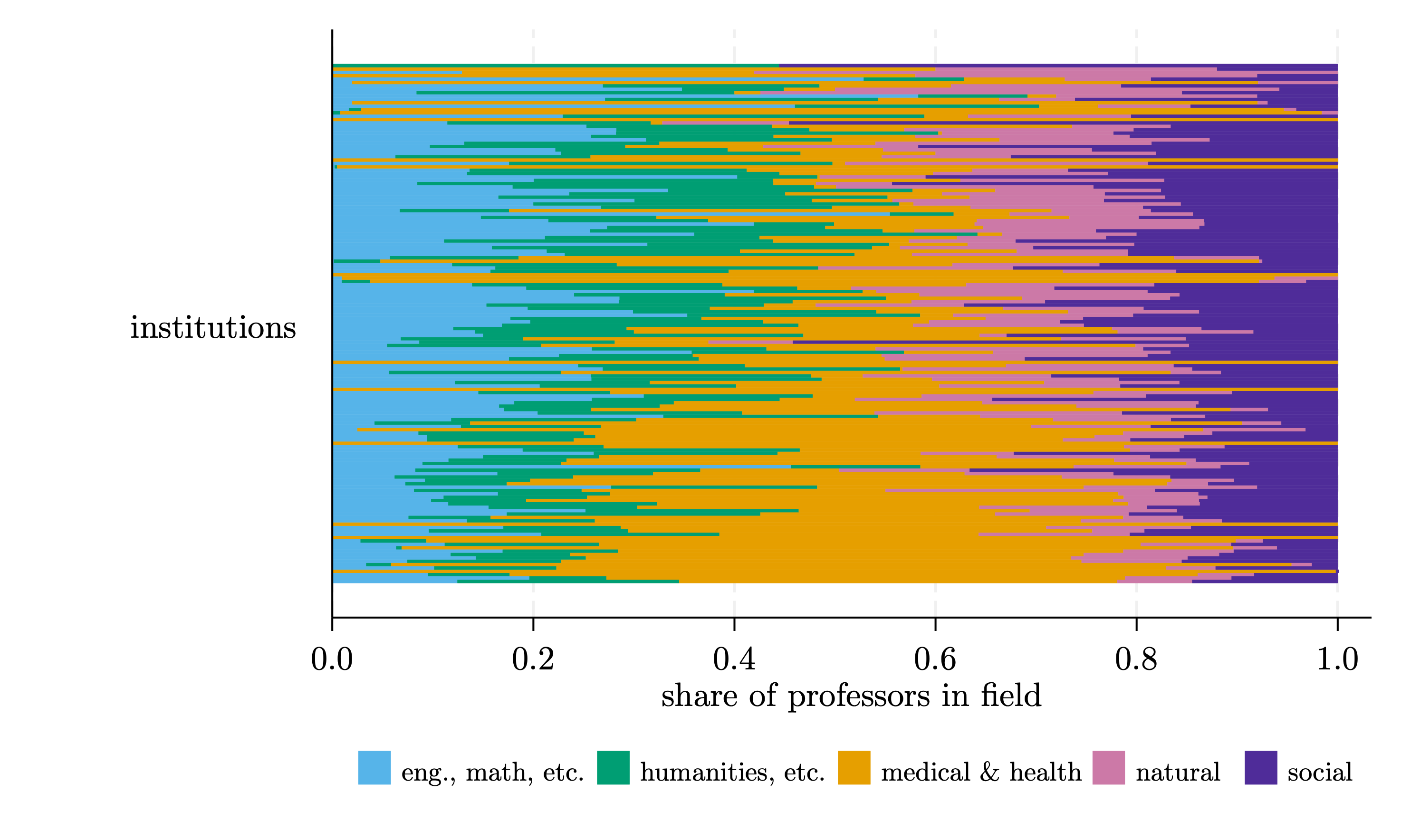}}\\
\subfloat[Rank composition of institutions]{
\includegraphics[width=0.65\linewidth, trim = 0mm 0mm 0mm 15mm , clip]{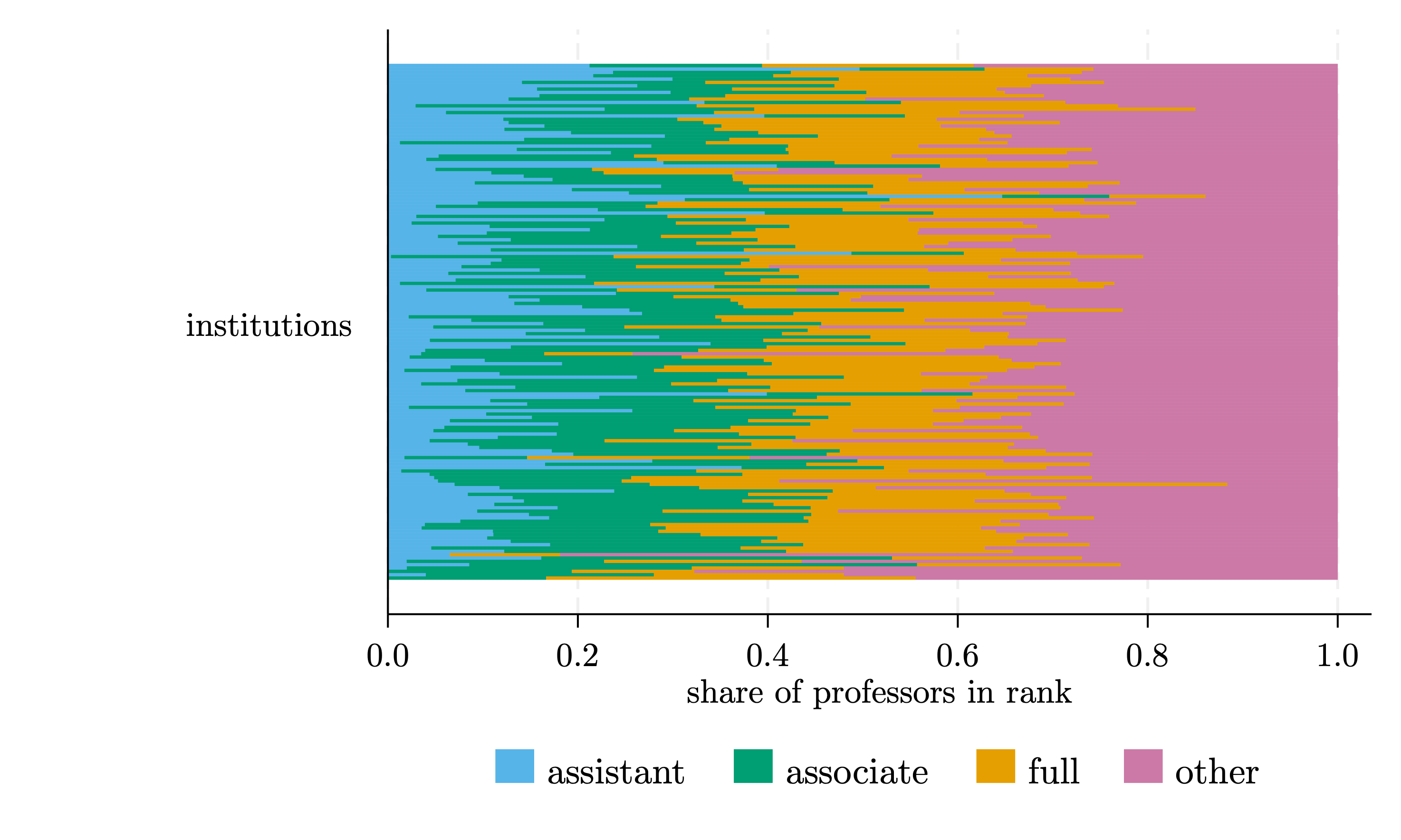}}\\
\subfloat[Rank composition of fields]{
\includegraphics[width=0.65\linewidth, trim = 0mm 0mm 0mm 15mm , clip]{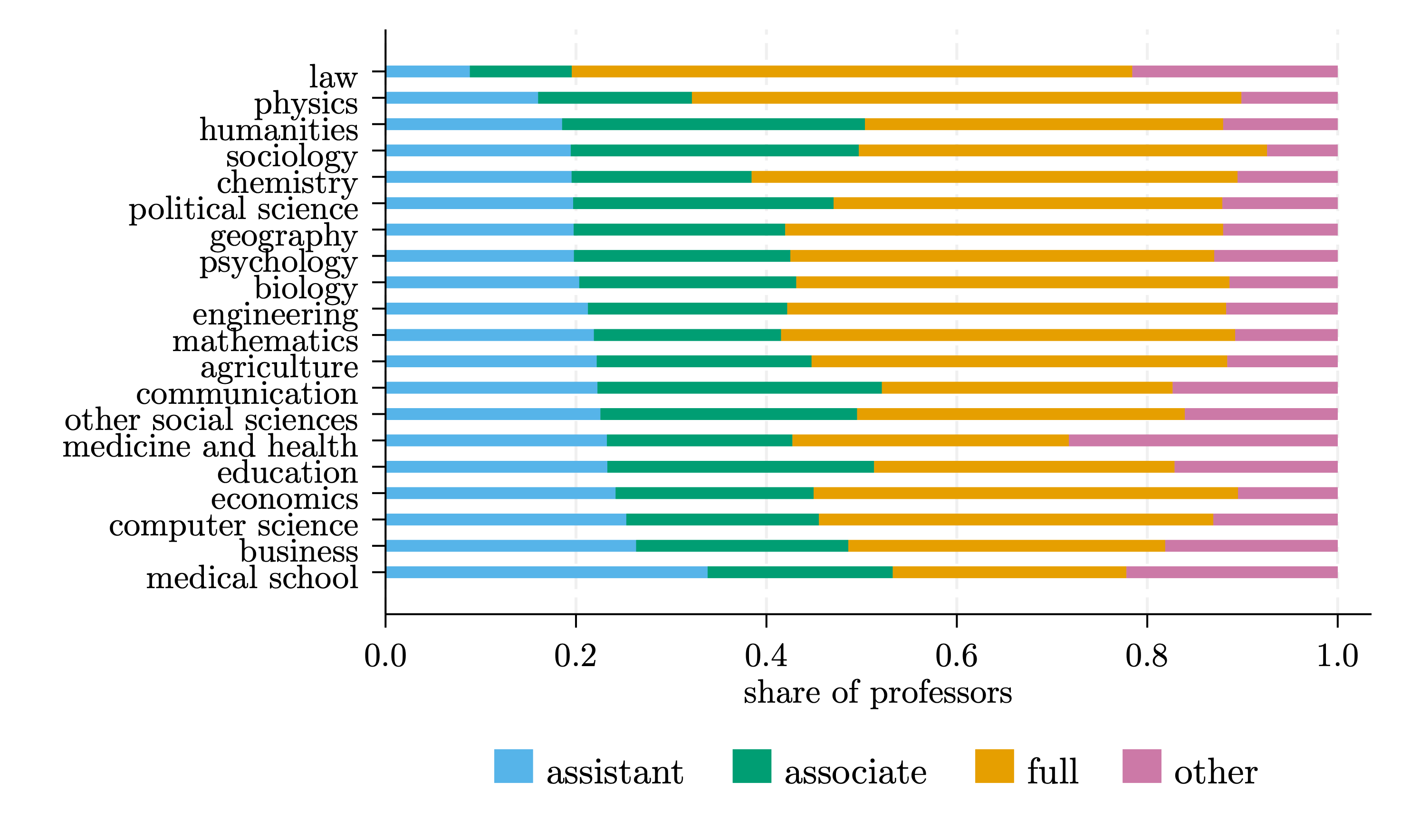}}
\begin{quote}
\emph{Note}: Panels (a--b) plot the share of each institution per field (Panel a) and rank (Panel b); institutions are not labeled, but are sorted by size in terms of number of professors. Panel (c) plots the share of professors in each field by rank.
\end{quote}
\end{figure}

%%%%%%%%%%%%%%%%%%%%%%%%%%%%%%%%%%%
%%%%%%%%%%%%%%%%%%%%%%%%%%%%%%%%%%%
\begin{figure}[htbp!]\centering\footnotesize
\caption{Additional earnings inequality results}
\label{fig_incomegini_extra}
\subfloat[Own earnings inequality, by field]{
\includegraphics[width=0.49\linewidth, trim = 0mm 0mm 0mm 0mm , clip]{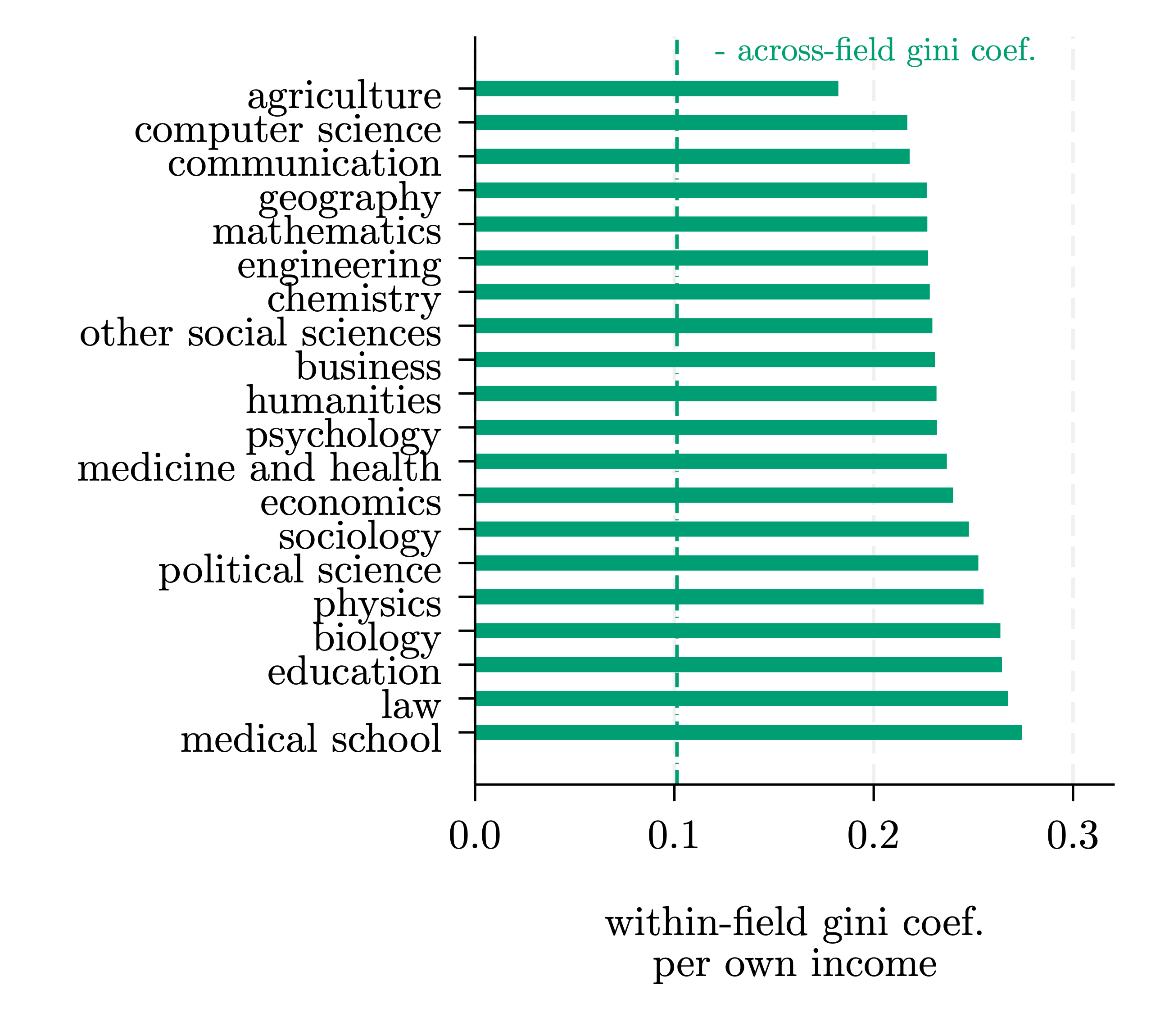}}
\subfloat[Household earnings inequality, by field]{
\includegraphics[width=0.49\linewidth, trim = 0mm 0mm 0mm 0mm , clip]{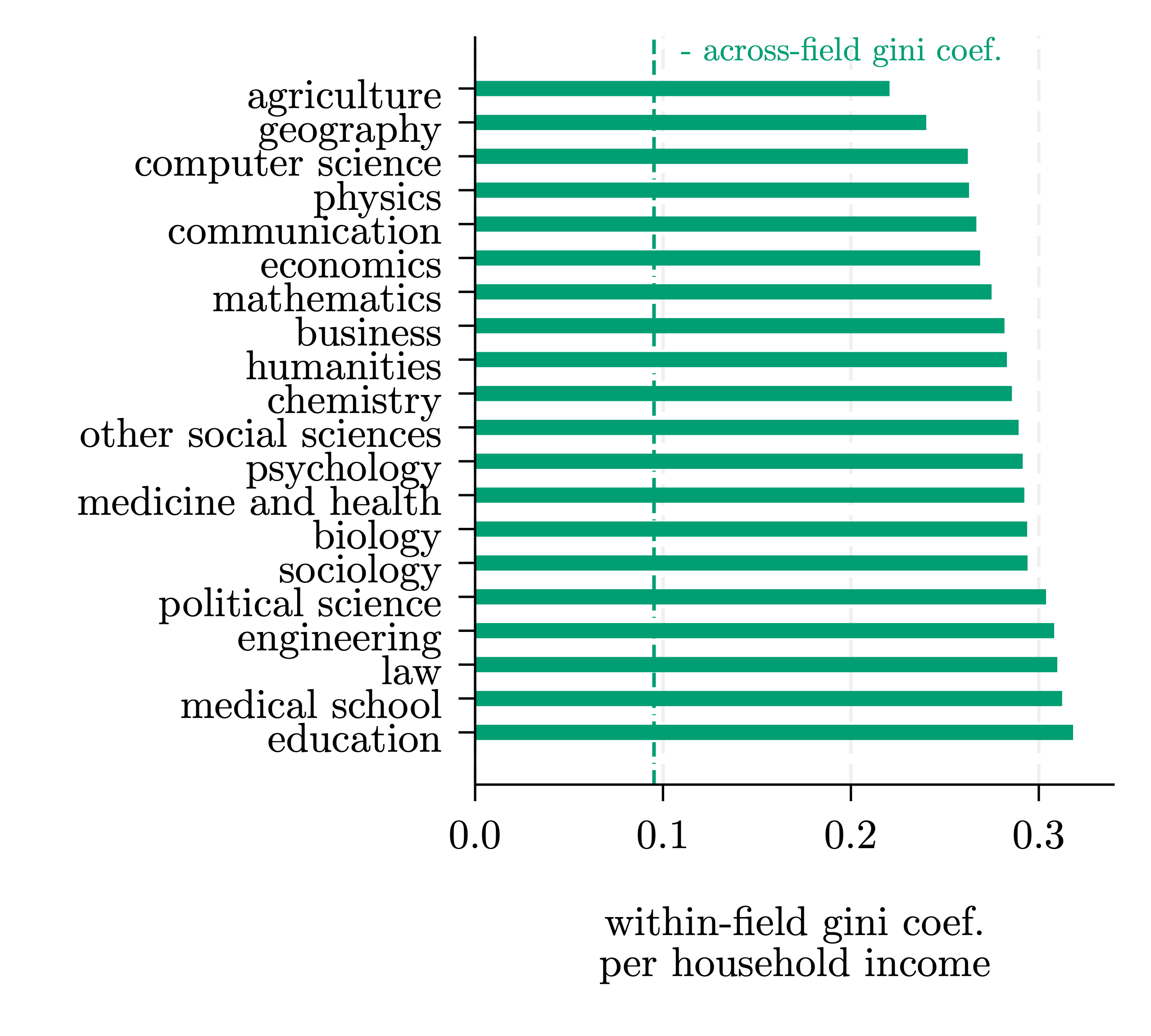}}
\\
\-\\ \-\\
\subfloat[Field level earnings and inequality]{
\includegraphics[width=0.49\linewidth, trim = 0mm 0mm 0mm 0mm , clip]{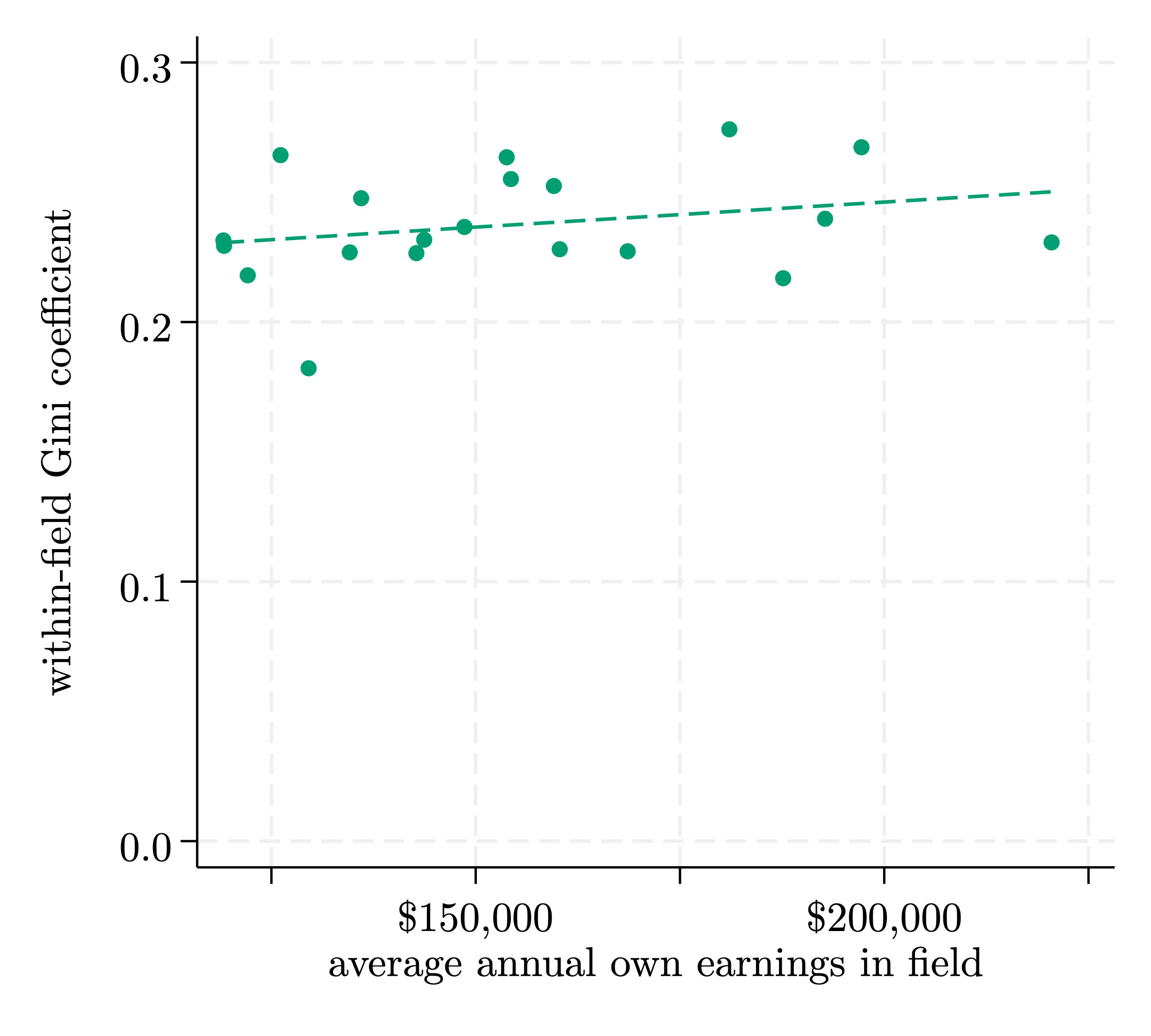}}
\begin{quote}
\emph{Note}: Panels (a--b) plot the within-field Gini coefficients using either own (a) or household (b) earnings. Panel (c) is a scatterplot of within-field Gini coefficients per average own earnings in the same field, showing no significant relationship between the two (i.e., inequality within a field is not correlated with the average earnings in the field).
\end{quote}
\end{figure}

%%%%%%%%%%%%%%%%%%%%%%%%%%%%%%%%%%%
%%%%%%%%%%%%%%%%%%%%%%%%%%%%%%%%%%%
\begin{figure}[htbp!]\centering\footnotesize
\caption{Tasks and earnings sources, by field}
\label{figs_app_extraresults_shrfields}
\subfloat[Work tasks, share of total]{
\includegraphics[width=0.65\linewidth, trim = 0mm 0mm 0mm 0mm , clip]{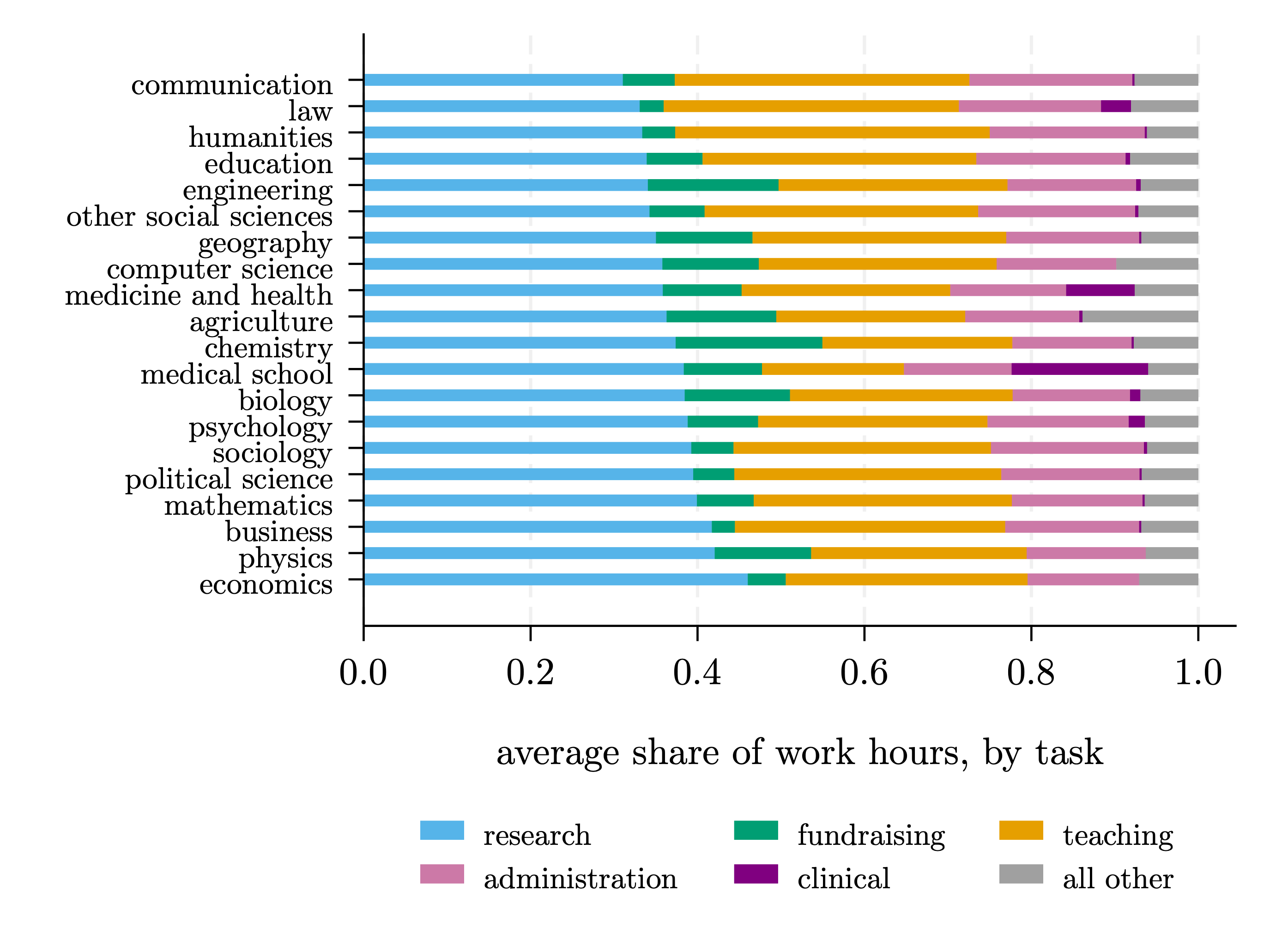}}\\ \-\\
\subfloat[Source of earnings, share of total]{
\includegraphics[width=0.65\linewidth, trim = 0mm 0mm 0mm 0mm , clip]{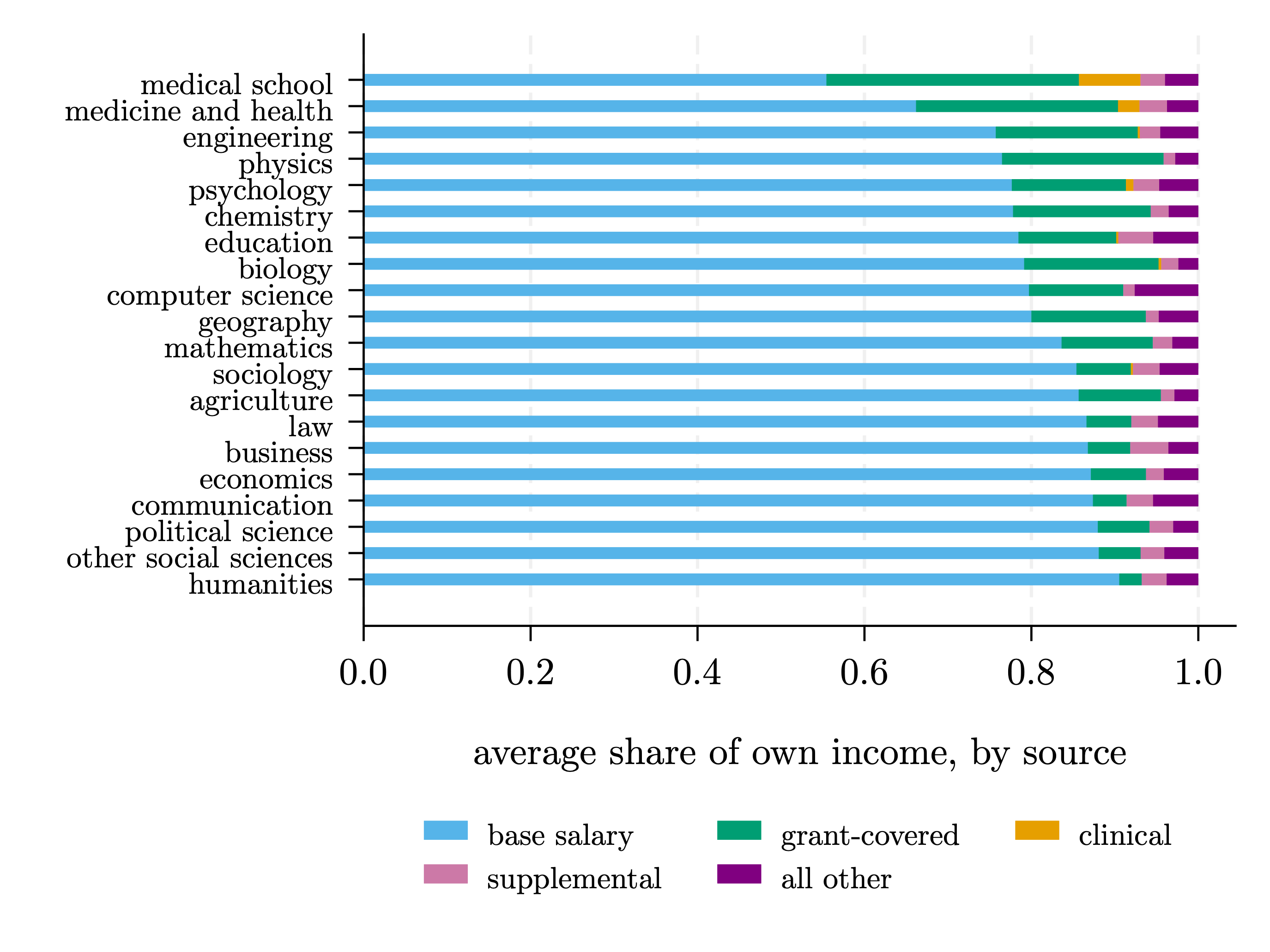}}\\ 
\begin{quote}
\emph{Note}: Plots the average share of professors' total work hours spent on each task (Panel a) and the share of their total annual earnings per each source (Panel b) averaged at the field level.
\end{quote}
\end{figure}

%%%%%%%%%%%%%%%%%%%%%%%%%%%%%%%%%%%
%%%%%%%%%%%%%%%%%%%%%%%%%%%%%%%%%%%
\clearpage
\begin{table}[htbp]\centering\small
\caption{Time use correlations}
\label{tab_pwcorr_hrstot_uncond_fullsamp}
{
\def\sym#1{\ifmmode^{#1}\else\(^{#1}\)\fi}
\begin{tabular}{l*{6}{c}}
\hline\hline
                &\multicolumn{6}{c}{}                                                                                             \\
                & research         &fundraising         & teaching         &administ.         & clinical         &    other         \\
\hline
research        &     1         &                  &                  &                  &                  &                  \\
fundraising     &     0.08\sym{***}&     1         &                  &                  &                  &                  \\
teaching        &    --0.23\sym{***}&    --0.13\sym{***}&     1         &                  &                  &                  \\
administ.       &    --0.27\sym{***}&    --0.07\sym{***}&     0.04\sym{**} &     1         &                  &                  \\
clinical        &    --0.24\sym{***}&    --0.15\sym{***}&    --0.19\sym{***}&    --0.07\sym{***}&     1         &                  \\
other           &    --0.11\sym{***}&     0.02         &    --0.00         &    --0.05\sym{***}&    --0.09\sym{***}&     1         \\
\hline\hline
\end{tabular}
}

\begin{quote}\footnotesize
\emph{Note}: Reports the pairwise unconditional correlations in hours worked per week on each task category.
\end{quote}
\end{table}

%%%%%%%%%%%%%%%%%%%%%%%%%%%%%%%%%%%
%%%%%%%%%%%%%%%%%%%%%%%%%%%%%%%%%%%
\begin{figure}[htbp!]\centering\footnotesize
\caption{Fundraising productivity and time allocations}
\label{figs_output_f}
\subfloat[Correlation between hourly and annual output]{
\includegraphics[width=0.475\linewidth, trim = 0mm 0mm 0mm 0mm , clip]{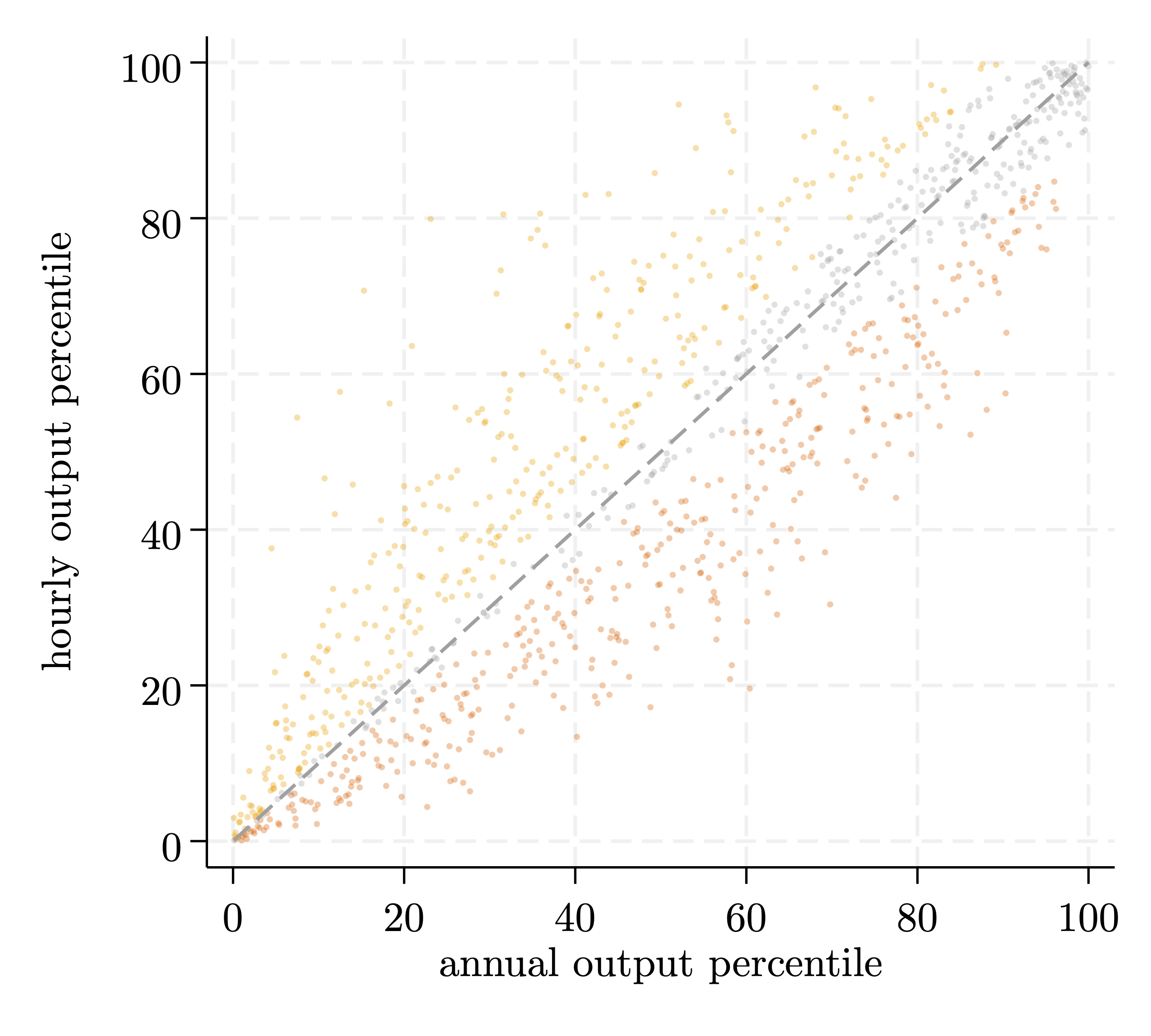}}
\subfloat[Ratio of hourly and annual output]{
\includegraphics[width=0.475\linewidth, trim = 0mm 0mm 0mm 0mm , clip]{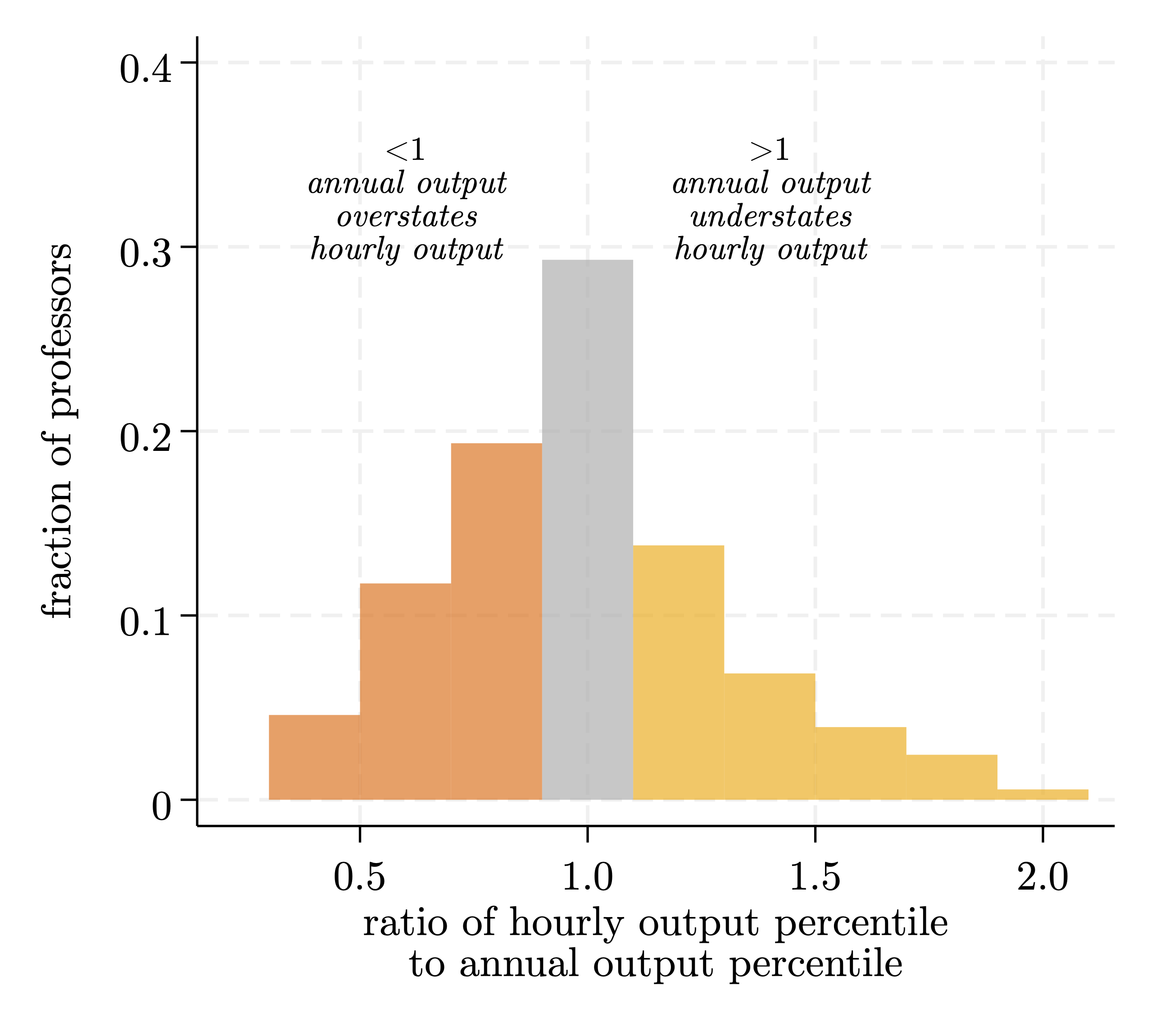}}\\
\begin{quote}
\emph{Note}: \input{figtab/stub_fig_output_ppratio_f.tex}
\end{quote}
\end{figure}

%%%%%%%%%%%%%%%%%%%%%%%%%%%%%%%%%%%
%%%%%%%%%%%%%%%%%%%%%%%%%%%%%%%%%%%
\begin{figure}[htbp!]\centering\footnotesize
\caption{Distribution of research risk beliefs}
\label{fig_hist_qriskresavg}
\includegraphics[width=0.6\linewidth, trim = 0mm 0mm 0mm 0mm , clip]{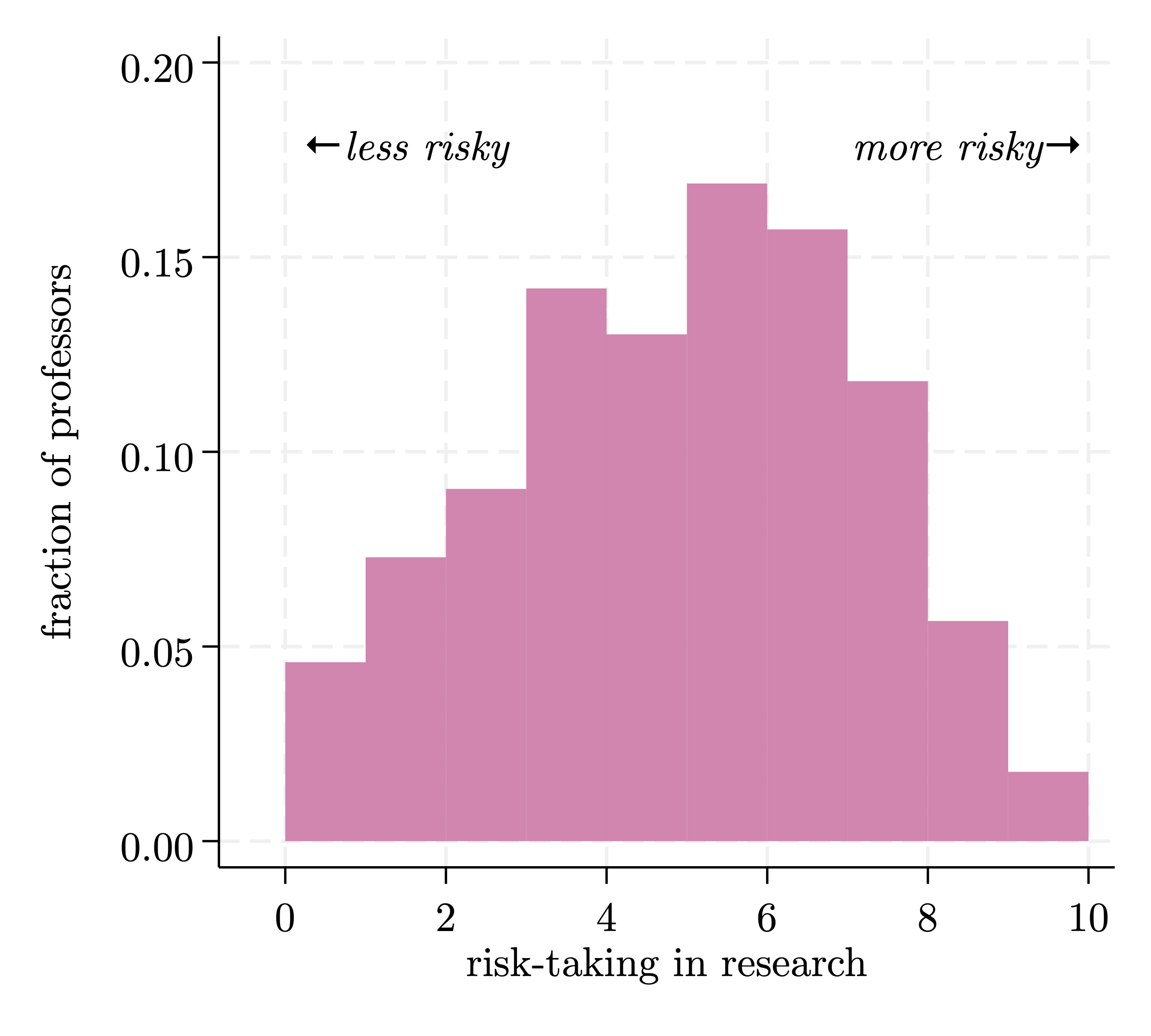}\\ 
\begin{quote}
\emph{Note}: Histogram of the average response to the two questions about researchers' own beliefs and second-order beliefs about peers' beliefs as to the riskiness of their research on a scale from 0 to 10.
\end{quote}
\end{figure}

\end{document}